\documentclass[
  aps,
  prx,
  reprint,
  superscriptaddress,
  amsfonts,amssymb,amsmath
]{revtex4-2}

\usepackage{newtxtext}
\usepackage{bm}
\usepackage{latexsym}
\usepackage{xcolor}
\usepackage{physics}
\usepackage{graphicx}
\usepackage{hyperref}
\hypersetup{
  setpagesize  = false,
  colorlinks   = true,
  urlcolor     = blue,
  linkcolor    = blue,
  citecolor    = blue
}

\begin{document}

\title{Stabilizer R\'{e}nyi Entropy and Conformal Field Theory}

\author{Masahiro Hoshino}
\email{hoshino-masahiro921@g.ecc.u-tokyo.ac.jp}
\affiliation{Department of Physics, The University of Tokyo, 7-3-1 Hongo, Bunkyo-ku, Tokyo 113-0033, Japan}

\author{Masaki Oshikawa}
\affiliation{Institute for Solid State Physics, University of Tokyo, Kashiwa, Chiba 277-8581, Japan}
\affiliation{Kavli Institute for the Physics and Mathematics of the Universe (WPI), University of Tokyo, Kashiwa, Chiba 277-8581, Japan}

\author{Yuto Ashida}
\affiliation{Department of Physics, The University of Tokyo, 7-3-1 Hongo, Bunkyo-ku, Tokyo 113-0033, Japan}
\affiliation{Institute for Physics of Intelligence, University of Tokyo, 7-3-1 Hongo, Bunkyo-ku, Tokyo 113-0033, Japan}

\date{\today}

\begin{abstract}
  Understanding universal aspects of many-body systems is one of the central themes in modern physics.
  Recently, the stabilizer R\'{e}nyi entropy (SRE) has emerged as a computationally tractable measure of nonstabilizerness, a crucial resource for fault-tolerant universal quantum computation.
  While numerical results suggested that the SRE in critical states can exhibit universal behavior, its comprehensive theoretical understanding has remained elusive.
  In this work, we develop a field-theoretical framework for the SRE in a $(1+1)$-dimensional many-body system and elucidate its universal aspects using boundary conformal field theory.
  We demonstrate that the SRE is equivalent to a participation entropy in the Bell basis of a doubled Hilbert space, which can be calculated from the partition function of a replicated field theory with the interlayer line defect created by the Bell-state measurements.
  This identification allows us to characterize the universal contributions to the SRE on the basis of the data of conformal boundary conditions imposed on the replicated theory.
  We find that the SRE of the entire system contains a universal size-independent term determined by the noninteger ground-state degeneracy known as the $g$-factor.
  In contrast, we show that the mutual SRE exhibits the logarithmic scaling with a universal coefficient given by the scaling dimension of a boundary-condition-changing operator, which elucidates the origin of universality previously observed in numerical results.
  As a concrete demonstration, we present a detailed analysis of the Ising criticality, where we analytically derive the universal quantities at arbitrary R\'{e}nyi indices and numerically validate them with high accuracy by employing tensor-network methods.
  These results establish a field-theoretical approach to understanding the universal features of nonstabilizerness in quantum many-body systems.
\end{abstract}

\maketitle


\section{Introduction}

\subsection{Background}

\begin{figure*}[tb]
  \centering
  \includegraphics[width=\linewidth, clip]{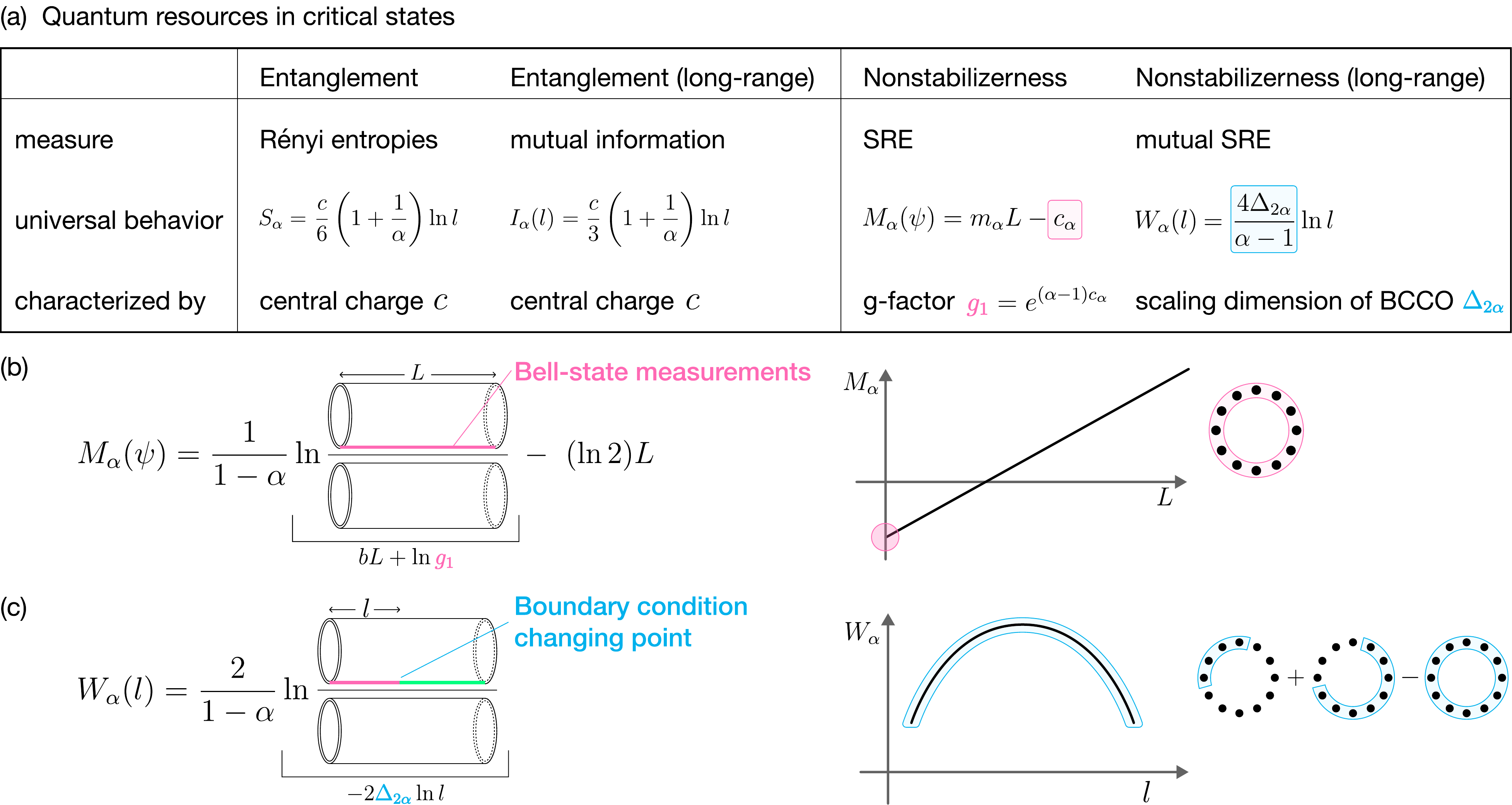}
  \caption{Summary of the main results.
    (a)~Comparison of the long-distance behavior of entanglement and nonstabilizerness in critical states.
    The universal feature of entanglement manifests itself as the logarithmic scaling with the coefficient given by the central charge $c$ of the corresponding CFT.
    In contrast, we show that the universality of nonstabilizerness is encoded in the size-independent term $c_\alpha$ in the SRE $M_{\alpha}$ and the coefficient of the logarithmic scaling in the mutual SRE $W_\alpha$, which are given by the $g$-factor $g_1$ and the scaling dimension $\Delta_{2\alpha}$ of a boundary-condition-changing operator (BCCO) $\mathcal{B}_{2\alpha}(x)$, respectively.
    Here, the R\'{e}nyi index is denoted by $\alpha$.
    (b)~From a field-theoretical perspective, the SRE of the entire system is obtained by the ratio between the partition functions with and without the measurement-induced interlayer line defect in the $2\alpha$-fold replicated field theory.
    The universal term $c_\alpha$ in the SRE can then be determined by the $g$-factor $g_1$ of the corresponding conformal boundary state.
    (c)~The long-distance behavior of the mutual SRE is governed by the partition function with two different boundary conditions (indicated by the red and green boundaries).
    This quantity can be expressed as a two-point correlation function $\langle\mathcal{B}_{2\alpha}(l)\mathcal{B}_{2\alpha}(0)\rangle$ of the corresponding BCCO.
    The resulting logarithmic scaling behavior is characterized by the universal coefficient determined by the scaling dimension $\Delta_{2\alpha}$ of $\mathcal{B}_{2\alpha}(x)$.}
  \label{fig:overview}
\end{figure*}

Understanding universality is a fundamental theme in physics, aiming to identify phenomena that emerge across different systems regardless of their microscopic details.
While universal behavior has traditionally been characterized by correlation functions and order parameters, quantum information quantities---most notably, entanglement entropy---have become powerful tools for understanding quantum many-body systems~\cite{amico2008entanglement,zeng2019quantum}.
A paradigmatic example is a one-dimensional critical state, where the entanglement entropy $S$ scales logarithmically with subsystem size $l$~\cite{holzhey1994geometric,calabrese2004entanglement,calabrese2009entanglement}:
\begin{equation}\label{CCscaling}
  S \sim \frac{c}{3}\ln l.
\end{equation}
This universal scaling is governed by the central charge $c$ of the underlying conformal field theory (CFT)~\cite{difrancesco1997conformal}, reflecting the universality class of a critical state.
Beyond its theoretical significance, the universal scaling relation can serve as a benchmark method for validating numerical approaches to quantum many-body systems. For instance,
by comparing numerical results with the analytical predictions, one can assess the accuracy of numerical methods such as tensor-networks~\cite{schollwock2011densitymatrix} and quantum Monte Carlo samplings~\cite{hastings2010measuring}.
The relation~\eqref{CCscaling} also provides a practical method to numerically extract the central charge of the critical states.

Besides entanglement, in quantum computation, it has been widely recognized that there exists yet another crucial quantum resource known as nonstabilizerness or magic, which quantifies how far a quantum state deviates from stabilizer states.
According to the Gottesman-Knill theorem~\cite{gottesman1998heisenberg,nielsen2010quantum}, stabilizer states---defined as common eigenstates of Pauli strings---can be efficiently simulated on classical computers.
Consequently, nonstabilizerness, together with entanglement, constitutes a necessary resource for achieving quantum advantage in computational tasks~\cite{horodecki2009quantum,campbell2017roads,chitambar2019quantum}.
An understanding of nonstabilizerness in many-body systems has direct practical implications for determining state preparation costs~\cite{gottesman1997stabilizer}, improving magic state distillation efficiency~\cite{gottesman1999demonstrating,bravyi2005universal,bravyi2012magicstate,litinski2019magic}, and assessing computational complexity in many-body simulations~\cite{zhang2025classical,lami2024quantum}.
Critical states are especially valuable in this context, as they are known to serve as substantial quantum resources in terms of, e.g., entanglement~\cite{vidal2003entanglement,latorre2004ground} and quantum Fisher information~\cite{zanardi2008quantum}.
Since a critical state corresponds to a fixed point of renormalization group (RG) flows~\cite{wilson1971renormalization,wilson1971renormalizationa}, of particular interest is the universal long-distance behavior associated with its nonstabilizerness.

To study the universality of nonstabilizerness, a proper and computationally tractable measure is essential.
In the case of entanglement, the entanglement entropy serves as an appropriate measure, which can be analyzed within the field-theoretical framework and efficiently evaluated by using tensor-networks~\cite{orus2014practical}.
Until recently, however, a corresponding measure for nonstabilizerness was lacking; the robustness of magic~\cite{howard2017application,heinrich2019robustness,sarkar2020characterization,hamaguchi2024handbook} defined through a minimization procedure is not computationally scalable for larger system sizes, and the negativity of the discrete Wigner function (known as mana)~\cite{veitch2012negative,veitch2014resource} is restricted to systems with odd prime-dimensional local Hilbert space~\cite{white2021conformal,tarabunga2024critical,nystrom2024harvesting,mao2024magic,zhang2024quantum}.

In recent years, the stabilizer R\'{e}nyi entropy (SRE)~\cite{leone2022stabilizer} has emerged as a computable measure for nonstabilizerness.
Unlike robustness of magic or mana, the SRE can be calculated directly from Pauli-string expectation values and is applicable to many-body systems with arbitrary local Hilbert-space dimensions.
Owing to its computational tractability, a variety of efficient numerical methods for calculating the SRE have been developed, including tensor-network approaches~\cite{haug2023quantifying,haug2023stabilizer,lami2023nonstabilizerness,tarabunga2024nonstabilizernessa,kozic2025computing} and sampling methods~\cite{tarabunga2023manybody,liu2024nonequilibrium,tarabunga2024magic,collura2024quantum,ding2025evaluating,sinibaldi2025nonstabilizerness}.
These numerical methods have allowed for investigating nonstabilizerness in various many-body settings, such as frustrated systems~\cite{odavic2023complexity}, Sachdev-Ye-Kitaev models~\cite{bera2025nonstabilizerness,russomanno2025efficient,jasser2025stabilizer}, and dynamical setups~\cite{rattacaso2023stabilizer,ahmadi2024quantifying,smith2024nonstabilizerness,haug2024probing,tirrito2024anticoncentration,szombathy2025independent,odavic2025stabilizer,szombathy2025spectral,turkeshi2025magic}.
In particular, it has been numerically observed that the SRE can exhibit universal behavior reminiscent of entanglement entropy~\cite{oliviero2022magicstate,piemontese2023entanglement,tarabunga2023manybody,cao2024gravitational,passarelli2024nonstabilizerness,frau2024nonstabilizerness,ding2025evaluating,falcao2025nonstabilizerness}.

Despite these significant developments, a comprehensive theoretical understanding of the origin and precise nature of the universality in nonstabilizerness has so far remained elusive.
This gap naturally raises the following questions:
\begin{itemize}
  \item[(A)]{Does nonstabilizerness in critical states exhibit universal phenomena, and if so, in what manner?}
  \item[(B)]{How can one understand universal aspects of nonstabilizerness from a field-theoretical perspective?}
\end{itemize}
The present work addresses these questions by establishing a field-theoretical framework to analytically study the SRE in $(1+1)$-dimensional critical many-body states (see Fig.~\ref{fig:overview}).
We answer question (A) in the affirmative way by showing that the SRE exhibits universal behavior in two distinct ways; first, through a size-independent constant term $c_\alpha$ in the SRE $M_\alpha$ of the entire system, and second, through the coefficient $\Delta_{2\alpha}$ of a logarithmic scaling in the mutual SRE $W_\alpha$ (Fig.~\ref{fig:overview}(a)).

To address question (B), we demonstrate that the universal contributions to the SRE can be analyzed by boundary conformal field theory (BCFT)~\cite{cardy1989boundary,cardy2008boundary,recknagel2013boundary}.
Specifically, we express the SRE as a participation entropy in the Bell basis of a doubled Hilbert space, i.e., the classical entropy associated with the Born probability of Bell-state measurements acting on the doubled state.
The long-distance behavior of the SRE can thus be characterized by the partition function of a replicated effective field theory with the interlayer line defect created by the Bell-state measurements.
From this perspective, the universal size-independent term $c_\alpha$ is determined by the noninteger ground-state degeneracy known as the $g$-factor~\cite{affleck1991universal}  (Fig.~\ref{fig:overview}(b)).
Meanwhile, the universal coefficient $\Delta_{2\alpha}$ in the logarithmic scaling stems from the scaling dimension of a boundary-condition-changing operator (BCCO)~\cite{cardy1989boundary,affleck1997boundary} associated with the change of boundary conditions between the subsystems (Fig.~\ref{fig:overview}(c)).
We note that both the $g$-factor and the scaling dimensions of BCCOs are universal quantities, as they are characterized by the infrared (IR) limit of the theory, independent of microscopic details of the models.

To validate these results, we provide a detailed analysis of the SRE in the Ising criticality as a concrete example, where we use bosonization techniques to identify the conformal boundary conditions induced by the Bell-state measurements.
The same boundary condition is derived via another expression of the SRE called the magic entropy.
We explicitly construct the corresponding boundary states in the multicomponent $S^1/\mathbb{Z}_2$ free-boson CFT and analytically derive the universal contributions to the SRE.
These predictions have been numerically confirmed by our tensor-network calculations with high accuracy.
Altogether, our results establish a field-theoretical framework for systematically characterizing the universality of nonstabilizerness in quantum critical states.

From a broader perspective, our results are related to previous studies at the intersection of quantum many-body physics and quantum information science.
First, we note that BCFT has previously proven successful in calculating various information quantities, such as entanglement entropy~\cite{holzhey1994geometric,calabrese2004entanglement,calabrese2009entanglement}, mutual information~\cite{furukawa2009mutual}, ground-state fidelity~\cite{camposvenuti2009universal}, and participation entropies in the computational basis~\cite{fradkin2006entanglement,hsu2009universal,stephan2009shannon,hsu2010universal,stephan2010renyi,zaletel2011logarithmic,stephan2011phase,alcaraz2013universal,stephan2014emptiness,stephan2014shannon,luitz2014participation,alcaraz2014universal,alcaraz2016universal}.
In particular, the participation entropy quantifies the localization of a state in a given configuration basis, and its universal subleading term can be used to characterize different quantum phases of matter.
It has also been demonstrated that the entanglement entropy for the Rokhsar-Kivelson wave functions in $d$ dimensions is exactly the participation entropy of a $(d-1)$-dimensional ground state in the computational basis~\cite{luitz2014participation}.
In this context, the present work extends previous studies by examining the participation entropy in an \textit{entangled} basis, namely, the Bell basis, which, to the best of our knowledge, has not been explored before.

Second, more recently, BCFT has been applied to describe critical states under measurements and decoherence~\cite{garratt2023measurements,sun2023new,ma2023exploring,zou2023channeling,weinstein2023nonlocality,lee2023quantum,yang2023entanglement,murciano2023measurementaltered,sala2024quantum,ashida2024systemenvironment,patil2024highly,sala2024decoherence,milekhin2024observableprojected,tang2024critical,liu2024boundary,kuno2025systemenvironmental}.
It was pointed out in Ref.~\cite{garratt2023measurements} that the effects of measurements on a Gibbs state with a path-integral description can be expressed as boundary perturbations in the action, and its long-distance behavior can be determined by the boundary RG flow.
This formalism was extended in Ref.~\cite{zou2023channeling} to include local quantum channels such as on-site decoherence.
Furthermore, in Ref.~\cite{hoshino2024entanglement}, the BCFT has been used to analyze many-body entanglement swapping, i.e., entanglement induced by Bell-state measurements across initially independent spin chains.
Building on these developments, the present work extends the applicability of BCFT to nonstabilizerness---an essential resource for quantum computation besides entanglement---and reveals its universal aspects, which might have implications for quantum information processing and quantum simulation.

\subsection{Overview of the key results}
Before delving into the detailed theoretical formulation in the subsequent sections, we provide a nontechnical summary of our main results.
Our primary focus is on the universal behavior of nonstabilizerness in critical ground states of quantum spin chains with periodic boundary conditions.
The first key result pertains to the full-state SRE denoted as $M_\alpha(\psi)$, which basically quantifies the extent to which a state $\psi$ of the entire chain possesses a quantum resource necessary to achieve a computational advantage over classical states in certain algorithms.
We establish that, for an $L$-qubit critical state, the SRE generally follows the scaling form
\begin{equation}\label{eq:sre_general}
  M_\alpha(\psi) = m_\alpha L - c_\alpha + o(1).
\end{equation}
Here, the leading term $m_\alpha L$ is nonuniversal, as its coefficient $m_\alpha$ depends on microscopic details such as the choice of ultraviolet (UV) momentum cutoff.
In contrast, the size-independent term $c_\alpha$ is universal and encodes intrinsic properties of the associated low-energy effective field theory.
We show that this term directly relates to the noninteger ground-state degeneracy of the BCFT characterizing the SRE, playing an analogous role to the central charge $c$ in the bulk CFT.

To understand the emergence of the universal subleading term $c_\alpha$, we note that the SRE can be interpreted as the R\'{e}nyi entropy of the Born probability for Bell-state measurements acting on the doubled state~\cite{haug2023scalable}.
In the path-integral formalism, the influence of such measurements can be described as the boundary perturbations localized in the imaginary-time direction~\cite{garratt2023measurements}.
Thus, this perspective naturally leads to a field-theoretical description wherein the universal contribution to $M_\alpha(\psi)$ arises from the presence of a line defect induced by those boundary perturbations. Crucially, this defect is not a bulk property but rather characterizes the universality of the intricate conformal boundary conditions imposed on the ($2\alpha$-component) replicated field theory.
This insight allows us to leverage BCFT techniques to analytically determine the value of $c_\alpha$, demonstrating that it corresponds to the logarithm of the $g$-factor.

We further investigate the mutual SRE defined as
\begin{equation}
  W_\alpha(A:B) = M_\alpha(\rho_A) + M_\alpha(\rho_B) - M_\alpha(\rho_{AB}),
\end{equation}
which serves as a direct analogue of the R\'{e}nyi mutual information $I_\alpha$ constructed from the R\'{e}nyi entropy $S_\alpha$.
Here, $\rho_{AB}$ represents the density matrix of the composite quantum system consisting of subsystems $A$ and $B$, with the corresponding reduced density matrices given by $\rho_A=\Tr_B[\rho_{AB}]$ and $\rho_B=\Tr_A[\rho_{AB}]$.
While the precise physical interpretation of the mutual SRE $W_\alpha$ is still under discussion~\cite{frau2024stabilizer}, it can be understood as a quantifier of the degree to which nonstabilizerness is distributed across subsystems in a correlated manner.
Intuitively, it probes the extent to which nonstabilizerness is long-ranged, i.e., cannot be eliminated by a finite-depth local unitary circuit, in analogy with how the mutual information captures long-range quantum correlations.
The mutual SRE $W_\alpha$ provides a natural diagnostic tool for identifying universal scaling behavior in nonstabilizerness, as it eliminates leading nonuniversal contributions and isolates intrinsic long-distance features.

To establish the scaling behavior of the mutual SRE, we consider the case where $AB$ forms the entire quantum system $\psi$, with subsystems defined as $A = \{1, 2, \ldots, l\}$ and $B = \{l+1,l+2, \ldots, L\}$.
Our second key result is that, in the long-distance limit, the mutual SRE exhibits a universal logarithmic scaling of the form
\begin{equation}\label{Wuniv}
  W_\alpha(l) = \frac{4\Delta_{2\alpha}}{\alpha-1}\ln l_c,
\end{equation}
where $l_c = (L/\pi) \sin(\pi l/L)$ is the chord length associated with subsystem size $l$.
This logarithmic scaling is deeply rooted in the underlying CFT description, where the coefficient of the logarithm is determined by the scaling dimension $\Delta_{2\alpha}$ of a BCCO $\mathcal{B}_{2\alpha}(x)$ appearing in the replicated theory.
This BCCO characterizes the change in boundary conditions at the interface between subsystems $A$ and $B$, playing a central role in the universal properties of the mutual SRE.

The universal relation~\eqref{Wuniv} bears a strong resemblance to the scaling behavior of the mutual information $I_\alpha$ in critical systems, where the coefficient of the universal logarithmic scaling is given by the central charge $c$ of the underlying CFT.
There, the bulk CFT played the fundamental role in determining the universal properties of quantum entanglement.
However, a key distinction arises in the case of nonstabilizerness, where the universal logarithmic scaling of the mutual SRE is not solely governed by the bulk properties but rather by the BCFT data, i.e., the scaling dimension $\Delta_{2\alpha}$ of the BCCO.
From a field-theoretical perspective, this distinction stems from the fact that the mutual SRE is intrinsically tied to nontrivial interlayer boundary conditions within the replicated field theory.
The appearance of the scaling dimension $\Delta_{2\alpha}$ in the logarithmic scaling follows naturally from the fact that the partition function for $W_{\alpha}(l)$ can be expressed as the logarithm of $\langle\mathcal{B}_{2\alpha}(l)\mathcal{B}_{2\alpha}(0)\rangle$, which is a two-point correlation function of the BCCOs. Its asymptotic behavior is governed by the scaling dimension $\Delta_{2\alpha}$ and leads to the relation~\eqref{Wuniv}. This characterization provides a systematic and physically intuitive explanation for the universal behavior in the mutual SRE, which has been previously observed in numerics~\cite{tarabunga2023manybody,tarabunga2024critical}.

As a concrete demonstration of our field-theoretical framework, we investigate the universal aspects of nonstabilizerness in the Ising criticality with the central charge $c = 1/2$, which serves as a paradigmatic example of a CFT.
To facilitate an analytical treatment, we employ bosonization techniques to map the doubled Ising CFT onto a single free-boson CFT compactified on an orbifold $S^1/\mathbb{Z}_2$.
This mapping significantly simplifies the computation of universal quantities, as the Gaussian nature of the free-boson theory makes the replica-trick calculations tractable.
Within this formalism, we explicitly determine the $g$-factor associated with the conformal boundary condition imposed by the Bell-state measurements and identify the scaling dimension of the relevant BCCO.
Our results reveal the following universal expressions for the size-independent term in the SRE and the coefficient of the logarithmic scaling in the mutual SRE:
\begin{align}
  c_\alpha         & = \frac{\ln \sqrt{\alpha}}{\alpha-1}, \\
  \Delta_{2\alpha} & = \frac{1}{16}.
\end{align}
These analytical predictions are confirmed with high accuracy by our numerical calculations of the transverse-field Ising model (TFIM) at criticality. Specifically, we utilize tensor-network simulations based on matrix product states to evaluate the SRE directly in the TFIM ground state, while we implement a Monte Carlo sampling approach to verify the scaling behavior of the mutual SRE.
The agreement between theoretical predictions and numerical results provides compelling evidence for the validity of our theoretical framework.

These results establish a direct connection between nonstabilizerness and BCFT, paving the way for a systematic understanding of universal aspects of quantum resources in many-body systems beyond entanglement.
The universality of the scaling behavior suggests that, similar to entanglement, nonstabilizerness could serve as a diagnostic tool for characterizing quantum phases and criticality in many-body systems.
Our identification also indicates that the study of nonstabilizerness can provide detailed information about the intricate conformal boundary states realized by measurements.
For instance, the universal scaling gives a practical way to extract the value of the $g$-factor from numerical calculations of nonstabilizerness, in a similar manner as entanglement entropy allows for extracting the central charge.

Beyond theoretical implications, our results should also have computational and experimental relevance.
The field-theoretical framework developed here enables the analytical determination of the universal contributions to nonstabilizerness in quantum many-body systems, offering a new avenue for benchmarking numerical methods of the SRE.
Owing to the universal nature of our results, the present framework also provides a predictive tool for analyzing nonstabilizerness in a wide range of quantum critical states, including those realizable in programmable quantum platforms such as Rydberg atom arrays~\cite{bernien2017probing}, superconducting qubits~\cite{kjaergaard2020superconducting}, and trapped ions~\cite{monroe2021programmable}.
The requirements for measuring the SRE are the Bell-state measurements and a preparation of critical states~\cite{haug2024efficient}, both of which have been within reach of current experimental techniques \cite{bluvstein2022quantum,fang2024probing,evered2025probing,huang2025exact}.
We note that the Bell magic~\cite{haug2023scalable}, a related measure of nonstabilizerness to the SRE, has been measured in neutral-atom arrays using Bell-state measurements across two copies of the state~\cite{bluvstein2024logical}.
This approach requires neither postselections nor mid-circuit measurements.
Importantly, our numerical results suggest that the universal behaviors can be observed even in small system sizes with $10$-$20$ lattice sites, which is a consequence of conformal invariance of the critical states.

The remainder of this paper is organized as follows.
In Sec.~\ref{sec:sre-as-pe}, we establish a connection between the SRE and the participation entropy in the Bell basis of the doubled Hilbert space, which allows us to analyze its universal aspects using BCFT.
This connection lays the basis for the subsequent sections, where we give an understanding of the SRE from a field-theoretical perspective.
We also use the magic entropy as an alternative expression of the SRE, which is useful in supporting our results based on the participation entropy.
In Sec.~\ref{sec:sre-in-criticalstates}, we present a general BCFT framework for calculating the SRE in critical states.
We introduce a replicated field theory with boundary conditions imposed by the Bell-state measurements and elucidate how universal features of the SRE emerge from this formulation.
Section~\ref{sec:example} demonstrates our framework through a detailed analysis of the SRE for the Ising criticality.
We use bosonization techniques to map the Ising CFTs onto a free-boson CFT and identify the conformal boundary conditions induced by the Bell-state measurements.
By constructing boundary states in the multicomponent $S^1/\mathbb{Z}_2$ free-boson CFT, we analytically derive the universal contributions to the SRE and validate them through tensor-network calculations.
Finally, in Sec.~\ref{sec:summary}, we discuss the broader implications of our findings and outline potential future directions.

Readers primarily interested in the conceptual aspects and universal properties of the SRE might focus on Secs.~\ref{sec:sre-as-pe} and \ref{sec:sre-in-criticalstates}, where we present the key theoretical insights in a general manner.
Those seeking detailed derivations and explicit calculations will find a comprehensive analysis of the Ising criticality in Sec.~\ref{sec:example} and Appendixes, which contain a step-by-step demonstration of our field-theoretical framework.

\section{\label{sec:sre-as-pe}Stabilizer R\'{e}nyi Entropy as Participation Entropy and Magic Entropy}
In this section, we demonstrate how SREs can be understood as participation entropies in the Bell basis.
We begin by summarizing the definition and key properties of the SRE.
We then employ the Choi-Jamiolkowski isomorphism~\cite{choi1975completely,jamiolkowski1972linear} to represent the expectation values of Pauli strings as Born probabilities associated with Bell-state measurements in the doubled Hilbert space.
We note that the relation between nonstabilizerness and Bell-state measurements across the doubled spin chain was pointed out in Ref.~\cite{montanaro2017learning} and further developed in Refs.~\cite{lai2022learning,haug2023scalable,haug2024efficient,yano2024quantum}.
This identification allows us to calculate the SRE using BCFT in the subsequent sections.

\subsection{Preliminaries}
For an $L$-qubit pure state $\ket{\psi}$, the $\alpha$-SRE is defined as follows:
\begin{equation}\label{eq:sre}
  M_{\alpha}(\psi) = \frac{1}{1-\alpha} \ln\sum_{\vec{m}}\frac{\Tr^{2\alpha}[\sigma^{\vec{m}}\psi]}{2^L}.
\end{equation}
Here, $\psi=\ketbra{\psi}{\psi}$ represents the density matrix corresponding to $\ket{\psi}$, and the $L$-qubit Pauli string $\sigma^{\vec{m}}$ is given by
\begin{equation}\label{eq:paulistring}
  \sigma^{\vec{m}} = \bigotimes_{j=1}^{L} \sigma^{m_{2j-1}m_{2j}}\quad(\vec{m}\in\qty{0,1}^{2L}),
\end{equation}
where the Pauli matrices are labeled as
\begin{align}\label{eq:paulimatrix}
  \sigma^{00} & =\mqty(1  & 0 \\0&1) ,& \sigma^{10} &= \mqty(0&1\\1&0),\notag\\
  \sigma^{01} & = \mqty(1 & 0 \\0&-1),& \sigma^{11} &=\mqty(0&-i\\i&0).
\end{align}
The SREs can serve as measures of nonstabilizerness for pure states and possess several key properties for all $\alpha$: (i) faithfulness: $M_{\alpha}(\psi)=0$ if and only if $\ket{\psi}$ is a stabilizer state, (ii) stability under Clifford unitaries $C\in C_L$: $M_{\alpha}(C\psi C^\dag)=M_{\alpha}(\psi)$, and (iii) additivity: $M_{\alpha}(\psi\otimes\phi)=M_{\alpha}(\psi)+M_{\alpha}(\phi)$; we here recall that the Clifford group $C_L$ consists of unitary transformations preserving the Pauli group, $C\sigma^{\vec{m}}C^\dag=\sigma^{\vec{m}'}$ for $C\in C_L$, and stabilizer states are a set of states constructed by applying the Clifford unitaries to the product state $\ket{0}^{\otimes L}$.
Since the calculation of the SRE requires no minimization procedure, it can serve as a computationally tractable measure for many-body systems.

The crucial property of monotonicity, the nonincreasing nature under stabilizer protocols, depends on the value of $\alpha$.
It has been shown in Ref.~\cite{leone2024stabilizer} that SREs are stabilizer monotones for $\alpha\geq2$, but not for $\alpha<2$.
Moreover, a related measure called the linear stabilizer entropy, defined as $M_{\alpha}^{\mathrm{lin}}(\psi) = 1 - \sum_{\vec{m}}2^{-L}\Tr^{2\alpha}[\sigma^{\vec{m}}\psi]$, has been proven to be a strong stabilizer monotone for $\alpha\geq2$~\cite{leone2024stabilizer}.
In this sense, using the SRE and linear stabilizer entropy as measures of nonstabilizerness is well founded for $\alpha\geq2$.
We note that, unlike other entropy measures such as entanglement entropy which require analytic continuation from integer Rényi indices, the SRE is directly meaningful at integer values $\alpha\geq2$.

It is desirable for informational quantities to have operational interpretations.
For example, the entanglement entropy characterizes the efficiency of entanglement distillation protocols for pure states.
An operational meaning for the SRE has recently been established in the context of stabilizer property testing~\cite{bittel2025complete,bu2025stabilizera,bittel2025operational}.
Intuitively, the magnitude of the SRE represents the indistinguishability between the Clifford orbit of the state and Haar random states.

\begin{figure}[tb]
  \centering
  \includegraphics[width=0.7\linewidth, clip]{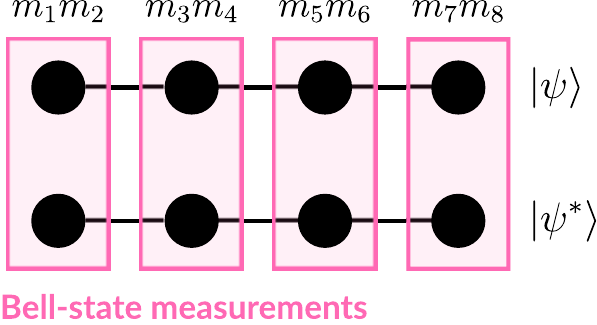}
  \caption{Schematic figure illustrating how the Bell basis is defined in the doubled spin chain. The SRE is calculated from the Born probability of the Bell-state measurements across the two chains. The measurement outcomes are labeled by $\vec{m}\in\qty{0,1}^{2L}$.}
  \label{fig:participation}
\end{figure}

\subsection{Participation entropy}
The SRE can be interpreted as the participation entropy in the Bell basis of the doubled spin chain.
This interpretation provides a powerful framework for analytical calculations: it enables us to systematically derive the line defect in the CFT partition function that is necessary to compute the universal terms in the SRE.
The essential insight of this approach is that expectation values of Pauli strings translate into projective measurements on the doubled system, which can then be rewritten as boundary perturbations in the replicated CFT.

In general, the participation entropy of a state $\ket*{\Psi}$ is defined as the classical entropy of the probability distribution $p_i=\abs{\braket{i}{\Psi}}^2$ in a certain orthonormal basis $\qty{\ket{i}}$.
The SRE of $\ket{\psi}$ corresponds to the case where $\ket{\Psi}=\ket{\psi}\!\ket{\psi^\ast}$ is the doubled state, and $\ket{i}$ is the Bell basis across the two chains (see Fig.~\ref{fig:participation}).
This connection arises from the following relation between the Pauli matrix and the Bell basis:
\begin{align}\label{eq:paulimatrix-bellstate}
  \ket*{\mathrm{vec}(\sigma^{00})} & = \sqrt{2}\ket*{\mathrm{Bell}^{00}},   \\
  \ket*{\mathrm{vec}(\sigma^{10})} & = \sqrt{2}\ket*{\mathrm{Bell}^{10}},   \\
  \ket*{\mathrm{vec}(\sigma^{01})} & = \sqrt{2}\ket*{\mathrm{Bell}^{01}},   \\
  \ket*{\mathrm{vec}(\sigma^{11})} & = -i\sqrt{2}\ket*{\mathrm{Bell}^{11}}.
\end{align}
Here, $\ket*{\mathrm{vec}(\cdot)}$ is a linear map defined as $\ket*{\mathrm{vec}(\ketbra{x}{y})}=\ket{x}\!\ket{y}$ in the computational basis, which establishes the Choi-Jamiolkowski isomorphism between operators and states in the doubled Hilbert space.
We also defined the Bell basis as
\begin{equation}
  \ket*{\mathrm{Bell}^{b_1b_2}}=\frac{\ket*{0b_1}+(-1)^{b_2}\ket{1\bar{b}_1}}{\sqrt{2}},
\end{equation}
which is labeled by the two binary digits $b_1b_2\in\qty{0,1}^2$.
Utilizing the property $\Tr[A^\dag B]=\braket{\mathrm{vec}(A)}{\mathrm{vec}(B)}$ of the linear map, we obtain the following expression for the expectation value of the Pauli string:
\begin{align}\label{eq:paulistring-bellbasis}
    & \Tr^2[\sigma^{\vec{m}}\psi]\notag                                                                            \\
  = & \abs{\braket{\mathrm{vec}(\sigma^{\vec{m}})}{\mathrm{vec}(\psi)}}^2\notag                                    \\
  = & 2^{L}\abs{\qty[\bigotimes_{j=1}^{L}\bra{\mathrm{Bell}^{m_{2j-1}m_{2j}}}]\ket{\psi}\!\ket{\psi^\ast}}^2\notag \\
  = & 2^{L} \Tr[P^{\vec{m}}(\psi\otimes\psi^\ast)].
\end{align}
In the last line, we have defined the projection operator onto the Bell basis as $P^{\vec{m}}=2^{-L}\ketbra{\mathrm{vec}(\sigma^{\vec{m}})}{\mathrm{vec}(\sigma^{\vec{m}})}$.
Using Eq.~\eqref{eq:paulistring-bellbasis} in  the definition~\eqref{eq:sre}, we obtain the following expression for the SRE:
\begin{equation}\label{eq:sre-bell-basis}
  M_\alpha(\psi)
  = \frac{1}{1-\alpha}\ln\sum_{\vec{m}}\underbrace{\Tr^\alpha[P^{\vec{m}}(\psi\otimes\psi^\ast)]}_{p_{\vec{m}}^\alpha} - (\ln2) L.
\end{equation}
Here, $p_{\vec{m}} := \Tr[P^{\vec{m}}(\psi\otimes\psi^\ast)]$ is the participation in the Bell basis, or equivalently, the Born probability associated with the Bell-state measurements performed on the doubled state.
Consequently, the $\alpha$-SRE can be interpreted as the $\alpha$-R\'{e}nyi entropy of the classical probability distribution $\{p_{\vec{m}}\}$, up to a constant offset chosen to ensure that $M_\alpha$ vanishes for stabilizer states.

We note that the connection between the SRE and the participation entropy is not limited to qubit systems.
For qudit systems with a local Hilbert-space dimension $d\geq3$, the SRE is defined using the expectation values of Heisenberg-Weyl operators~\cite{gross2006hudsons,wang2023stabilizer}.
Using the Choi-Jamiolkowski isomorphism, we can map these expectation values to the Born probabilities of measurements in a maximally entangled state across the doubled space.
Thus, the SRE in qudit systems can also be interpreted as participation entropies in an orthonormal maximally entangled basis of the doubled state.

\subsection{Magic entropy}
In addition to expressing the SRE as participation entropy in the Bell basis, there is an alternative expression of the SRE as magic entropy, introduced in Ref.~\cite{bu2025stabilizera}.
This perspective is also valuable for calculating the SRE with BCFT, as it explicitly provides the configuration of the line defect in the partition function.
We see in the following that these two approaches for determining the line defect lead to consistent results, providing compelling evidence for the validity of our results.

To define the magic entropy, we introduce the notion of convolution of a quantum state.
The key unitary for convolution is the operator $V$ acting on $K$ quantum systems of $L$ qubits, defined as
\begin{equation}\label{eq:key-unitary}
  V := U^{\otimes L},\;\; U = \qty(\prod_{j=2}^{K}\text{CNOT}_{j\to 1})\qty(\prod_{i=2}^{K}\text{CNOT}_{1\to i}),
\end{equation}
where $U$ is a unitary acting on $K$ qubits, and $\text{CNOT}_{2\to1}\ket*{x}\!\ket*{y} = \ket*{x{+}y}\!\ket*{y},\;(x,y\in\mathbb{Z}_2)$ is the CNOT gate.
To relate the magic entropy to the SRE, we choose $K$ to be odd.
The unitary $V$ satisfies the following properties:
\begin{align}\label{eq:key-unitary-properties}
  V\qty(\otimes_{i=1}^{K}\ket*{\vec{x}_i})
   & = \ket*{\vec{x}_{\text{CM}}}\otimes_{i=2}^{K}\ket*{\vec{x}_i+\vec{x}_1},           \\
  V^\dag\qty(\otimes_{i=1}^{K}\ket*{\vec{x}_i})
   & = \ket*{\vec{x}_{\text{CM}}}\otimes_{i=2}^{K}\ket*{\vec{x}_{\text{CM}}-\vec{x}_i}, \\
  V(\sigma^{\vec{m}}\otimes I^{\otimes K-1})V^\dag
   & \propto (\sigma^{\vec{m}})^{\otimes K},                                            \\
  V^\dag(\sigma^{\vec{m}}\otimes I^{\otimes K-1})V\label{eq:V_to_pauli}
   & \propto (\sigma^{\vec{m}})^{\otimes K}.
\end{align}
Here, $\vec{x}_{\text{CM}}:=\sum_{i=1}^{K}\vec{x}_i\;(\vec{x}_i{\in}\mathbb{Z}_2^L)$ and the proportionality constants in the last two lines are given by $(-1)^{\frac{K-1}{2}\sum_{j=1}^{2L}m_{2j{-}1}m_{2j}}$.
The convolution channel is then defined as follows:
\begin{equation}\label{eq:convolution}
  \boxtimes_K\rho := \Tr_{1^c}[V(\underbrace{\rho\otimes\rho\otimes\cdots\otimes\rho}_{K})V^\dag].
\end{equation}
Here, $\Tr_{1^c}[\cdot]$ denotes the partial trace over subsystems $2,3,\ldots,K$.
Precisely speaking, the map $\boxtimes_K$ is a quantum channel from $\mathcal{H}^{\otimes K}$ to $\mathcal{H}$, where $\mathcal{H}$ is a single copy of the Hilbert space.
This notation is justified since we only consider identical copies $\rho^{\otimes K}$ as the input state.

The magic entropy is then defined as the entropy of the state after convolution:
\begin{equation}
  \text{ME}_{n}^{(K)}(\psi) := S_{n}(\boxtimes_K\psi).
\end{equation}
Here, $S_{n}(\rho) := (1-n)^{-1}\ln\Tr[\rho^n]$ is the Rényi entropy.
The magic entropy for $n=2$ turns out to be proportional to the SRE with Rényi index $\alpha=K$ odd.
We can easily confirm this relation using the swap trick, i.e., $\Tr[AB] = \Tr[\mathbb{F}(A\otimes B)]$, with the swap operator $\mathbb{F}=2^{-L}\sum_{\vec{m}}(\sigma^{\vec{m}})^{\otimes 2}$.
\begin{align}
    & S_2(\boxtimes_K\psi)\notag                                                                                                   \\
  = & -\ln\Tr[\mathbb{F}\;(\boxtimes_K\psi)^{\otimes 2}] \notag                                                                    \\
  = & -\ln \Tr[(\boxtimes_K^\dag)^{\otimes 2}(\mathbb{F})\;\psi^{\otimes 2K}]\notag                                                \\
  = & -\ln\Tr[2^{{-}L}\sum_{\vec{m}}(V^\dag(\sigma^{\vec{m}}\otimes I^{\otimes K{-}1})V)^{\otimes 2}\;\psi^{\otimes 2K}]\notag     \\
  = & -\ln \sum_{\vec{m}}\frac{\Tr^{2K}[\sigma^{\vec{m}}\psi]}{2^L}\quad \quad (\because \; \text{Eq.~\ref{eq:V_to_pauli}}) \notag \\
  = & (K-1)M_K(\psi).
\end{align}
Here, we used the dual channel $\boxtimes_K^\dag(\rho) = V^\dag(\rho\otimes I^{\otimes K-1})V$ in the second line.
This alternative expression for the SRE as magic entropy will play a complementary role to the approach using Bell-state measurements in identifying the line defect in the replicated CFT.

It is noteworthy that there exist resource measures for quantum computation defined in the same manner as magic entropy.
For example, measures of non-Gaussianity, which is a resource for universal quantum computation using fermionic/bosonic systems, can similarly be defined through the entanglement created by convolution~\cite{lyu2024fermionic,coffman2025measuring,bu2025efficient,hahn2025measuring}.
This construction is quite general: there exists a certain correspondence between measures that involve the commutant in the trace like the SRE and fermionic anti-flatness~\cite{sierant2025fermionic} and measures based on convolutions, which can relate to each other through the dual channel of the convolution.
Therefore, our theoretical framework could potentially be extended to handle a broader class of measures beyond qubit systems (see Sec.~\ref{sec:unified} for further discussions).

\section{\label{sec:sre-in-criticalstates}Stabilizer R\'{e}nyi Entropy in Critical States}
We now employ BCFT to analyze the SRE in critical quantum spin chains.
BCFT has been useful in characterizing universal properties of one-dimensional critical states and successfully applied to various information-theoretic quantities, including entanglement entropy~\cite{holzhey1994geometric,calabrese2004entanglement,calabrese2009entanglement}, mutual information~\cite{furukawa2009mutual}, and participation entropies in the computational basis~\cite{fradkin2006entanglement,hsu2009universal,stephan2009shannon,hsu2010universal,stephan2010renyi,zaletel2011logarithmic,stephan2011phase,alcaraz2013universal,stephan2014emptiness,stephan2014shannon,luitz2014participation,alcaraz2014universal,alcaraz2016universal}.
Moreover, recent studies have demonstrated the utility of BCFT in describing critical states under measurements and decoherence~\cite{garratt2023measurements,sun2023new,ma2023exploring,zou2023channeling,weinstein2023nonlocality,lee2023quantum,yang2023entanglement,murciano2023measurementaltered,sala2024quantum,ashida2024systemenvironment,patil2024highly,sala2024decoherence,milekhin2024observableprojected,tang2024critical,liu2024boundary,kuno2025systemenvironmental}, including Bell-state measurements acting on the two spin chains~\cite{hoshino2024entanglement}.

Given the identification of the SRE as a participation entropy as outlined in the previous section, BCFT provides a natural and powerful framework for analyzing the SRE in critical spin chains.
Within this perspective, the universal properties of the SRE can be extracted from the partition function of a replicated field theory with a nontrivial interlayer line defect created by the Bell-state measurements.
This approach allows us to systematically characterize the universal contributions to the SRE in terms of the BCFT data, such as the $g$-factor and the scaling dimensions of boundary-condition-changing operators.

\subsection{Measurements in CFT}
We present a brief introduction to our methodology for a CFT description of quantum measurements.
We begin with a Gibbs state $\rho$ of a Hamiltonian $H$:
\begin{equation}\label{gibbs}
  \rho = \frac{e^{-\beta H}}{\Tr e^{-\beta H}}.
\end{equation}
In the path-integral formalism, the expectation value of an observable $A$ in this state can be expressed as an imaginary-time evolution through a path integral over eigenstates of a quantum field $\varphi$:
\begin{equation}
  \langle A\rangle = \Tr[A\rho] = \frac{1}{Z}\int\mathcal{D}\varphi\; A[\varphi]\,e^{-\mathcal{S}[\varphi]}.
\end{equation}
Here, $Z$ is the partition function,
\begin{equation}
  Z = \Tr e^{-\beta H} = \int\mathcal{D}\varphi\; e^{-\mathcal{S}[\varphi]},
\end{equation}
and $\mathcal{S}[\varphi]$ is the Euclidean action defined on a two-dimensional sheet.
When the Hamiltonian $H$ is defined on a periodic chain of length $L$, the theory is defined on a torus of size $L\times \beta$ with spatial coordinate $x$ and imaginary time $\tau$.

Now, let us consider projective local measurements acting on $\rho$ in Eq.~\eqref{gibbs}.
We label the outcomes by $\vec{m}=(m_1,m_2,\ldots)$, a string of labels for local measurement outcomes, and write the corresponding projection operators as $P^{\vec{m}}$.
The postmeasurement state becomes $\rho^{\vec{m}}=P^{\vec{m}}\rho P^{\vec{m}}/p_{\vec{m}}$, where the Born probability $p_{\vec{m}}$ is given by
\begin{equation}\label{born}
  p_{\vec{m}} = \Tr[P^{\vec{m}}\rho] = \frac{1}{Z}\int\mathcal{D}\varphi\; P^{\vec{m}}\, e^{-\mathcal{S}[\varphi]}.
\end{equation}
Throughout this paper, we assume that the operator $P^{\vec{m}}$ is independent of imaginary time $\tau$, i.e., diagonal in $\varphi$, and describe the projection operator as a boundary perturbation $P^{\vec{m}} =e^{-\delta\mathcal{S}[\varphi]}$ at a time slice $\tau=0$.
We note that this condition can be satisfied in the Bell-state measurements~\cite{hoshino2024entanglement}, which are the primary focus of the present work.
The projective nature manifests itself as an infinitely large coupling constant in $\delta\mathcal{S}[\varphi]$, effectively fixing the field configuration and leading to a line defect at the time slice~\footnote{Precisely speaking, it must be understood that a boundary action acts on the $\tau=0$ edge, and its complex conjugate acts also on the $\tau=\beta$. These edges are glued together and turn into the single $\tau=0$ line when discussing the trace in Eq.~\eqref{born}.}.

To calculate the participation entropy, which is a nonlinear function of the Born probability, we employ the replica trick.
Specifically, we compute the sum of the $n$th power of the Born probability $p_{\vec{m}}$ through
\begin{equation}\label{eq:sum_born_prob}
  \sum_{\vec{m}} p_{\vec{m}}^n = \Tr[\sum_{\vec{m}}(P^{\vec{m}})^{\otimes n}\rho^{\otimes n}].
\end{equation}
This reveals that the participation entropy can be obtained from the partition function of an $n$-replicated theory with the operator $\sum_{\vec{m}}(P^{\vec{m}})^{\otimes n}$ acting on the time slice of the torus.
Since each component $(P^{\vec{m}})^{\otimes n}$ in the sum is orthogonal to all others, the composite operator $\sum_{\vec{m}}(P^{\vec{m}})^{\otimes n}$ remains a projection operator.
As detailed later, when the measurement basis is a stabilizer state, the projection operator $\sum_{\vec{m}}(P^{\vec{m}})^{\otimes n}$ is guaranteed to be diagonal in $\varphi$, given that each $P^{\vec{m}}$ is diagonal in $\varphi$.
We can then describe the composite projection operator as a certain boundary perturbation: $\sum_{\vec{m}}(P^{\vec{m}})^{\otimes n}= e^{-\delta\mathcal{S}_{n}[\vec{\varphi}]}$, which allows us to rewrite Eq.~\eqref{eq:sum_born_prob} in the form of a path integral as follows:
\begin{equation}\label{eq:participation-path-integral}
  \sum_{\vec{m}}p_{\vec{m}}^n = \frac{1}{Z^n}\int\!\qty[\prod_{i=1}^{n}\mathcal{D}\varphi_i]\, \exp(-\sum_{i=1}^{n}\mathcal{S}[\varphi_i] - \delta\mathcal{S}_n[\vec{\varphi}]).
\end{equation}
The field configuration at the time slice is constrained to minimize the interlayer boundary action $\delta\mathcal{S}_n[\vec{\varphi}]$.
This formalism provides a systematic way to determine the universal contributions to the SRE, as we will demonstrate in the case of the Ising criticality later.

At this point, we can examine what kind of line defect will be imposed by the boundary perturbation arising from the measurements.
In general, when an extended operator $A$ in the partition function $\Tr[A\rho^{\otimes n}]$ commutes with the Hamiltonian, it manifests as a topological defect~\cite{quella2007reflection}.
Meanwhile, when the operator is a rank-one projection operator, it is expected to impose factorizing defects, i.e., defects that split into two conformal boundaries on each side.
The corresponding boundary perturbation in this case takes the form of a relevant perturbation with infinitely large coupling constant, which is known to flow to factorizing defects~\cite{popov2025factorizing}, consistent with our physical intuition.
This latter scenario is realized for a single round of measurement, described by the boundary perturbation $P^{\vec{m}}= e^{-\delta\mathcal{S}[\varphi]}$.
A crucial assumption for the line defect to maintain conformal invariance is that the measurement outcomes are uniformly distributed over sufficiently large intervals.

We now turn to the case of participation entropy, described by the boundary perturbation $\sum_{\vec{m}}(P^{\vec{m}})^{\otimes n} = e^{-\delta\mathcal{S}_n[\vec{\varphi}]}$ acting on the replicated theory.
The translation invariance of this perturbation is automatically ensured by the summation over $\vec{m}$.
Since the projection operator is not guaranteed to be rank one in this case, the boundary perturbation can be irrelevant.
Nevertheless, the infinitely large coupling constant can still drive the system toward nontrivial line defects.
This behavior is consistent with our observation in the following section that certain boundary conditions lead to $g$-factors greater than unity~\cite{affleck1991universal,affleck1993exact,friedan2004boundary}.
Indeed, the $g$-factor starts from the unity $g=1$ in the UV limit and, according to the g-theorem, it must monotonically decrease under boundary RG flows if the perturbation is relevant.

\subsection{Full-state SRE}
\begin{figure}[tb]
  \centering
  \includegraphics[width=0.85\linewidth, clip]{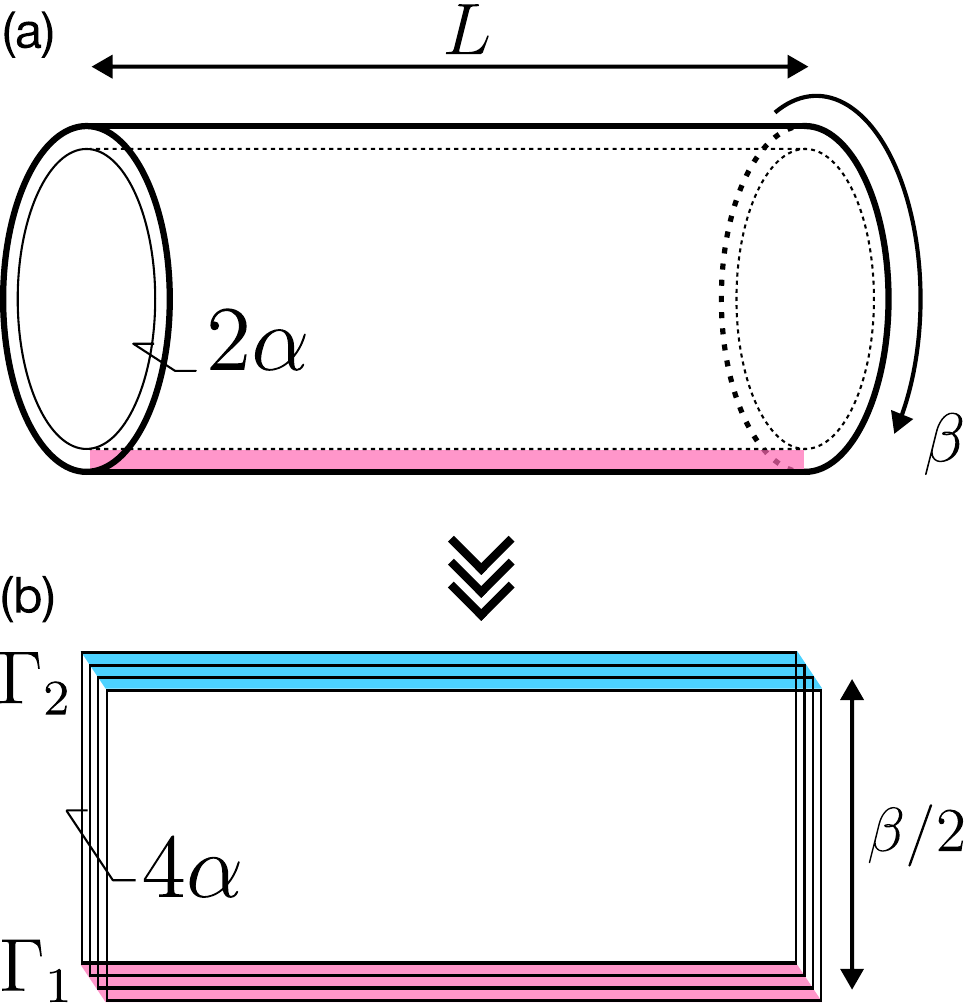}
  \caption{(a)~The partition function $Z_{2\alpha}$ of the $2\alpha$-component theory is defined on the torus of size $L\times \beta$.
    The figure only illustrates periodicity in the imaginary-time direction.
    (b)~The system is folded into the $4\alpha$-component theory defined on the cylinder of circumference $L$ and length $\beta/2$ with boundary conditions $\Gamma_1$ at $\tau=0$ and $\Gamma_2$ at $\tau=\beta/2$.
    If we write the original fields as $\varphi_i\,(i=1,2,\ldots,2\alpha)$, then the remaining $2\alpha$ fields $\varphi_j\,(j=2\alpha+1,2\alpha+2,\ldots,4\alpha)$ satisfy the sewing condition $\varphi_i=\varphi_{i+2\alpha}$ at the boundaries $\Gamma_1$ and $\Gamma_2$.}
  \label{fig:partition-function-full}
\end{figure}

We here discuss how the universal size-independent term in the full-state SRE is determined from the $g$-factor of the replicated theory with the interlayer line defect.
We consider a one-dimensional critical chain of length $L$, whose low-energy behavior is described by a ($1+1$)-dimensional CFT with the central charge $c$.
We assume time-reversal symmetry ($\psi=\psi^\ast$) for the sake of simplicity.
As outlined in the previous section, we use the replica trick to calculate the SRE of an entire chain for integer $\alpha\geq2$.
While the analytical continuation to arbitrary $\alpha$ is not required for our purpose, we numerically verified that the continued results are consistent with the SRE obtained for integer $\alpha$ in the Ising CFT.
Let us rewrite Eq.~\eqref{eq:sre-bell-basis} for integer $\alpha$ in the following form:
\begin{equation}\label{eq:sre-replica}
  M_{\alpha}(\psi) = \frac{1}{1-\alpha}\ln\Tr[\sum_{\vec{m}}(P^{\vec{m}})^{\otimes \alpha}(\psi\otimes\psi^\ast)^{\otimes \alpha}]-(\ln 2)L.
\end{equation}
The operator $\sum_{\vec{m}}(P^{\vec{m}})^{\otimes \alpha}$ acts as a projection operator in the $2\alpha$-component replicated theory and induces an interlayer line defect at the time slice $\tau=0$ on the two-dimensional sheet.
We can define the partition function of this theory as follows:
\begin{equation}\label{eq:partition-function-full}
  \frac{Z_{2\alpha}}{Z^{2\alpha}} = \Tr[\sum_{\vec{m}}(P^{\vec{m}})^{\otimes \alpha}(\psi\otimes\psi^\ast)^{\otimes \alpha}].
\end{equation}
Here, $Z$ is the original partition function of a single component theory without a defect.
The partition function $Z_{2\alpha}$ is calculated by a path integral of $2\alpha$ components on a torus with a line defect (Fig.~\ref{fig:partition-function-full}(a)).
We employ a folding trick to rewrite $Z_{2\alpha}$ as a path integral of $4\alpha$ components on a cylinder with different boundary conditions imposed at $\tau=0$ and $\tau=\beta/2$ (Fig.~\ref{fig:partition-function-full}(b)).
The boundary condition imposed by the projection operator $\sum_{\vec{m}}(P^{\vec{m}})^{\otimes \alpha}$ at $\tau=0$ is denoted by  $\Gamma_1$ (red), while the artificial boundary condition at $\tau=\beta/2$ arising from the folding is expressed by $\Gamma_2$ (blue).
In the IR limit, these boundary conditions flow to conformally invariant boundary conditions.

In BCFT, partition functions of CFTs on a cylinder with conformal boundary conditions at the ends are expressed as amplitudes between boundary states as follows:
\begin{equation}\label{eq:partition-function-amplitude}
  Z_{2\alpha} = \mel*{\Gamma_2}{e^{-\frac{\beta}{2}H}}{\Gamma_1}.
\end{equation}
Here, $\ket*{\Gamma_1}$ and $\ket*{\Gamma_2}$ are the boundary states corresponding to $\Gamma_1$ and $\Gamma_2$, respectively, and $H$ is the CFT Hamiltonian of the $4\alpha$ independent components on the cylinder.
In general, partition functions with boundaries consist of contributions from bulk energy, line energy proportional to the length of the boundary, and boundary entropy that is a universal constant assigned to each conformally invariant boundary condition~\cite{affleck1991universal}.
Since the bulk energy cancels between $Z_{2\alpha}$ and $Z^{2\alpha}$, Eq.~\eqref{eq:partition-function-full} in the limit $\beta\to\infty$ generally behaves as
\begin{equation}\label{eq:free-energy-general}
  \ln\frac{Z_{2\alpha}}{Z^{2\alpha}} = bL + \ln g + o(1),
\end{equation}
where $b$ is a nonuniversal line energy density contributing to the SRE density $m_{\alpha}$ in Eq.~\eqref{eq:sre_general}, and the universal constant $g=g_1g_2$ is the product of $g$-factors associated with $\Gamma_1$ and $\Gamma_2$.
We note that the boundary condition $\Gamma_2$ must have a trivial $g$-factor $g_2=1$ since it is merely an artificial boundary created from the folding.
Therefore, from Eqs.~\eqref{eq:sre-replica}, \eqref{eq:partition-function-full}, and~\eqref{eq:free-energy-general}, the constant term of the SRE (cf. Eq.~\eqref{eq:sre_general}) is written as
\begin{equation}\label{eq:result1}
  c_{\alpha} = \frac{\ln g_1}{\alpha - 1}.
\end{equation}

In practice, the $g$-factors $g_{1,2}$ can be calculated from the overlap of the boundary states $\ket*{\Gamma_{1,2}}$ with the ground state $\ket*{\mathrm{GS}}$ of $H$.
To see this, we consider the leading contribution to the partition function $Z_{2\alpha}$ in Eq.~\eqref{eq:partition-function-amplitude} in the limit $\beta\gg L$
\begin{equation}
  Z_{2\alpha} \simeq \braket*{\Gamma_1}{\mathrm{GS}}\braket*{\mathrm{GS}}{\Gamma_2} e^{-\frac{\beta}{2}E_{\mathrm{GS}}(L)},
\end{equation}
where $E_{\mathrm{GS}}(L)$ is the ground-state energy of the Hamiltonian on a cylinder of circumference $L$: $E_{\mathrm{GS}}(L) = - \pi c /(6L)$.
The phase factor of $\ket*{\mathrm{GS}}$ can be chosen such that $\braket*{\mathrm{GS}}{\Gamma}$ is real and positive for all boundary states $\ket*{\Gamma}$.
Comparing this behavior with Eq.~\eqref{eq:free-energy-general}, we obtain $g_1=\braket*{\mathrm{GS}}{\Gamma_1}$ and $g_2=\braket*{\mathrm{GS}}{\Gamma_2}$.

In the infrared limit, a fine detail about informational quantities is typically encoded in the boundary rather than the bulk of the theory.
While entanglement entropy reflects the properties of a special topological defect with universal behavior determined solely by the central charge, more general informational quantities require the extensive boundary data of replicated theories.
For instance, when degrees of freedom at cuts of the Hilbert space appear (such as in gauge theories), the $g$-factor of the conformal boundary to which the cut flows contributes as a subleading correction~\cite{ohmori2015physics}.
The bulk CFT, characterized merely by the central charge and bulk primary correlations, cannot capture such detailed information.
In contrast, replicated theories possess numerous (typically infinite) conformal boundaries with large degrees of freedom, enabling the encoding of finer details.

Before closing this section, we comment on the SRE of critical spin chains subject to open boundary conditions in the spatial direction.
In this case, there appears a logarithmic term in $L$ to appear in the SRE~\cite{hoshino2025stabilizera}.
This is the contribution from corners on the two-dimensional sheet, and the coefficient of the logarithmic term is a universal quantity determined by the central charge and the scaling dimensions of BCCOs that connect the boundaries on both sides~\cite{cardy1988finitesize}.
Such a subleading logarithmic term has been discussed also in the context of the participation entropy in the computational basis~\cite{luitz2014participation}.

\subsection{Mutual SRE}
\begin{figure}[tb]
  \centering
  \includegraphics[width=0.95\linewidth, clip]{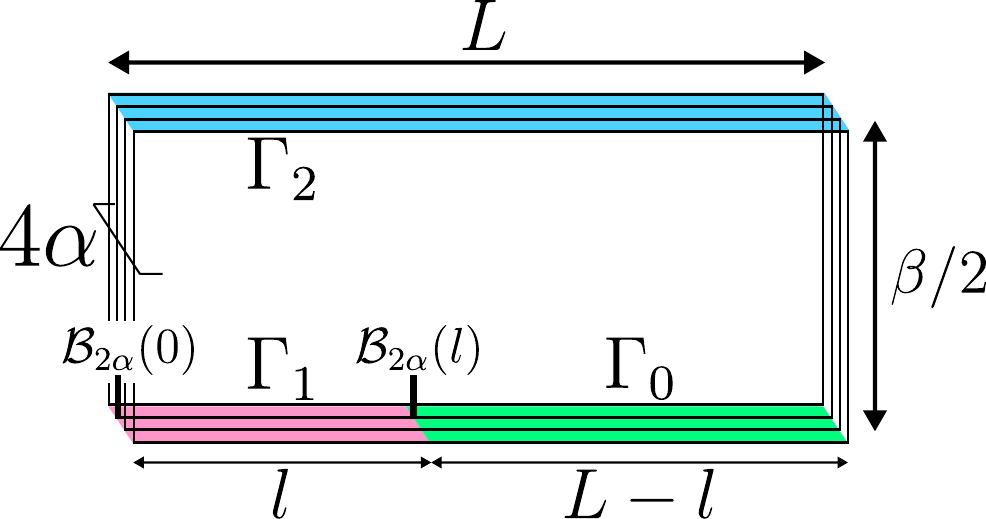}
  \caption{The partition function $Z_{2\alpha}(A)$ is calculated from the path-integral on a cylinder of circumference $L$ and length $\beta/2$ with the boundary condition $\Gamma_2$ at $\tau=\beta/2$ and two different boundary conditions at $\tau=0$: $\Gamma_1$ in $0\leq x\leq l$ and $\Gamma_0$ in $l\leq x\leq L$.
    There are two boundary-condition-changing points at $x=0,l$, and the partition function $Z_{2\alpha}(A)$ can be calculated from a two-point correlation function of the BCCOs $\mathcal{B}_{2\alpha}(x)$ inserted at these points.}
  \label{fig:partition-function-mutual}
\end{figure}

The path-integral formalism can also be applied to the mutual SRE by adding a boundary-condition-changing point in the replicated field theory.
Let $\vec{m}_A\in\qty{0,1}^{2l}$ denote the label of Pauli strings in the $l$-qubit subsystem $A=\{1,2,\ldots,l\}$.
We then obtain the following relation:
\begin{equation}\label{eq:trace-relation}
  \Tr_A[\sigma^{\vec{m}_A}\rho_A] = \Tr[(\sigma^{\vec{m}_A}\otimes I_B)\psi],
\end{equation}
where $I_B=(\sigma^{00})^{\otimes (L-l)}$ is the identity operator in the $(L{-}l)$-qubit subsystem $B=\{l+1,l+2,\ldots,L\}$.
Since $\sigma^{\vec{m}_A}\otimes I_B$ is still a Pauli string in the total system $\psi$, we can use Eq.~\eqref{eq:paulistring-bellbasis} to write
\begin{equation}\label{eq:paulistring-partial}
  \Tr_A^2[\sigma^{\vec{m}_A}\rho_A] = 2^L \Tr[(P^{\vec{m}_A}\otimes (P^{00})^{\otimes (L-l)})(\psi\otimes\psi^\ast)].
\end{equation}
In the similar manner as in the full-state SRE, the SRE of the subsystem $A$ reads
\begin{align}\label{eq:sre-subsystem}
    & M_{\alpha}(\rho_A)\notag                                                                                                                                              \\
  = & \frac{(\alpha L -l)\ln 2}{1 - \alpha} + \frac{1}{1 - \alpha}\ln\Tr[\sum_{\vec{m}_A}({\tilde P}^{\vec{m}_A})^{\otimes \alpha}(\psi\otimes\psi^\ast)^{\otimes \alpha}],
\end{align}
where $\tilde{P}^{\vec{m}_A}=P^{\vec{m}_A}\otimes(P^{00})^{\otimes (L-l)}$.
While a modified definition of the SRE has been proposed for mixed states~\cite{leone2022stabilizer}, for the purpose of our analysis, we adopt the conventional definition of the SRE in Eq.~\eqref{eq:sre} to ensure consistency in our field-theoretical framework.
The projection operator $\sum_{\vec{m}_A}({\tilde P}^{\vec{m}_A})^{\otimes \alpha}$ creates an interlayer line defect in the replicated $2\alpha$-component theory, where the type of the defect is different between the subsystems $A$ and $B$.
The corresponding partition function in the path-integral formalism can be read as follows:
\begin{equation}\label{eq:partition-function-mutual}
  \frac{Z_{2\alpha}(A)}{Z^{2\alpha}} = \Tr[\sum_{\vec{m}_A}({\tilde P}^{\vec{m}_A})^{\otimes \alpha}(\psi\otimes\psi^\ast)^{\otimes \alpha}].
\end{equation}
As shown in Fig.~\ref{fig:partition-function-mutual}, after the folding, the partition function $Z_{2\alpha}(A)$ is calculated by a path integral of the $4\alpha$-component theory on a cylinder with the boundary condition $\Gamma_2$ at $\tau=\beta/2$ (blue) and two different boundary conditions at $\tau=0$: $\Gamma_1$ in $A$ with $x\in[0,l]$ (red) and $\Gamma_0$ in $B$ with $x\in[l,L]$ (green).
Here, $\Gamma_2$ is the artificial boundary condition created by the folding, $\Gamma_1$ is induced by the projector $\sum_{\vec{m}_A}(P^{\vec{m}_A})^{\otimes \alpha}$ acting on $A$, and $\Gamma_0$ is imposed by the product $(P^{00})^{\otimes \alpha(L-l)}$ of the projection onto $\ket*{\mathrm{Bell}^{00}}$ across the local doubled Hilbert space in $B$.

One can calculate the partition function $Z_{2\alpha}(A)$ as a two-point correlation function of the BCCO $\mathcal{B}_{2\alpha}(x)$ associated with the change between $\Gamma_1$ and $\Gamma_0$ as follows \cite{cardy1989boundary,affleck1997boundary} (see Fig.~\ref{fig:partition-function-mutual}):
\begin{equation}\label{eq:partition-function-bcco}
  Z_{2\alpha}(A) = \langle \mathcal{B}_{2\alpha}(l)\mathcal{B}_{2\alpha}(0) \rangle.
\end{equation}
If we write the scaling dimension of $\mathcal{B}_{2\alpha}(x)$ as $\Delta_{2\alpha}$ and consider the ground state $\beta\to\infty$, the partition function behaves as $Z_{2\alpha}(A)/Z^{2\alpha}\sim 2^{-\alpha(L-l)}e^{-\epsilon l}l_c^{-2\Delta_{2\alpha}}$, where $\epsilon$ is the nonuniversal line-energy density associated with $\Gamma_1$ and $l_c=(L/\pi)\sin(\pi l/L)$ is the chord length.
The factor $2^{-\alpha(L-l)}$ is the line-energy contribution from the $\Gamma_0$ segment and follows from the exact identity $\Tr[(P^{00})^{\otimes L}(\psi\otimes\psi^\ast)]=2^{-L}$ for a normalized state $\psi$.
From Eqs.~\eqref{eq:sre-subsystem}, \eqref{eq:partition-function-mutual}, and~\eqref{eq:partition-function-bcco}, the SRE of the subsystem $A$ is expressed by
\begin{align}\label{eq:result2}
  M_{\alpha}(\rho_A) & = \left(\frac{\epsilon}{\alpha-1}-\ln 2\right)l + \frac{2\Delta_{2\alpha}}{\alpha-1}\ln l_c+\cdots,
\end{align}
and the corresponding leading extensive contributions to $M_\alpha(\rho_A)$, $M_\alpha(\rho_B)$, and $M_\alpha(\rho_{AB})$ are proportional to $l$, $L-l$, and $L$, respectively, and thus cancel in the mutual SRE.
This leads to the long-distance behavior
\begin{align}
  W_{\alpha}(l)
   & =  M_\alpha(\rho_A) + M_\alpha(\rho_B) - M_\alpha(\rho_{AB})\label{Walpha} \\
   & =  \frac{4\Delta_{2\alpha}}{\alpha-1}\ln l_c.
\end{align}
It is now evident that the coefficient of the logarithmic scaling in the mutual SRE is characterized by the universal quantity $\Delta_{2\alpha}$ in the IR theory.

\subsection{Mutual information}
We remark that there is yet another definition of the $2$-SRE that involves the R\'{e}nyi-2 entropy as a normalization contribution~\cite{leone2022stabilizer}:
\begin{equation}
  \tilde{M}_2(\rho) = M_2(\rho) - S_2(\rho),
\end{equation}
where $S_\alpha(\rho) = (1-\alpha)^{-1}\ln\Tr[\rho^\alpha]$ is the R\'{e}nyi entropy; we note that, for a pure state $\rho$, this definition reduces to Eq.~\eqref{eq:sre} since $S_2(\rho)=0$.
The mutual $2$-SRE based on $\tilde{M}_2$, denoted as $\tilde{W}_2$, can be related to the above mutual SRE~\eqref{Walpha} via~\footnote{For the sake of notational consistency, we define $\tilde{W}_2$ such that its sign is opposite to that of the mutual SRE introduced in Ref.~\cite{tarabunga2023manybody}.}
\begin{align}\label{tildeW}
  \tilde{W}_2(A:B)=W_2(A:B)-I_2(A:B),
\end{align}
where the R\'{e}nyi mutual information $I_\alpha(A:B)$ is defined as
\begin{equation}\label{eq:mutual-info}
  I_{\alpha}(A:B) = S_{\alpha}(\rho_A) + S_{\alpha}(\rho_B) - S_{\alpha}(\rho_{AB}).
\end{equation}
In fact, the R\'{e}nyi-$2$ entropy $S_2$ can also be calculated from the Pauli-string expectation values~\cite{klich2024swap}.
Thus, our framework can be equally applicable for calculating the R\'{e}nyi-2 mutual information $I_2$ and, consequently, another version of the mutual $2$-SRE $\tilde{W}_2$ in Eq.~\eqref{tildeW}.

This observation motivates us to calculate the R\'{e}nyi-$2$ mutual information $I_2$ based on BCFT formalism as a further check of the validity of our theoretical framework.
Specifically, using the relation
\begin{equation}\label{eq:purity}
  \Tr[\rho_A^2] = 2^{L-l}\Tr[\sum_{\vec{m}_A}{\tilde P}^{\vec{m}_A}(\psi\otimes\psi^\ast)],
\end{equation}
we may express the R\'enyi-2 entropy $S_2$ in terms of the partition function $Z_{2\alpha}(A)$ evaluated at $\alpha=1$ (cf. Eq.~\eqref{eq:partition-function-mutual}), even though $Z_{2\alpha}(A)$ was originally defined for $\alpha\geq2$.
In this case, we have $\sum_{\vec{m}_A}{\tilde P}^{\vec{m}_A}=I_{A}\otimes (P^{00})^{\otimes L-l}$, and the boundary condition $\Gamma_1$ in $A$ becomes the trivial one equivalent to $\Gamma_2$, while the boundary condition $\Gamma_0$ in $B$ corresponds to the projection onto the product of $\ket*{\mathrm{Bell}^{00}}$ across the doubled state.
Accordingly, if we denote the scaling dimension of the corresponding BCCO as $\Delta_2$ (extending our notation to include $\alpha=1$), the mutual information exhibits the universal logarithmic scaling,
\begin{equation}\label{eq:mutual_info}
  I_2(l) = 4\Delta_2\ln l_c.
\end{equation}
For the ground state of a CFT with the central charge $c$, it has been known that the R\'{e}nyi entropy scales as $S_{\alpha}(l)=(c/6)(1+\alpha^{-1})\ln l_c$ at long distances~\cite{calabrese2009entanglement}, which leads to $I_2(l)=(c/2)\ln l_c$.
Comparing these two results yields $\Delta_2=c/8$.
Below we will reproduce this result for the Ising criticality, though this agreement is a priori not obvious because the scaling dimension $\Delta_{2\alpha}$ was originally defined for $\alpha\geq2$, and analytical continuation to $\alpha=1$ is not ensured to give a correct value.

\section{\label{sec:example}Ising criticality}
In the previous sections, we have shown that calculations of the universal contributions to the SRE boil down to construction of the conformal boundary states corresponding to the Bell-state measurements.
In the present section, we demonstrate this through a detailed analysis of the Ising CFT as a case study.
A standard lattice model that realizes the Ising criticality is the TFIM, whose Hamiltonian is given by
\begin{equation}\label{eq:tfim}
  H_{\mathrm{TFIM}} = -\sum_{j=1}^{L}\qty(Z_jZ_{j+1} + \lambda X_j),
\end{equation}
where we write the Pauli matrices as $Z=\sigma^{01}, X=\sigma^{10}$ for later convenience, $\lambda$ represents the transverse field, and the critical point is at $\lambda=1$.
We note that, when $\lambda\to0$ or $\lambda\to\infty$, the ground state of the TFIM reduces to a stabilizer state.

To analytically calculate the $g$-factor and scaling dimension, we employ a bosonization approach and map two Ising CFTs to a single free-boson CFT for simplifying the analysis of multicomponent, replicated field theory.
Our procedure consists of two main steps:
first, we determine the conformal boundary conditions by expressing the projection operator as a boundary perturbation in the Euclidean action.
Second, we calculate the $g$-factor and scaling dimension by constructing the corresponding boundary states within the free-boson theory.
As shown later, the analytical results obtained through this approach perfectly agree with our numerical calculations.

\subsection{Boundary perturbation formalism}
We first rewrite the projection operator $\sum_{\vec{m}}(P^{\vec{m}})^{\otimes \alpha}$ acting on the $\alpha$-copies of the doubled state as a Euclidean boundary action in terms of bosonic fields.
To this end, we note that the projection operator on replicated qubits at a local lattice site can be written by Pauli matrices as follows \footnote{In general, we note that a projection operator onto a stabilizer state can be expressed as a product of $(I+s)/2$; each acts as a projection operator onto the eigenstate of the generator $s$ with eigenvalue $+1$~\cite{nielsen2010quantum}.}:
\begin{align}\label{eq:projection-exponential}
    & \sum_{b_1,b_2}(P^{b_1b_2})^{\otimes \alpha}\notag                                           \\
  = & \sum_{b_1,b_2}\qty(\frac{I+(-1)^{b_1}ZZ}{2}\frac{I+(-1)^{b_2}XX}{2})^{\otimes \alpha}\notag \\
  = & \prod_{s\in\tilde{S}^{(\alpha)}}\frac{I+s}{2}.
\end{align}
Here, $\tilde{S}^{(\alpha)}$ is a set of the generators that stabilize the projected subspace and spanned by the following set of Pauli strings:
\begin{equation}\label{eq:generator-stabilizer}
  \tilde{S}^{(\alpha)} {=} \qty{XX^{(1)}XX^{(2)},ZZ^{(1)}ZZ^{(2)},\ldots,ZZ^{(\alpha{-}1)}ZZ^{(\alpha)}}.
\end{equation}
The superscript of $XX^{(i)}\,(i=1,2,{\ldots},\alpha)$ labels the $i$-th pair of qubits in the $\alpha$-fold doubled Hilbert space, resulting in $2\alpha$ spins at each site in total.
We may write $(I+s)/2=\lim_{\mu\to\infty}e^{\mu\,s}/(2\cosh\mu)$, which enables us to express the whole projection operator in the exponential form, $\sum_{\vec{m}}(P^{\vec{m}})^{\otimes \alpha} \propto e^{-\delta\mathcal{S}_{2\alpha}}$, where we introduce the $2\alpha$-interlayer boundary coupling by
\begin{equation}\label{eq:boundary-perturbation}
  \delta\mathcal{S}_{2\alpha} = -\mu\int \!d\tau\,\delta(\tau)\!\int\!dx\sum_{s\in\tilde{S}^{(\alpha)}}s.
\end{equation}
The coupling constant $\mu$ is understood to be taken infinitely large, reflecting the projective nature of the measurement considering here.
This path-integral representation of the projection operator leads to the following expression for the partition function in Eq.~\eqref{eq:partition-function-full}:
\begin{equation}\label{eq:path-integral}
  Z_{2\alpha} = \int\exp(-\sum_{i=1}^{2\alpha}\mathcal{S}^{(i)}-\delta\mathcal{S}_{2\alpha}).
\end{equation}
Here, $\mathcal{S}^{(i)}$ is the bulk Ising CFT action of the $i$-th component in the $2\alpha$ replicated theory.
Since the coupling constant $\mu$ tends to be infinite, we can identify the boundary condition as the field configuration that minimizes the boundary action $\delta\mathcal{S}_{2\alpha}$.

\begin{figure}[tb]
  \centering
  \includegraphics[width=0.7\linewidth, clip]{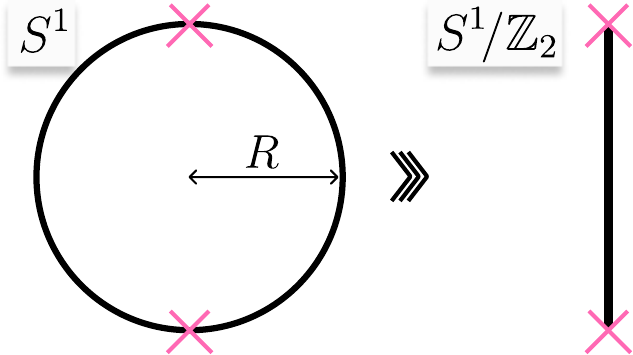}
  \caption{Schematic illustration of the $\mathbb{Z}_2$ orbifold construction.
    Left: The $S^1$ free boson is defined on a circle with radius $R$.
    Right: The $S^1\!/\mathbb{Z}_2$ free-boson CFT is obtained by identifying points related by the $\mathbb{Z}_2$ transformation $\phi\to-\phi$, effectively folding the circle into the line segment $[0,\pi]$.
    The crosses indicate the fixed points ($\phi=0,\pi$) of the $\mathbb{Z}_2$ orbifold.}
  \label{fig:orbifold}
\end{figure}

\subsection{Bosonization}

We next map a pair of the Ising CFTs into a single free-boson theory and determine the boundary conditions in terms of the bosonic fields.
In the Ising CFT, there are three primary fields $I$, $\varepsilon$, and $\sigma$, with conformal weights $0$, $1/2$, and $1/16$, respectively~\cite{difrancesco1997conformal}.
The correspondence between lattice operators and these primary fields is known to be $Z\sim \sigma$ and $X-\langle X\rangle\sim \varepsilon$, where $\langle X\rangle$ is the nonzero expectation value of $X$ in the critical ground state~\cite{zou2020conformal}.
The fields $\sigma_1\sigma_2,\varepsilon_1\varepsilon_2,\varepsilon_1+\varepsilon_2$ in the pair of the Ising CFTs are mapped to a bosonic field as follows~\cite{zuber1977quantum,difrancesco1987critical}:
\begin{align}\label{eq:bosonization}
  \sigma_1\sigma_2            & = \cos\phi,                                      \\
  \varepsilon_1\varepsilon_2  & = (\partial_\mu\phi)^2, \label{eq:bosonization2} \\
  \varepsilon_1+\varepsilon_2 & = \cos2\phi.\label{eq:bosonization3}
\end{align}
The resulting bosonized theory is the $\mathbb{Z}_2$ orbifold of the free-boson CFT compactified on $S^1$~\cite{dixon1987conformal,ginsparg1988curiosities}. We recall that the $S^1$ theory has the Lagrangian density
\begin{equation}
  \mathcal{L}= \frac{1}{2\pi}(\partial_\mu\phi)^2\quad (\mu=x,\tau),
\end{equation}
where the field $\phi$ is compactified on a circle as $\phi\sim\phi+2\pi$, i.e., the compactification radius is $R=1$.
The $\mathbb{Z}_2$ orbifold theory, or equivalently, $S^1\!/\mathbb{Z}_2$ free boson is then obtained from the $S^1$ theory through the identification $\phi\sim-\phi$ (see Fig.~\ref{fig:orbifold}).

This mapping to the $S^1\!/\mathbb{Z}_2$ theory enables us to express the boundary action~\eqref{eq:boundary-perturbation} by the $\alpha$-component bosonic fields.
Specifically, using the relations~\eqref{eq:bosonization} and \eqref{eq:bosonization2}, the generators $ZZ^{(i)}ZZ^{(i{+}1)}$ and $XX^{(i)}XX^{(i{+}1)}$ in $\tilde{S}^{(\alpha)}$ can be expressed by the bosonic fields $\phi_i$ and $\phi_{i{+}1}$ as follows:
\begin{align}
  ZZ^{(i)}ZZ^{(i{+}1)}
   & \sim \cos\phi_i\cos\phi_{i{+}1}\notag                                           \\
   & = \cos(\phi_i+\phi_{i{+}1}) + \cos(\phi_i-\phi_{i{+}1})\notag                   \\
   & = 2\cos(\phi_i-\phi_{i{+}1}),                                                   \\
  XX^{(i)}XX^{(i{+}1)}
   & \sim \langle X\rangle^2[(\partial_\mu\phi_i)^2 + (\partial_\mu\phi_{i{+}1})^2].
\end{align}
Here, we consider the terms with the lowest scaling dimensions within each generator while imposing the identification $\phi\sim-\phi$.
We note that, among the terms in $ZZ$, the potential term $\cos(\phi_i -\phi_{i{+}1})$ should be the most relevant, imposing the Dirichlet boundary conditions (DBCs) on $\phi_i -\phi_{i{+}1}$ in the infrared limit.

\subsection{Conformal boundary conditions}

After the folding trick, the number of fields is doubled, and there exist $2\alpha$-component bosonic fields in total; see Fig.~\ref{fig:partition-function-full}, where we note that the number of the fields differ by a factor of two because two Ising CFTs are bosonized into a single free-boson CFT here.
The above boundary action $\delta\mathcal{S}_{2\alpha}$ then leads to the conditions $\phi_i=\phi_{i{+}1}$ and $\phi_{i+\alpha}=\phi_{i{+}1+\alpha}$ for $i=1,2,\ldots,\alpha-1$.
Meanwhile, the folding procedure imposes the sewing condition $\phi_i=\phi_{i+\alpha}$ for $i=1,2,\ldots,\alpha$, reflecting the periodic boundary condition along the imaginary-time axis in the original torus geometry.
Consequently, the resulting boundary condition $\Gamma_1$ due to the projection $\sum_{\vec{m}}(P^{\vec{m}})^{\otimes \alpha}$ consists of the DBC for all the relative fields $\phi_i-\phi_{i{+}1}$ and the Neumann boundary condition (NBC) for the center of mass field $\sum_i\phi_i$ as follows:
\begin{equation}\label{eq:gamma1_description}
  \Gamma_1:\left\{
  \begin{array}{lc}
    \displaystyle\sum_{i=1}^{2\alpha} \phi_i & \text{NBC} \\
    \phi_1-\phi_2=0                          & \text{DBC} \\
    \phi_2-\phi_3=0                          & \text{DBC} \\
    \quad\vdots                              &            \\
    \phi_{2\alpha-1} - \phi_{2\alpha}=0      & \text{DBC}
  \end{array}
  \right..
\end{equation}
Below we shall write the corresponding boundary state in the $2\alpha$-component $S^1\!/\mathbb{Z}_2$ free-boson CFT as $\ket*{\Gamma_1}_{\mathrm{orb}}$, which has a nontrivial $g$-factor $g_1$ as shown later.

In contrast, the boundary condition $\Gamma_2$ artificially created by the folding procedure (cf. the blue boundary in Fig.~\ref{fig:partition-function-full}(b)) simply corresponds to the sewing condition, leading to
\begin{equation}\label{eq:gamma2_description}
  \Gamma_2:\left\{
  \begin{array}{lc}
    \phi_i-\phi_{i+\alpha}=0 \quad(i=1,2,\ldots,\alpha) & \text{DBC} \\
    \phi_i+\phi_{i+\alpha} \quad(i=1,2,\ldots,\alpha)   & \text{NBC} \\
  \end{array}
  \right..
\end{equation}
As above, we write the boundary state corresponding to $\Gamma_2$ in the $S^1\!/\mathbb{Z}_2$ theory as $\ket*{\Gamma_2}_{\mathrm{orb}}$.
We will explicitly check that the $g$-factor of this boundary state is trivial $g_2=1$ as it should be.
We note that the boundary condition $\Gamma_1$ reduces to the trivial boundary condition $\Gamma_2$ in the case of $\alpha=1$.

Finally, we consider the boundary condition $\Gamma_0$ corresponding to the projection onto the product of $\ket*{\mathrm{Bell}^{00}}$, which is necessary when evaluating the mutual SRE (cf. the product of $P^{00}$ in Eq.~\eqref{eq:paulistring-partial} and the green boundary in Fig.~\ref{fig:partition-function-mutual}).
Since this projection preserves the independence of replicas, the resulting boundary state must consist of a $2\alpha$-fold tensor product of single-component $S^1\!/\mathbb{Z}_2$ free-boson boundary states.
We also note that the projection on $\ket*{\mathrm{Bell}^{00}}$ forces the two qubits in the local doubled Hilbert space to occupy the same state (either $\ket{0}$ or $\ket{1}$) without constraining any specific spin configuration, which, in the field-theoretical language, should correspond to the free boundary condition.
It has been known that such a free boundary state can be realized by the DBC with $\phi=\pi/2$ in the single-component $S^1\!/\mathbb{Z}_2$ free-boson theory, which, in the paired Ising CFTs, corresponds to a tensor product of two free boundary states $\ket*{f}\!\otimes\!\ket*{f}$~\cite{oshikawa1997boundary}.
Thus, the resulting expression of $\Gamma_0$ in terms of the bosonic fields is
\begin{equation}\label{eq:gamma0_description}
  \Gamma_0:\left\{
  \begin{array}{lc}
    \phi_1 = \pi/2         & \text{DBC} \\
    \phi_2 = \pi/2         & \text{DBC} \\
    \quad\vdots            &            \\
    \phi_{2\alpha} = \pi/2 & \text{DBC}
  \end{array}
  \right..
\end{equation}
Again, we write the corresponding boundary state in the multicomponent $S^1\!/\mathbb{Z}_2$ theory as $\ket*{\Gamma_0}_{\mathrm{orb}}$, whose $g$-factor is denoted by $g_0$.
In fact, from Eq.~\eqref{eq:paulistring-bellbasis}, we can deduce that this $g$-factor must be trivial, $g_0=1$, since the Born probability for obtaining the uniform measurement outcome corresponding to $\ket*{\mathrm{Bell}^{00}}$ is nothing but the square of the expectation value of the identity operator:
\begin{align}
  \Tr[(P^{00})^{\otimes L}(\psi\otimes\psi^\ast)]
   & = \frac{1}{2^L}\Tr^2[(\sigma^{00})^{\otimes L}\psi]\notag \\
   & = \underbrace{g_0}_{1}e^{-(\ln 2)L}.
\end{align}
We will explicitly validate the above expression~\eqref{eq:gamma0_description} of the boundary condition $\Gamma_0$ by showing that, within the single-component theory, the only boundary condition yielding the unit $g$-factor is indeed the DBC at a non-fixed point $\phi\neq 0,\pi$ (see Eq.~\eqref{eq:unity_of_g0} and the corresponding discussions).

\subsection{Alternative derivation based on magic entropy}
The line defect configuration consisting of a Neumann boundary condition for the center-of-mass field and Dirichlet boundary conditions for the remaining relative fields can be derived through an alternative approach using magic entropy.
Recall that the SRE is equivalent to the magic entropy $M_2^{(K)}(\psi) = S_2(\boxtimes_K\psi)$, which requires calculating the second Rényi entropy of the convolution $\boxtimes_K\psi$.
To proceed, we employ the Choi-Jamiolkowski isomorphism to map the density matrix to a state:
\begin{equation}
  |\boxtimes_K\psi) := \langle\underbrace{\Phi\cdots\Phi}_{K-1}|V\otimes V^\ast|\underbrace{\psi\cdots\psi}_{K}\rangle|\underbrace{\psi^\ast\cdots\psi^\ast}_{K}\rangle,
\end{equation}
where $|\Phi\rangle=2^{-L/2}\sum_{\vec{x}}|\vec{x}\rangle|\vec{x}\rangle$ is the maximally entangled state in the computational basis $\vec{x}\in\mathbb{Z}_2^L$.
We use a rounded bracket $|\cdot)$ to represent a density operator as a vector in the doubled Hilbert space.
The second Rényi entropy can then be calculated as follows:
\begin{align}
  S_2(\boxtimes_K\psi)
   & = -\ln\text{Tr}[|\boxtimes_K\psi)(\boxtimes_K\psi|]\notag                                                       \\
   & = -\ln\text{Tr}\left[\sum_{\vec{x}_1,\vec{y}_1}P(\vec{x}_1,\vec{y}_1)(\psi\otimes\psi^\ast)^{\otimes K}\right].
\end{align}
Here, $P(\vec{x}_1,\vec{y}_1)=|\Psi(\vec{x}_1,\vec{y}_1)\rangle\langle\Psi(\vec{x}_1,\vec{y}_1)|$ is a projection operator onto the state
\begin{align}\label{eq:projection_V_basis}
    & |\Psi(\vec{x}_1,\vec{y}_1)\rangle\notag                                                                                                                               \\
  = & (V^\dagger\otimes V^{\mathsf{T}})|\vec{x}_1,\vec{y}_1\rangle|\underbrace{\Phi\cdots\Phi}_{K-1}\rangle\notag                                                           \\
  = & 2^{-L(K-1)/2} \sum_{\Gamma}|\vec{x}_{\text{CM}}\,\vec{y}_{\text{CM}}\rangle\otimes_{i=2}^{K}|\vec{x}_{\text{CM}}{-}\vec{x}_i\,\vec{y}_{\text{CM}}{-}\vec{y}_i\rangle.
\end{align}
The summation $\sum_{\Gamma}$ is taken under the constraint
\begin{equation}
  \Gamma:\; \vec{x}_{\text{CM}}-\vec{x}_i = \vec{y}_{\text{CM}} - \vec{y}_i\quad (i=2,3,\ldots,K).
\end{equation}
Note that this constraint $\Gamma$ leaves the configurations of $\vec{x}_{\text{CM}}$ and $\vec{y}_{\text{CM}}$ unconstrained.
Thus, after bosonization and folding, the operator $\sum_{\vec{x}_1,\vec{y}_1}P(\vec{x}_1,\vec{y}_1)$ imposes the same boundary condition as $\Gamma_1$: a Neumann condition on the center-of-mass field and Dirichlet conditions on the relative fields.
This argument can be considered as alternative derivation of the boundary condition arising from the Bell-state measurements.

\subsection{Boundary states in multicomponent \texorpdfstring{$S^1\!/\mathbb{Z}_2$}{orbifold} free-boson CFT}
We here construct a class of boundary states in the multicomponent $S^1\!/\mathbb{Z}_2$ free-boson CFT.
While the $g$-factors of boundary states and the scaling dimensions of BCCOs can be determined from the spectrum of the theory with boundaries, they can be analyzed more concisely and systematically using the boundary-state formalism.
Our construction builds upon the boundary states in the multicomponent $S^1$ free boson, from which we derive the corresponding boundary states in the multicomponent $S^1\!/\mathbb{Z}_2$ free boson through symmetrization.
To this end, below we begin by introducing a class of the boundary states in the multicomponent $S^1$ theory corresponding to the mixed Dirichlet-Neumann boundary conditions~\cite{oshikawa2006junctions,furukawa2011entanglement,oshikawa2010boundary}.
The boundary states obtained through this construction are not guaranteed to exhaust all possible boundary states in the theory.
Nevertheless, this class of boundary states contains all the boundary states $\ket*{\Gamma_{0,1,2}}_{\mathrm{orb}}$ that are necessary to study the SRE and is thus sufficient for our purpose.

\subsubsection{Bulk CFT}
The Lagrangian density of the bulk theory is
\begin{equation}
  \mathcal{L} = \frac{\kappa}{4\pi}(\partial_\mu\vec{\phi})^2,
\end{equation}
where $\vec{\phi}=(\phi_1,\phi_2,\ldots,\phi_N)$ is the $N$-component boson field, and $\kappa$ is the coupling constant that we keep as a variable for the sake of generality; to express the pairs of the Ising CFTs, we will later focus on the case of $\kappa=2$ and the compactification radius $R=1$.
We describe the dual field of $\vec{\phi}$ as $\vec{\theta}=(\theta_1,\theta_2,\ldots,\theta_N)$.
These boson fields are compactified as
\begin{equation}
  \vec{\phi}\sim\vec{\phi}+2\pi\vec{R},\quad \vec{\theta}\sim\vec{\theta}+2\pi\vec{K}.
\end{equation}
The vectors $\vec{R}$ and $\vec{K}$ belong to the following compactification lattice,
\begin{align}
  \Lambda
   & = \qty{\vec{x}\;|\; \vec{x}=\sum_{i=1}^{N}n_i\vec{e}_i,\;n_i\in\mathbb{Z}}, \\
  \Lambda^\ast
   & = \qty{\vec{y}\;|\; \vec{y}=\sum_{i=1}^{N}m_i\vec{f}_i,\;m_i\in\mathbb{Z}}, \\
  \vec{e}_i\cdot\vec{f}_j
   & = \delta_{ij},
\end{align}
respectively, where the lattice $\Lambda^\ast$ is the dual of $\Lambda$, and $\vec{R}\cdot\vec{K}\in\mathbb{Z}$ is satisfied.
For instance, when all the field components are decoupled, the compactification lattice is simply a square lattice.
The fields $\vec{\phi}$ and $\vec{\theta}$ defined on the cylinder of circumference $L$ can be expanded in the Fourier modes as follows:
\begin{align}
    & \vec{\phi}(x,t)\notag                                                                                                                    \\
  = & \vec{\phi_0} + \frac{2\pi}{L}\qty(\vec{R}x + \frac{\vec{K}}{\kappa}t) \notag                                                             \\
    & + \frac{1}{\sqrt{2\kappa}}\sum_{n=1}^\infty\frac{1}{\sqrt{n}}\qty(\vec{a}_n^L e^{-ik_n(x+t)} + \vec{a}_n^R e^{ik_n(x-t)} + \text{H.c.}), \\
    & \vec{\theta}(x,t)\notag                                                                                                                  \\
  = & \vec{\theta_0} + \frac{2\pi}{L}\qty(\vec{K}x + \kappa\vec{R}t) \notag                                                                    \\
    & + \sqrt{\frac{\kappa}{2}}\sum_{n=1}^\infty\frac{1}{\sqrt{n}}\qty(\vec{a}_n^L e^{-ik_n(x+t)} - \vec{a}_n^R e^{ik_n(x-t)} + \text{H.c.}),
\end{align}
where $\vec{\phi}_0$ and $\vec{\theta}_0$ are the zero-mode operators, $\vec{a}_n^{L(R)}$ is a vector of the annihilation operators of left- (right-) moving oscillator modes having quantum number $n$, and $k_n=2\pi n/L$.
These operators satisfy the following commutation relations:
\begin{align}
  [(\vec{a}_n^s)_i,(\vec{a}_m^t)_j^\dag] & = \delta_{nm}\delta_{st}\delta_{ij}\;\;(s,t\in\qty{L,R}), \\
  [(\vec{\phi}_0)_i,(\vec{K})_j]         & = i\delta_{ij},                                           \\
  [(\vec{\theta}_0)_i,(\vec{R})_j]       & = i\delta_{ij}.
\end{align}
The Hamiltonian on the cylinder can be expressed in terms of these operators as follows:
\begin{equation}
  H = \frac{2\pi}{L}\qty(\frac{\kappa\vec{R}^2}{2} {+} \frac{\vec{K}^2}{2\kappa} {+} \sum_{n=1}^\infty n\qty[(\vec{a}_n^L)^\dag{\cdot} \vec{a}_n^L {+} (\vec{a}_n^R)^\dag{\cdot}\vec{a}_n^R] {-} \frac{N}{12}).
\end{equation}
Its ground state is the simultaneous eigenstate of the winding modes $\vec{R},\vec{K}$ and the oscillators $\vec{a}_n^{L/R}$ with zero eigenvalues, and the ground-state energy is the Casimir energy $E_{\mathrm{GS}}(L) = -\pi N/(6L)$.

\subsubsection{Conformal invariance of a boundary state}
The condition of conformal invariance of a boundary state $\ket*{\Gamma}$ is given by
\begin{equation}\label{eq:conformal-invariance}
  (L_m - \bar{L}_{-m})\ket*{\Gamma}=0,
\end{equation}
where $L_m,\bar{L}_m$ are the Virasoro generators, expressed in terms of the Fourier modes as follows:
\begin{equation}\label{eq:virasoro-generator}
  L_m = \frac{1}{2}\sum_l :\vec{\alpha}_{m-l}^L\cdot \vec{\alpha}_l^L:,\quad\bar{L}_m = \frac{1}{2}\sum_l :\vec{\alpha}_{m-l}^R\cdot\vec{\alpha}_l^R:.
\end{equation}
Here, $:{\cdots}:$ is the operator normal ordering, and the operators $\vec{\alpha}_n^{L,R}$ are defined as
\begin{align}
  \vec{\alpha}_n^L & = \left\{\begin{array}{cc}
                                -i\sqrt{n}\vec{a}_n^L                                                         & (n>0) \\\\
                                \displaystyle\frac{1}{\sqrt{2\kappa}}\vec{K} + \sqrt{\frac{\kappa}{2}}\vec{R} & (n=0) \\\\
                                i\sqrt{n}(\vec{a}_{-n}^L)^\dag                                                & (n<0)
                              \end{array}\right.,
  \\
  \vec{\alpha}_n^R & = \left\{\begin{array}{cc}
                                -i\sqrt{n}\vec{a}_n^R                                                         & (n>0) \\\\
                                \displaystyle\frac{1}{\sqrt{2\kappa}}\vec{K} - \sqrt{\frac{\kappa}{2}}\vec{R} & (n=0) \\\\
                                i\sqrt{n}(\vec{a}_{-n}^R)^\dag                                                & (n<0)
                              \end{array}\right..
\end{align}
A sufficient condition to satisfy Eq.~\eqref{eq:conformal-invariance} is
\begin{equation}\label{eq:sufficient-condition}
  (\vec{\alpha}_m^L - \mathcal{R}\vec{\alpha}_{-m}^R)\ket{\Gamma} = 0,
\end{equation}
where $\mathcal{R}$ is an $N\times N$ orthogonal matrix.
This matrix $\mathcal{R}$ becomes symmetric if each of the components satisfies either the DBC or the NBC.
The states satisfying the sufficient condition~\eqref{eq:sufficient-condition} for $m\neq0$ can be given by the coherent states
\begin{equation}\label{coherent}
  S(\mathcal{R})\ket*{\vec{R},\vec{K}},
\end{equation}
where
\begin{equation}
  S(\mathcal{R}) = \exp[-\sum_{n=1}^{\infty}(\vec{a}_n^L)^\dag\cdot\mathcal{R}(\vec{a}_n^R)^\dag]
\end{equation}
is the squeezing operator, and $\ket*{\vec{R},\vec{K}}$ are the oscillator vacua being the eigenstates of the winding modes $\vec{R},\vec{K}$.
These coherent states are known as the Ishibashi states in BCFT~\cite{ishibashi1989boundary}.
The condition~\eqref{eq:sufficient-condition} for $m=0$ restricts the allowed states $\ket*{\vec{R},\vec{K}}$ to satisfy
\begin{equation}\label{eq:winding-condition}
  \vec{K} + \kappa \vec{R} = \mathcal{R}(\vec{K} - \kappa\vec{R}).
\end{equation}
Thus, any linear combination of the coherent states $S(\mathcal{R})\ket*{\vec{R},\vec{K}}$ that satisfy~\eqref{eq:winding-condition} are conformally invariant.

\subsubsection{Cardy's consistency condition}
\begin{figure}[tb]
  \centering
  \includegraphics[width=\linewidth, clip]{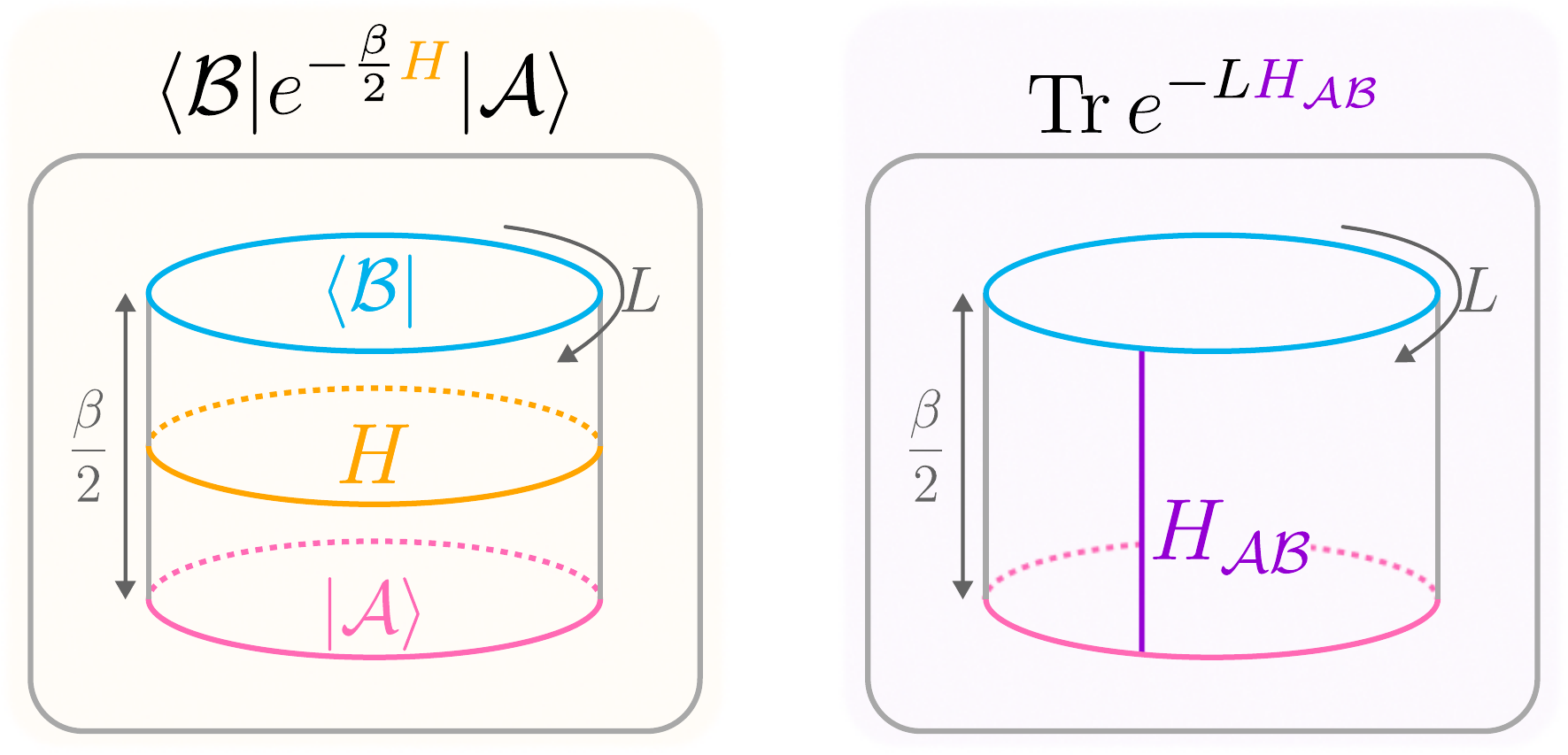}
  \caption{Schematic illustration of the relationship between two representations of the partition function.
    The left panel represents it as the transition amplitude between the boundary states $\ket*{\mathcal{A}}$ and $\ket*{\mathcal{B}}$ evolved through the imaginary-time evolution governed by the Hamiltonian on a cylinder.
    The right panel represents the same quantity as a trace of the Gibbs state, $e^{-LH_{\mathcal{A}\mathcal{B}}}$, where $H_{\mathcal{A}\mathcal{B}}$ is the Hamiltonian on a strip with boundary conditions $\mathcal{A}$ or $\mathcal{B}$ at each side.
    These two expressions are related by the modular transformation $\tau\to-1/\tau$, which exchanges the direction of imaginary time and space.}
  \label{fig:consistency}
\end{figure}

To correctly describe the theory with boundaries, physical boundary states must also satisfy the consistency condition known as the Cardy's consistency condition.
To explain this, consider the transition amplitude between a pair of boundary states $\ket*{\mathcal{A}},\ket*{\mathcal{B}}$ as follows:
\begin{equation}
  Z_{\mathcal{A}\mathcal{B}}(q)=\mel*{\mathcal{B}}{e^{-\frac{\beta}{2}H}}{\mathcal{A}},
\end{equation}
where $q=e^{2\pi i\tau}$ with $\tau=i\beta/L$ being the modular parameter.
Under the modular transformation $\tau\to-1/\tau\;(q\to\tilde{q}=e^{-2\pi i/\tau})$, which exchanges the direction of imaginary time and space, this amplitude is written as the trace over the Hilbert space $\mathcal{H}_{\mathcal{A}\mathcal{B}}$ of the theory on a strip with boundary conditions $\mathcal{A}$ and $\mathcal{B}$:
\begin{equation}
  Z_{\mathcal{A}\mathcal{B}}(q) = \Tr e^{-LH_{\mathcal{A}\mathcal{B}}},
\end{equation}
where $H_{\mathcal{A}\mathcal{B}}$ is the corresponding CFT Hamiltonian (see Fig.~\ref{fig:consistency}).
Given that the boundary conditions are conformally invariant, this Hilbert space decomposes into irreducible representations $\mathcal{H}_h$ of the Virasoro algebra.
Thus, the amplitude must be written as
\begin{equation}
  Z_{\mathcal{A}\mathcal{B}}(q)=\sum_{h}n_{\mathcal{A}\mathcal{B}}^{h}\chi_h(\tilde{q}),
\end{equation}
where
\begin{equation}
  n_{\mathcal{A}\mathcal{B}}^{h}\in\mathbb{N}_0\label{cardy}
\end{equation}
is a nonnegative integer interpreted as the number of primary fields with conformal weight $h$ in the spectrum, and $\chi_h(\tilde{q})$ is the Virasoro character.
When $\mathcal{A}=\mathcal{B}$ ($\mathcal{A}\neq\mathcal{B}$), we call the condition~\eqref{cardy} self (mutual) consistency condition.
If the ground state of $H_{\mathcal{A}\mathcal{A}}$ is unique, the boundary condition $\mathcal{A}$ satisfies $n_{\mathcal{A}\mathcal{A}}^0=1$.
We note that the consistency conditions cannot be satisfied by a single coherent state in Eq.~\eqref{coherent} alone.
Thus, to construct physical boundary states, we must take appropriate linear combinations of the coherent states.

\subsubsection{Construction of the consistent boundary states in the multicomponent \texorpdfstring{$S^1$}{compactified} free-boson CFT}
We now construct a class of the consistent boundary states for general mixed Dirichlet-Neumann boundary conditions.
To begin with, we consider the simplest case of the fully DBC or fully NBC boundary states.
The fully DBC corresponds to taking $\mathcal{R}=I$, and Eq.~\eqref{eq:winding-condition} translates to $\vec{R}=\vec{0}$.
The consistent boundary state is then
\begin{equation}\label{fullyD}
  \ket*{D(\vec{\phi}_D)}_c := g_D \sum_{\vec{K}\in\Lambda^\ast} e^{-i\vec{K}\cdot\vec{\phi}_D}S(I)\ket*{\vec{0},\vec{K}},
\end{equation}
where the coefficient $e^{-i\vec{K}\cdot\vec{\phi}_D}$ is introduced to ensure that this state is an eigenstate of $\vec{\phi}$ with eigenvalue $\vec{\phi}_D$.
The subscript in $\ket*{\cdots}_c$ stands for {\it circle} of the $S^1$ theory, which we use to distinguish it from the boundary states in the $S^1\!/\mathbb{Z}_2$ orbifold theory.
The overall coefficient $g_D$ plays the role of the $g$-factor, which can be given by the overlap of the boundary state with the ground state $\ket*{\mathrm{GS}}=\ket*{\vec{0},\vec{0}}$.

Similarly, the fully NBC corresponds to taking $\mathcal{R}={-}I$, for which the boundary state is
\begin{equation}
  \ket*{N(\vec{\theta}_D)}_c := g_N \sum_{\vec{R}\in\Lambda} e^{-i\vec{R}\cdot\vec{\theta}_D} S(-I)\ket*{\vec{R},\vec{0}},
\end{equation}
where $g_N$ is the corresponding $g$-factor.
Since the NBC for $\vec{\phi}$ is equivalent to DBC for its dual $\vec{\theta}$, these states are labeled by the eigenvalue $\vec{\theta}_D$ of $\vec{\theta}$.
The $g$-factors $g_D$ and $g_N$ are determined from the self-consistency condition and given by
\begin{align}
  g_D
   & = (2\kappa)^{-N/4}v_0(\Lambda)^{-1/2},        \label{eq:$g$-factor-dbc}    \\
  g_N
   & = \qty(\frac{\kappa}{2})^{N/4}v_0(\Lambda)^{1/2}\label{eq:$g$-factor-nbc},
\end{align}
where $v_0(\Lambda)$ is the unit-cell volume of the lattice $\Lambda$ (see Appendix~\ref{appendix:derivation-$g$-factor} for the derivation).

A general boundary state for mixed Dirichlet-Neumann boundary conditions can be expressed as
\begin{align}\label{eq:boundary-state-general}
     & \ket*{\mathcal{R}(\vec{\phi}_D,\vec{\theta}_D)}_c \notag                                                                                                                                      \\
  := & \,g_\mathcal{R} \sum_{\vec{R}\in\Lambda_\mathcal{R}}\sum_{\vec{K}\in\Lambda^\ast_\mathcal{R}} e^{-i\vec{R}\cdot\vec{\theta}_D-i\vec{K}\cdot\vec{\phi}_D}S(\mathcal{R})\ket*{\vec{R},\vec{K}}.
\end{align}
Here, $\Lambda_\mathcal{R}$ and $\Lambda_\mathcal{R}^\ast$ are subspaces of $\Lambda$ and $\Lambda^\ast$ that satisfy Eq.~\eqref{eq:winding-condition}.
When each of the components satisfies either the DBC or the NBC, we can write $\mathcal{R}=\mathcal{P}_D-\mathcal{P}_N$, where $\mathcal{P}_{D/N}$ is the projection matrix onto the subspace $\mathcal{V}_{D/N}$ satisfying DBC/NBC.
In terms of the projection matrices, Eq.~\eqref{eq:winding-condition} becomes
\begin{equation}
  \mathcal{P}_N\vec{K}=0,\;\mathcal{P}_D\vec{R}=0,
\end{equation}
and the subspaces $\Lambda_\mathcal{R}$ and $\Lambda_\mathcal{R}^\ast$ can be expressed as
\begin{equation}
  \Lambda_\mathcal{R} = \Lambda\cap\mathcal{V}_N,\;\Lambda_\mathcal{R}^\ast = \Lambda^\ast\cap\mathcal{V}_D.
\end{equation}
We note that the $g$-factor $g_{\mathcal{R}}$ does not depend on the zero-mode phases $\vec{\phi}_D,\vec{\theta}_D$.

\begin{center}
  {\fontsize{9pt}{12pt}\selectfont \textsf{ $g$-factor of $|\Gamma_1\rangle_{c}$}}
\end{center}

We now focus on the mixed boundary condition $\Gamma_1$ in Eq.~\eqref{eq:gamma1_description} and construct the corresponding boundary state $|\Gamma_1\rangle_c$ of the $S^1$ theory.
In this case, the subspace $\mathcal{V}_N$ is a one-dimensional space spanned by the vector $\vec{d}=(1,1,\ldots,1)^{\mathrm{T}}$, which corresponds to the NBC of the center of mass field $\sum_{i=1}^{N}\phi_i$.
The complement of $\mathcal{V}_N$ is the subspace $\mathcal{V}_D$, which corresponds to the DBC of the phase differences $\phi_i-\phi_{i{+}1}$.
The zero-mode phases are simply given by $\vec{\phi}_D=0$ and $\vec{\theta}_D=0$.
Since all the field components are decoupled in the bulk theory, the compactification lattice $\Lambda$ is a square lattice of lattice constant $R$, and the sublattices are written as
\begin{align}
  \Lambda_\mathcal{R}      & = \qty{nR\vec{d}\;\vert\;n\in\mathbb{Z}},                      \\
  \Lambda_\mathcal{R}^\ast & = \qty{\vec{K}\in\Lambda^\ast \;\vert\;\vec{d}\cdot\vec{K}=0}.
\end{align}
To calculate the $g$-factor of $\ket*{\Gamma_1}_{c}$, we consider the mutual consistency with the boundary state $\ket*{D(\vec{\phi}_D)}_c$ in Eq.~\eqref{fullyD}.
The amplitude between the two states reads
\begin{align}\label{eq:d_gamma1_amplitude}
  Z_{D\Gamma_1}(q)
   & = {}_c\!\mel*{D(\vec{\phi}_D)}{e^{-\frac{\beta}{2}H}}{\Gamma_1}_{c}\notag                                                                                                                         \\
   & = \frac{\sqrt{2}\,g_Dg_1^{\mathrm{circ}}}{(\eta(q))^{N-1}}\sqrt{\frac{\eta(q)}{\theta_2(q)}}\sum_{\vec{K}\in\Lambda_\mathcal{R}^\ast} e^{i\vec{K}\cdot\vec{\phi}_D}q^{\frac{\vec{K}^2}{4\kappa}},
\end{align}
where we express the $g$-factor of $\ket*{\Gamma_1}_{c}$ by $g_1^{\mathrm{circ}}$ and use the Dedekind eta function $\eta(q)$ and the theta function $\theta_2(q)$ (see Appendix~\ref{appendix:theta-functions} for details).
We note that the contribution $z_{DN}:=\sqrt{\eta(q)/\theta_2(q)}$ originates from the amplitude of the center of mass field, and it is equivalent to the amplitude between the Dirichlet and Neumann boundary states of the single-component $S^1$ theory.
By modular transformation $\tau\to-1/\tau$, the amplitude~\eqref{eq:d_gamma1_amplitude} is rewritten in terms of $\tilde{q}$ as
\begin{align}\label{eq:d_gamma1_amplitude_tilde}
    & Z_{D\Gamma_1}(q) \notag                                                                                                                                                                                                                                                     \\
  = & \frac{\sqrt{2}\,g_Dg_1^{\mathrm{circ}}(2\kappa)^{(N-1)/2}}{v_0(\Lambda^\ast_\mathcal{R})(\eta(\tilde{q}))^{N-1}}\sqrt{\frac{\eta(\tilde{q})}{\theta_4(\tilde{q})}}\sum_{\vec{R}\in\tilde{\Lambda}_\mathcal{R}} \tilde{q}^{\kappa\qty(\vec{R}-\frac{\vec{\phi}_D}{2\pi})^2},
\end{align}
where $\tilde{\Lambda}_{\mathcal{R}}$ is the $(N-1)$-dimensional lattice dual to $\Lambda_{\mathcal{R}}^\ast$.
The mutual consistency leads to the following condition on the $g$-factors:
\begin{equation}
  g_Dg_1^{\mathrm{circ}}\frac{\sqrt{2}(2\kappa)^{(N-1)/2}}{v_0(\Lambda^\ast_\mathcal{R})}=1.
\end{equation}
The unit cell volume of $\Lambda_\mathcal{R}^\ast$ can be calculated as $v_0(\Lambda_\mathcal{R}^\ast)=\sqrt{N} R^{-N+1}$~\cite{oshikawa2010boundary}.
Thus, using Eq.~\eqref{eq:$g$-factor-dbc} with $v_0(\Lambda)=R^N$ for a square lattice, we obtain
\begin{equation}\label{eq:g_1_circ}
  g_1^{\mathrm{circ}} = \sqrt{\frac{N}{2}}(\sqrt{2\kappa}R)^{-N/2+1}.
\end{equation}

\subsubsection{Construction of the consistent boundary states in the multicomponent \texorpdfstring{$S^1\!/\mathbb{Z}_2$}{orbifold} free-boson CFT}
We are now in a position to construct the boundary states in the multicomponent $S^1\!/\mathbb{Z}_2$ free-boson CFT.
To this end, we symmetrize the boundary states of the $S^1$ theory constructed above so that the resulting states are invariant under the $\mathbb{Z}_2$ transformation: $\phi\to-\phi$.
In the case of an $N$-component theory, the $\mathbb{Z}_2$ transformations form a group $G$ with $\abs{G}=2^N$ elements, whose matrix representation in the $N$-dimensional vector space reads $a=\mathrm{diag}(\pm1,\pm1\ldots,\pm1)$ for $a\in G$.
The action of the transformation $a$ on the boundary state is expressed as a unitary transformation $D(a)$ in the following manner
\begin{align}
    & D(a)\ket*{\mathcal{R}(\vec{\phi}_D,\vec{\theta}_D)}_c\notag                                                                                                                                               \\
  = & g_\mathcal{R}\sum_{\vec{R}\in\Lambda_\mathcal{R}}\sum_{\vec{K}\in\Lambda_\mathcal{R}^\ast} e^{-i\vec{R}\cdot\vec{\theta}_D-i\vec{K}\cdot\vec{\phi}_D} S(a\mathcal{R}a)\ket*{a\vec{R},a\vec{K}}\notag      \\
  = & g_\mathcal{R}\sum_{\vec{R}\in a\Lambda_\mathcal{R}}\sum_{\vec{K}\in a\Lambda_\mathcal{R}^\ast} e^{-i\vec{R}\cdot a\vec{\theta}_D-i\vec{K}\cdot a\vec{\phi}_D}S(a\mathcal{R}a)\ket*{\vec{R},\vec{K}}\notag \\
  = & \ket*{a\mathcal{R}a(a\vec{\phi}_D,a\vec{\theta}_D)}_c,
\end{align}
where we use $g_\mathcal{R}=g_{a\mathcal{R}a}$ and $a\Lambda_\mathcal{R}=\Lambda_{a\mathcal{R}a}$ to obtain the last line.

We start from the simplest case of the single-component ($N=1$) $S^1\!/\mathbb{Z}_2$ free-boson CFT~\cite{oshikawa1997boundary}.
In this case, the orthogonal matrix $\mathcal{R}=\pm 1$ remains invariant under any $a\in G=\qty{\pm 1}$, and the zero modes transform as $\phi_D\to-\phi_D,\theta_D\to-\theta_D$ under the action of $a=-1$.
Thus, when the zero-mode parameters $\phi_D$ and $\theta_D$ take generic values, the symmetrized boundary states for $\mathcal{R}=\pm 1$ in the single-component $S^1\!/\mathbb{Z}_2$ theory are given by
\begin{align}\label{sDirichlet}
  \ket*{D(\phi_D)}_{\mathrm{orb}}   & = \frac{1}{\sqrt{2}}\qty(\ket*{D(\phi_D)}_c + \ket*{D(-\phi_D)}_c),                     \\
  \ket*{N(\theta_D)}_{\mathrm{orb}} & = \frac{1}{\sqrt{2}}\qty(\ket*{N(\theta_D)}_c + \ket*{N(-\theta_D)}_c).\label{sNeumann}
\end{align}
The coefficient $1/\sqrt{2}$ is determined from the self-consistency condition.

Meanwhile, when the zero-mode parameters take the fixed-point values of the transformation $G$, i.e., $\phi_D=\phi_E\in\qty{0,\pi R}$ or $\theta_D=\theta_E\in\qty{0,\pi/R}$, the boundary states $\ket*{D(\phi_E)}_c,\ket*{D(-\phi_E)}_c$ and $\ket*{N(\theta_E)}_c,\ket*{N(-\theta_E)}_c$ coincide  and cannot satisfy the consistency condition.
To address this issue, one can introduce the boundary states in the twisted sector, i.e., the $S^1$ free-boson CFT with anti-periodic boundary condition $\phi(x+L,t)=-\phi(x,t)$ on the cylinder.
The boundary states in the twisted sector are labeled by the fixed-point values $\phi_E,\theta_E$, which are the only allowed eigenvalues due to the anti-periodicity.
We denote the boundary states in the twisted sector by the subscript in $\ket*{\cdots}_t$.
Accordingly, the consistent boundary states within the $S^1\!/\mathbb{Z}_2$ theory can be constructed by the following combination of the fixed-point boundary states in the untwisted and twisted sectors:
\begin{align}
  \ket*{D(\phi_E)}_{\mathrm{orb}}   & = \frac{1}{\sqrt{2}}\ket*{D(\phi_E)}_c \pm 2^{-1/4}\ket*{D(\phi_E)}_t,     \\
  \ket*{N(\theta_E)}_{\mathrm{orb}} & = \frac{1}{\sqrt{2}}\ket*{N(\theta_E)}_c \pm 2^{-1/4}\ket*{N(\theta_E)}_t.
\end{align}
Here, the coefficient $1/\sqrt{2}$ for the untwisted sector is determined from the mutual consistency with the boundary states in Eqs.~\eqref{sDirichlet} and \eqref{sNeumann}, and the coefficient $2^{-1/4}$ for the twisted sector is determined from the self-consistency.

We next consider a multicomponent theory with $N\geq2$, for which the orthogonal matrix $\mathcal{R}$ transforms nontrivially under the action of $G$, requiring a more careful analysis of boundary states.
We can categorize these boundary states based on both the structure of $\mathcal{R}$ and the values of $\vec{\phi}_D,\vec{\theta}_D$ into the following cases:
\begin{enumerate}
  \item Block-diagonal $\mathcal{R}=\mathcal{R}_1\oplus\mathcal{R}_2\oplus\cdots\oplus\mathcal{R}_k$ (with appropriate exchange of the components).
  \item Not block-diagonal and,
        \begin{enumerate}
          \item for all $a\in G$, $a\vec{\phi}_D\neq\vec{\phi}_D$ and $a\vec{\theta}_D\neq\vec{\theta}_D$,
          \item $\vec{\phi}_D=-\vec{\phi}_D$ and $\vec{\theta}_D=-\vec{\theta}_D$ under the compactification.
        \end{enumerate}
\end{enumerate}
Case 1 corresponds to the situation where the action of $G$ is disconnected, and the boundary states are factorized into boundary states of the connected subspace of $G$.
Case 2(a) corresponds to the boundary states with generic zero modes, which can be expressed as
\begin{equation}\label{case1orb}
  \ket*{\mathcal{R}(\vec{\phi}_D,\vec{\theta}_D)}_{\mathrm{orb}} = \frac{1}{\sqrt{\abs{G}}}\sum_{a\in G}D(a)\ket*{\mathcal{R}(\vec{\phi}_D,\vec{\theta}_D)}_c.
\end{equation}
In contrast, case 2(b) corresponds to the zero-mode parameters at the fixed points $\vec{\phi}_D=\vec{\phi}_E\in\pi\Lambda\cap\mathcal{V}_D$ and $\vec{\theta}_D=\vec{\theta}_E\in\pi\Lambda^\ast\cap\mathcal{V}_N$, for which the boundary states transform identically under both $a$ and $-a$, allowing us to make $G$-invariant states by symmetrizing over the subgroup $G_0=G/\qty{\pm I}$.
In the similar manner as in the single-component case, one can construct the consistent boundary state in this case by including the boundary state in the twisted sector as follows:
\begin{align}\label{eq:orbifold_boundary_fixed}
    & \ket*{\mathcal{R}(\vec{\phi}_E,\vec{\theta}_E)}_{\mathrm{orb}}\notag                                                                                              \\
  = & \sum_{b\in G_0}D(b)\qty[\frac{1}{\sqrt{\abs{G}}}\ket*{\mathcal{R}(\vec{\phi}_E,\vec{\theta}_E)}_c \pm 2^{-N/4}\ket*{\mathcal{R}(\vec{\phi}_E,\vec{\theta}_E)}_t],
\end{align}
where the coefficients are determined from the consistency conditions (see Appendix~\ref{appendix:consistency-check} for the derivation).
We mention that our construction correctly reproduces the results previously obtained for $N=2$ in Ref.~\cite{becker2017conformal}.

Building on these general constructions, we can now determine the boundary states $\ket*{\Gamma_{0,1,2}}_{\mathrm{orb}}$, which are necessary to analyze the SRE.
First, we recall that $\ket*{\Gamma_1}_{\mathrm{orb}}$ corresponding to the mixed boundary condition in Eq.~\eqref{eq:gamma1_description} has the zero modes $\vec{\phi}_D=\vec{\phi}_E=\vec{0}$ and $\vec{\theta}_D=\vec{\theta}_E=\vec{0}$, which are at the fixed points.
We can thus use Eq.~\eqref{eq:orbifold_boundary_fixed} to obtain
\begin{align}\label{eq:boundary_state_gamma_1_orb}
    & \ket*{\Gamma_1}_{\mathrm{orb}} \notag                                                                            \\
  = & \frac{1}{\sqrt{\abs{G}}}\sum_{b\in G_0}D(b)\ket*{\Gamma_1}_{c} \pm 2^{-N/4}\sum_{b\in G_0}D(b)\ket*{\Gamma_1}_t,
\end{align}
where $\ket*{\Gamma_1}_t$ represents the corresponding boundary state in the twisted sector.
Second, the boundary state $\ket*{\Gamma_0}_{\mathrm{orb}}$ corresponding to the boundary condition in Eq.~\eqref{eq:gamma0_description} takes a simpler form as it factorizes into a tensor product of identical single-component boundary states of DBC as follows:
\begin{equation}\label{gamma0orbd}
  \ket*{\Gamma_0}_{\mathrm{orb}} = \ket*{D(\pi/2)}_{\mathrm{orb}}^{\otimes N}.
\end{equation}
Since the zero modes $\vec{\phi}_D=(\pi/2)\vec{d}$ in this case does not lie at a fixed point, we can express this boundary state in terms of the boundary states in the $N=2\alpha$-component $S^1$ theory by (cf. Eq.~\eqref{case1orb})
\begin{equation}\label{gamma0orb}
  \ket*{\Gamma_0}_{\mathrm{orb}} = \frac{1}{\sqrt{\abs{G}}}\sum_{a\in G}D(a)\ket*{D((\pi/2)\vec{d})}_c.
\end{equation}
Finally, we note that the boundary state $\ket*{\Gamma_2}_{\mathrm{orb}}$ corresponding to the artificially created boundary is factorized into two-component boundary states, each of which is equivalent to $\ket*{\Gamma_1}_{\mathrm{orb}}$ at $\alpha=1$ (see Eq.~\eqref{eq:gamma2_description} and the corresponding discussion).

\begin{center}
  {\fontsize{9pt}{12pt}\selectfont \textsf{$g$-factors of $|\Gamma_{0,1,2}\rangle_{\mathrm{orb}}$ and the full-state SRE}}
\end{center}

The $g$-factor can be calculated from the inner product between the ground state $\ket*{\mathrm{GS}}$ and the boundary state of interest; we note that the ground state of the $S^1\!/\mathbb{Z}_2$ free-boson CFT is the same as the $S^1$ free-boson CFT, i.e., $\ket*{\mathrm{GS}}=\ket*{\vec{0},\vec{0}}$.
To calculate a nontrivial $g$-factor $g_1$ of $|\Gamma_{1}\rangle_{\mathrm{orb}}$ in Eq.~\eqref{eq:boundary_state_gamma_1_orb}, we note that the action $D(b)$ on the boundary state $\ket*{\Gamma_1}_{c}$ does not affect the $g$-factor, and the twisted sector has no overlap with the ground state.
We thus obtain
\begin{align}\label{g1val}
  g_1 & =  \langle\mathrm{GS}|\Gamma_{1}\rangle_{\mathrm{orb}}\notag                           \\
      & =  \frac{1}{\sqrt{\abs{G}}}\times \abs{G_0}\times g_1^{\mathrm{circ}} = \sqrt{\alpha},
\end{align}
where we recall $N=2\alpha$ and use Eq.~\eqref{eq:g_1_circ} with the parameters $\kappa=2$ and $R=1$ that correspond to the pairs of the Ising CFTs.

Similarly, the $g$-factor $g_0$ of $\ket*{\Gamma_0}_{\mathrm{orb}}$ in Eq.~\eqref{gamma0orb} can be obtained by evaluating the inner product as follows:
\begin{align}
  g_0 & =  \langle\mathrm{GS}|\Gamma_{0}\rangle_{\mathrm{orb}}\notag                   \\
      & =  \frac{1}{\sqrt{\abs{G}}}\times \abs{G}\times g_D = 1,\label{eq:unity_of_g0}
\end{align}
showing that the $g$-factor is equal to unity and thus trivial as argued before.
Here, we use Eq.~\eqref{eq:$g$-factor-dbc} with $\kappa=2,\,R=1$.
The result~\eqref{eq:unity_of_g0} can be also inferred from Eq.~\eqref{gamma0orbd} and the fact that, within the single-component $N=1$ orbifold theory, the Dirichlet state $|D(\phi_D)\rangle_{\mathrm{orb}}$ with $\phi_{D}\neq 0,\pi$ is the only state that yields the unit $g$-factor $\langle\mathrm{GS}|D(\phi_D)\rangle_{\mathrm{orb}}=1$.

One can also readily check that the $g$-factor $g_2$ of the boundary state $\ket*{\Gamma_2}_{\mathrm{orb}}$ at the artificially created boundary is indeed equal to unity.
To see this, we recall that $\ket*{\Gamma_2}_{\mathrm{orb}}$ is an $\alpha$-fold product of $\ket*{\Gamma_1}_{\mathrm{orb}}\,\vert_{\alpha=1}$, leading to $g_2=(g_1\,\vert_{\alpha=1})^\alpha=1$, where we use Eq.~\eqref{g1val}.

Using these results, the size-independent contribution to the full-state SRE (cf. Eq.~\eqref{eq:result1}) is given by
\begin{equation}\label{eq:result3}
  c_\alpha = \frac{\ln\sqrt{\alpha}}{\alpha - 1}.
\end{equation}
We again emphasize that this result is universal in that it is determined from the IR property of the theory and independent of microscopic details of the model.
We will also check this result numerically in the following section.

\begin{center}
  {\fontsize{9pt}{12pt}\selectfont \textsf{Scaling dimension of $\mathcal{B}_{2\alpha}$ and the mutual SRE}}
\end{center}

The scaling dimension $\Delta_{2\alpha}$ of the BCCO $\mathcal{B}_{2\alpha}$ that changes the boundary conditions between $\Gamma_1$ and $\Gamma_0$ can be determined by calculating the transition amplitude between $\ket*{\Gamma_1}_{\mathrm{orb}}$ and $\ket*{\Gamma_0}_{\mathrm{orb}}$.
The scaling dimension of the BCCO corresponds to the lowest conformal weight of the Virasoro character included in the decomposition of this amplitude~\cite{recknagel2013boundary}.
We begin with the amplitude between the corresponding $S^1$ free-boson boundary states $\ket*{\Gamma_1}_{c}$ and $\ket*{D((\pi/2)\vec{d})}_c$, which has been already calculated in Eq.~\eqref{eq:d_gamma1_amplitude} and~\eqref{eq:d_gamma1_amplitude_tilde}.
The zero mode in the present case is $\vec{\phi}_D=(\pi/2)\vec{d}$, ensuring $\vec{K}\cdot\vec{\phi}_D=0$, and the amplitude simplifies to
\begin{align}
  Z_{\Gamma_0\Gamma_1}(q)
   & = \frac{ \sqrt{2} \, g_D g_1^{\mathrm{circ} }}{ (\eta(q))^{N-1} }\sqrt{ \frac{\eta(q)}{\theta_2(q)} } \sum_{ \vec{K}\in\Lambda_{\mathcal{R}}^\ast } q^{\frac{\vec{K}^2}{4\kappa}}\notag \\
   & = \frac{1}{(\eta(\tilde{q}))^N}\frac{\theta_2(\tilde{q}^{1/2})}{2}\sum_{\vec{R}\in\tilde{\Lambda}_\mathcal{R}} \tilde{q}^{\kappa \vec{R}^2},
\end{align}
where we use the identity
\begin{equation}
  \sqrt{\frac{\eta(\tilde{q})}{\theta_4(\tilde{q})}} = \frac{\theta_2({\tilde{q}^{1/2}})}{2\eta(\tilde{q})}.
\end{equation}
By combining the relations $\frac{1}{2}\theta_2(\tilde{q}^{1/2}) = \sum_{n=1}^{\infty}\tilde{q}^{\frac{1}{4}\qty(n-\frac{1}{2})^2}$ and $\chi_h(\tilde{q})=\tilde{q}^h/(\eta(\tilde{q}))^N$ for the $N$-component $S^1$ free boson, we find that the lowest conformal weight in the amplitude is $h=1/16$.
This result extends to the $S^1\!/\mathbb{Z}_2$ boundary states $\ket*{\Gamma_1}_{\mathrm{orb}}$ and $\ket*{\Gamma_0}_{\mathrm{orb}}$, whose amplitude is given by:
\begin{align}
     & Z_{\Gamma_0\Gamma_1}^{\,\mathrm{orb}}(q) \notag                                                                           \\
  := & \frac{1}{\abs{G}}\sum_{a\in G}\sum_{b\in G_0}{}_c\!\mel*{D((\pi/2)\vec{d})}{D(a)e^{-\frac{\beta}{2}H}D(b)}{\Gamma_1}_{c}.
\end{align}
The Virasoro character with conformal weight $h=1/16$ appears in this amplitude only when $b=\pm a$.
Since this occurs $\abs{G}$ times in the sum and the amplitude includes a factor of $\abs{G}^{-1}$, the character with $h=1/16$ appears exactly once.
Consequently, the scaling dimension of the BCCO $\mathcal{B}_{2\alpha}$ is
\begin{equation}
  \Delta_{2\alpha}=\frac{1}{16},
\end{equation}
which is independent of $\alpha$.
This implies that, in the long-distance limit, the mutual SRE $W_{\alpha}(l)$ in Eq.~\eqref{eq:result2} and the R\'{e}nyi-$2$ mutual information $I_2(l)$ in Eq.~\eqref{eq:mutual_info} scale as
\begin{align}
  W_{\alpha}(l) & = \frac{1}{4(\alpha-1)}\ln l_c, \label{eq:result4} \\
  I_2(l)        & = \frac{1}{4}\ln l_c,\label{eq:result5}
\end{align}
respectively.
These results reveal that the universal coefficients of the logarithmic scalings in the mutual SRE and R\'{e}nyi mutual information both have the common origin, namely, the scaling dimension of the boundary operator changing the boundaries between $\Gamma_{1}$ and $\Gamma_{0}$.
We mention that, if we define the mutual SRE as $\tilde{W}_2=W_2-I_2$ (cf. Eq.~\eqref{tildeW}), the universal contributions cancel out, and the logarithmic scaling term should be absent, at least in the case of the Ising critical states.

\subsection{Numerical results}
\subsubsection{Full-state SRE}

\begin{figure}[tb]
  \centering
  \includegraphics[width=\linewidth, clip]{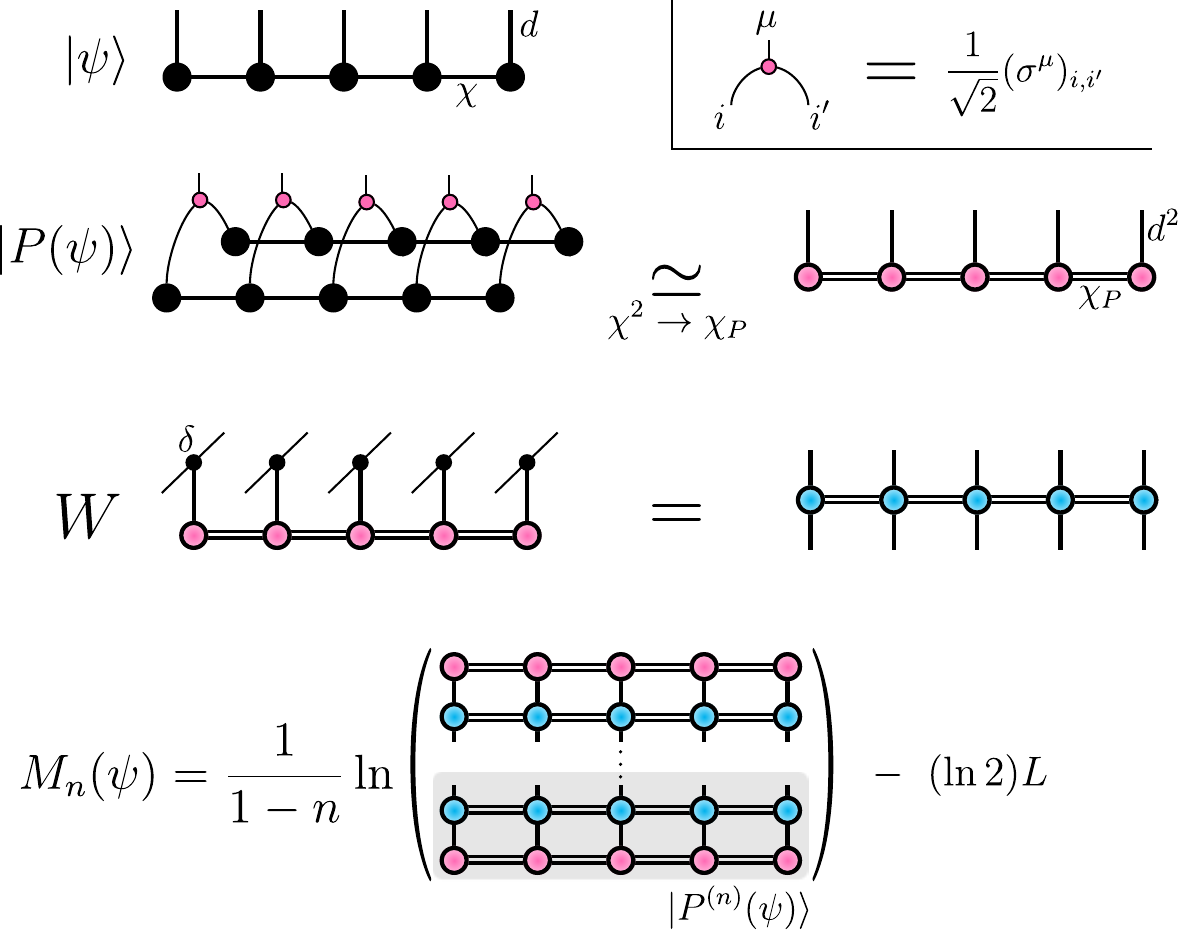}
  \caption{Graphical representation of the tensors used in the replica-Pauli MPS method.
    The doubled spin chain $\ket*{P(\psi)}$ is expressed in the Pauli basis labeled by $\vec{m}$.
    We truncate the bond dimension of $\ket*{P(\psi)}$ from $\chi^2$ to $\chi_P$, where $\chi$ is the bond dimension of $\ket*{\psi}$.
    The MPO $W$ is constructed from $\ket*{P(\psi)}$ by contracting with an identity tensor $\delta$.
    The SRE $M_n(\psi)\;(n=2,3,\ldots,)$ is calculated from the inner product of $\ket*{P^{(n)}(\psi)}=W^{n-1}\ket*{P(\psi)}$.}
  \label{fig:paulibasis}
\end{figure}

\begin{figure}[tb]
  \centering
  \includegraphics[width=0.94\linewidth, clip]{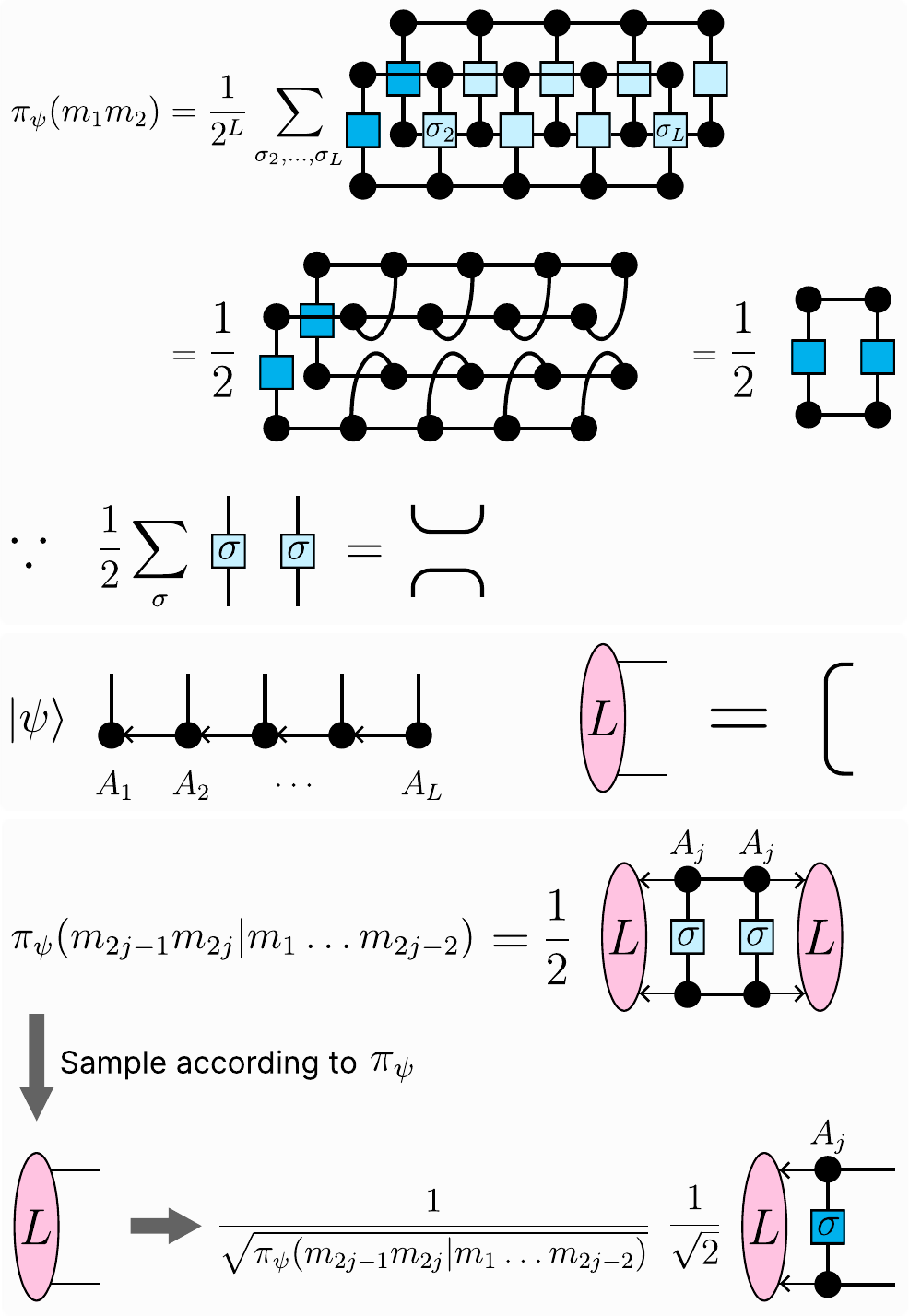}
  \caption{(Top)~Evaluation of the marginal probability $\pi_\psi(m_1m_2)$.
    The orthogonality center of the ground-state MPS $\ket*{\psi}$ is at site $1$.
    The contraction simplifies due to the property $\frac{1}{2}\sum_{\sigma}(\sigma\otimes\sigma)_{ij,kl}=\delta_{ij}\delta_{kl}$.
    (Middle)~The input of the algorithm.
    The environment tensor $L$ is initialized as an identity tensor.
    (Bottom)~The conditional probability is computed iteratively, encoding each outcome $m_{2j{-}1}m_{2j}$ in the environment tensor.
    After a sweep, we obtain a Pauli string $\sigma^{\vec{m}}$ with probability $\Pi_\psi(\vec{m})$.}
  \label{fig:perfectsampling}
\end{figure}

To validate the analytical result obtained in Eq.~\eqref{eq:result3}, we numerically calculate the full-state SRE of the TFIM critical ground state using two distinct methods.
The first approach is the \textit{replica-Pauli MPS} method developed in Ref.~\cite{tarabunga2024nonstabilizernessa} (see Fig.~\ref{fig:paulibasis}).
This method begins with an MPS representation of the ground state $\ket*{\psi}$ and constructs a doubled spin chain $\ket*{P(\psi)}$ expressed in the Pauli basis $\ket*{\vec{m}}\,(\vec{m}\in\qty{0,1}^{2L})$.
This Pauli basis is essentially the Bell basis of the doubled spin chain, and the replica-Pauli MPS method provides a direct numerical implementation of the analytical framework presented in this paper.
Each element of $\ket*{P(\psi)}$ corresponds to the expectation value of a Pauli string: $\braket*{\vec{m}}{P(\psi)} = \mel*{\psi}{\sigma^{\vec{m}}}{\psi}/\sqrt{2^L}$.
To compute the SRE, we use a diagonal matrix product operator (MPO) $W$ defined by $\mel*{\vec{m}'}{W}{\vec{m}}=\delta_{\vec{m}'\vec{m}}\braket*{\vec{m}}{P(\psi)}$.
Through repeated application of this operator, we construct $\ket*{P^{(n)}(\psi)}:=W^{n-1}\ket*{P(\psi)}$, which yields a vector with elements $\braket*{\vec{m}}{P^{(n)}(\psi)}=\mel*{\psi}{\sigma^{\vec{m}}}{\psi}^n/\sqrt{2^{Ln}}$ (corresponding to the Hadamard product of $\ket*{P(\psi)}$).
This leads to the relation:
\begin{equation}
  \frac{1}{2^{Ln}} \sum_{\vec{m}}\mel*{\psi}{\sigma^{\vec{m}}}{\psi}^{2n} = \braket*{P^{(n)}(\psi)}{P^{(n)}(\psi)},
\end{equation}
allowing for the full-state SRE to be expressed as:
\begin{equation}
  M_n(\psi) = \frac{1}{1-n}\ln\braket*{P^{(n)}(\psi)}{P^{(n)}(\psi)} - (\ln2)L.
\end{equation}
This method calculates the full-state SRE as a tensor-network contraction, which allows for reducing the computational cost with controlled error by truncating the bond dimensions.

\begin{figure}[tb]
  \centering
  \includegraphics[width=0.95\linewidth, clip]{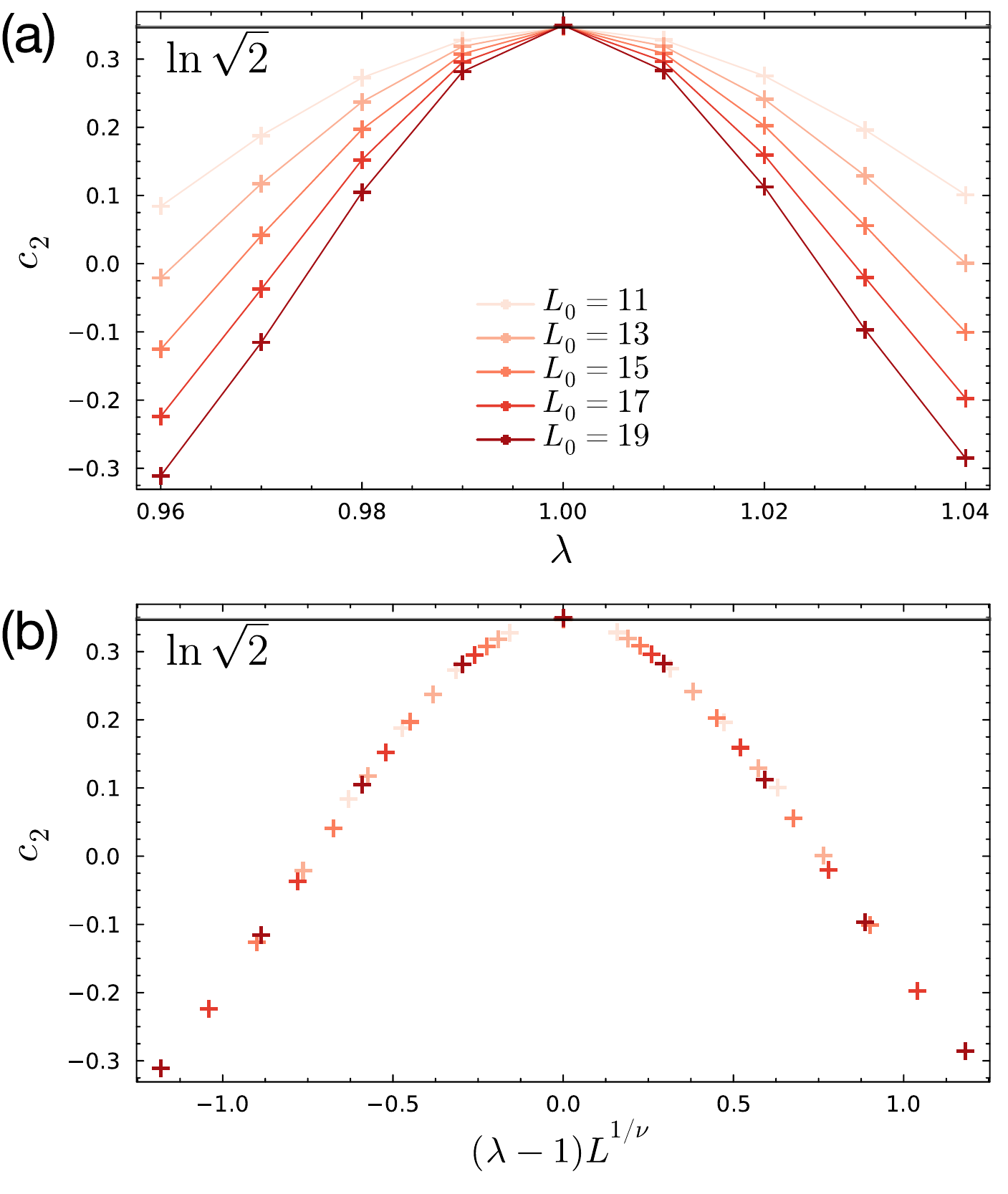}
  \caption{Numerical calculations of the size-independent term $c_2$ in the full-state SRE of the TFIM with $\alpha=2$.
    The size-independent contribution $c_2$ is extracted from the full-state SRE $M_2$ fitted to $M_2=m_2L-c_2+rL^{-1}$ with $L\in\qty{L_0-5,L_0-3,\ldots,L_0+5}$ and $L_0=11,13,\ldots,19$.
    (a)~The size-independent term $c_2$ plotted as a function of the transverse field strength $\lambda$ for different system sizes $L_0$.
    (b)~Plot of the corresponding data collapse with $\nu=1.1$.
    The data converge to the analytical prediction $\ln\sqrt{2}$ at the critical point $\lambda=1$, which is indicated by the solid horizontal line in both panels.}
  \label{fig:c_2_ising}
\end{figure}

\begin{figure}[tb]
  \centering
  \includegraphics[width=0.95\linewidth, clip]{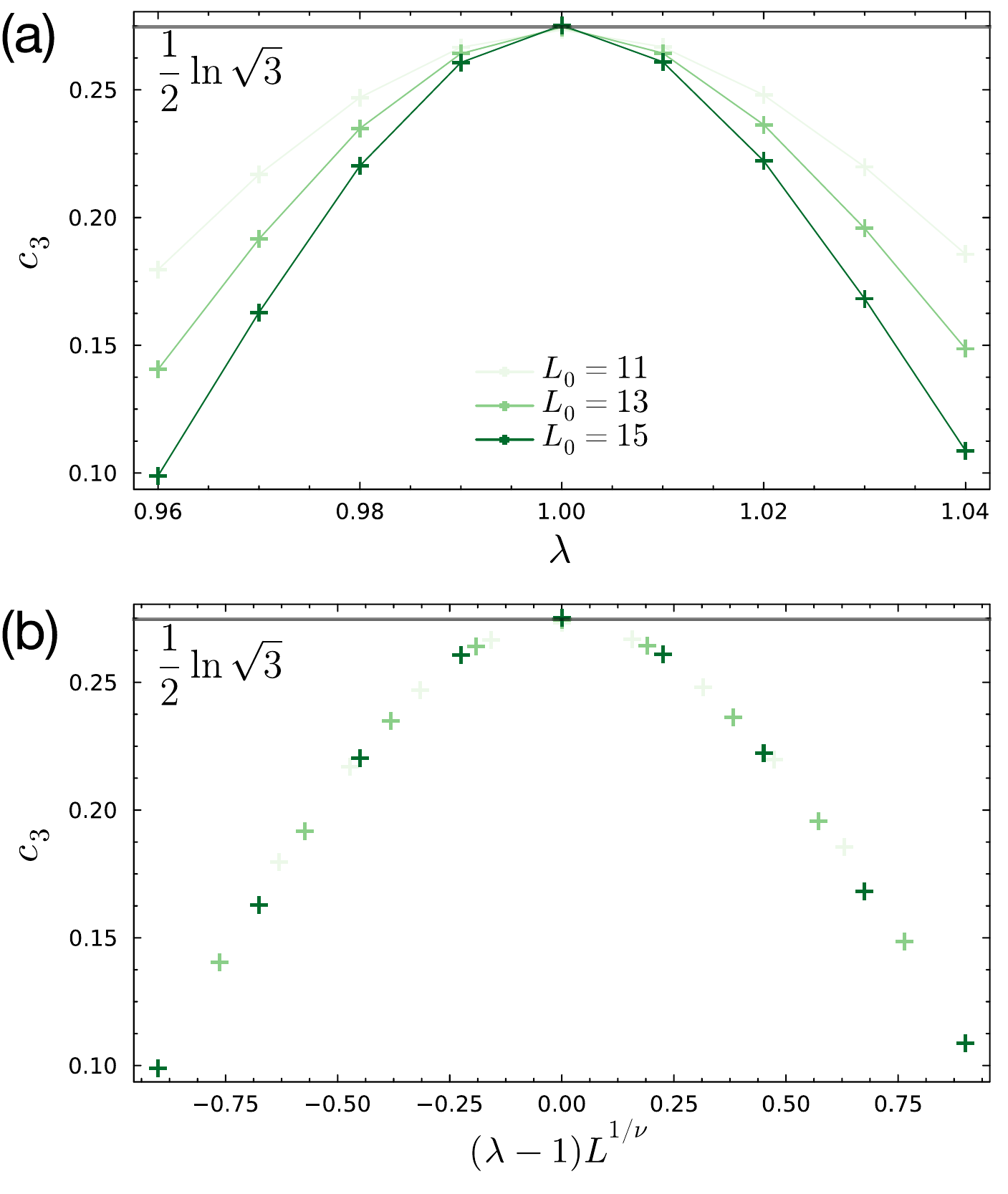}
  \caption{Numerical calculations of the size-independent term $c_3$ in the full-state SRE of the TFIM with $\alpha=3$.
    The size-independent contribution $c_3$ is extracted from the full-state SRE $M_3$ fitted to $M_3=m_3L-c_3$ with $L\in\qty{L_0-5,L_0-3,\ldots,L_0+5}$ and $L_0=11,13,15$.
    (a)~The size-independent term $c_3$ plotted as a function of the transverse field strength $\lambda$ for different system sizes $L_0$.
    (b)~Plot of the corresponding data collapse with $\nu=1.1$.
    The data converge to the analytical prediction $\frac{1}{2}\ln\sqrt{3}$ at the critical point $\lambda=1$, which is indicated by the solid horizontal line in both panels.}
  \label{fig:c_3_ising}
\end{figure}

The second method is the \textit{perfect Pauli sampling} technique introduced in Ref.~\cite{lami2023nonstabilizerness}.
Unlike the replica-Pauli MPS method, this approach enables calculation of the full-state SREs at arbitrary R\'{e}nyi indices $\alpha$ with favorable computational scaling.
The insight is that the SREs can be estimated through direct sampling of Pauli strings from their target probability distribution without requiring Markov chain processes.
Given $\mathcal{N}$ samples of Pauli strings $\{\sigma^{\vec{m}_i}\}_{i=1}^{\mathcal{N}}$ drawn with probability $\Pi_\psi(\vec{m}) = 2^{-L}\Tr^2[\sigma^{\vec{m}}\psi]$, the SREs can be estimated as
\begin{equation}
  M_\alpha(\psi) = \frac{1}{1-\alpha}\ln\qty(\frac{1}{\mathcal{N}}\sum_{i=1}^{\mathcal{N}}\Pi_\psi(\vec{m}_i)^{\alpha-1}) - (\ln 2)L
\end{equation}
for $\alpha\neq 1$, and
\begin{equation}
  M_1(\psi)       = -\frac{1}{\mathcal{N}}\sum_{i=1}^{\mathcal{N}}\ln\Pi_\psi(\vec{m}_i) -(\ln 2)L
\end{equation}
for $\alpha=1$.

The key to the algorithm's efficiency lies in decomposing the full probability $\Pi_\psi(\vec{m})$ into conditional probabilities:
\begin{align}
  \Pi_\psi(\vec{m}) = & \pi_\psi(m_1m_2)\pi_\psi(m_3m_4|m_1m_2)\notag         \\
                      & \cdots\pi_\psi(m_{2L-1}m_{2L}|m_1m_2\ldots m_{2L-2}),
\end{align}
where $\pi_\psi(m_{2j-1}m_{2j}|m_1m_2\ldots m_{2j-2})$ is the probability of Pauli matrix $\sigma^{m_{2j-1}m_{2j}}$ occurring at position $j$ conditioned on the string $\sigma^{m_1m_2}\ldots\sigma^{m_{2j-3}m_{2j-2}}$ having occurred at positions $1,2,\ldots,j{-}1$.
For MPS representations, these conditional probabilities can be computed efficiently through local tensor contractions involving only the MPS tensors at site $j$, the Pauli operator, and an environment tensor that encodes previous outcomes (see Fig.~\ref{fig:perfectsampling}).
The algorithm proceeds sequentially through the sites, sampling the Pauli matrix $\sigma^{m_{2j-1}m_{2j}}$ with probability $\pi_\psi(m_{2j-1}m_{2j}|m_1m_2\ldots m_{2j-2})$.
After a sweep along the system, we obtain a Pauli string $\sigma^{\vec{m}}$ with probability $\Pi_\psi(\vec{m})$, establishing a perfect sampling.

\begin{figure}[tb]
  \centering
  \includegraphics[width=0.9\linewidth, clip]{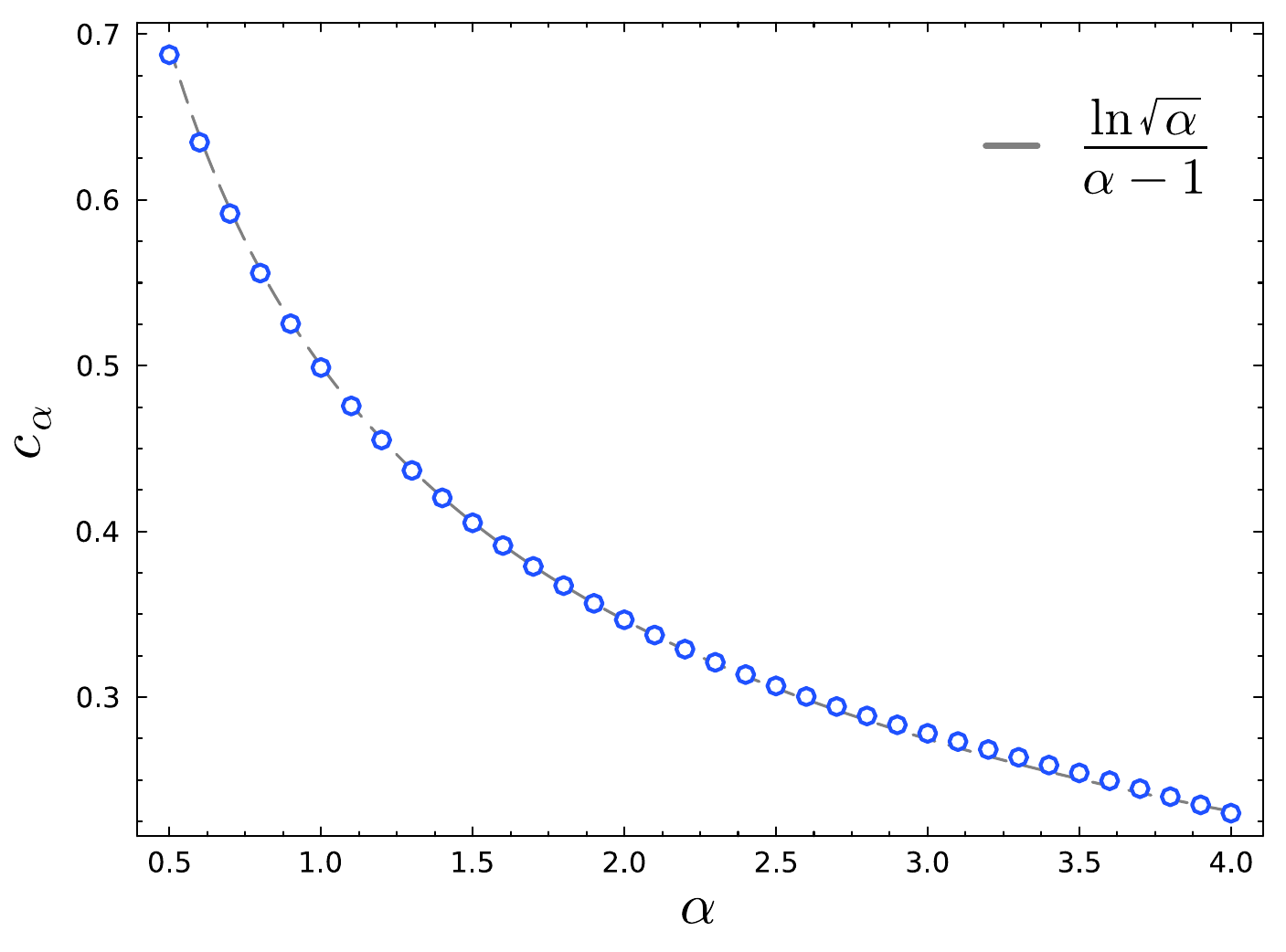}
  \caption{Numerical calculations of the size-independent term $c_\alpha$ in the full-state SRE of the TFIM, as a function of the R\'{e}nyi index $\alpha$.
    The number of samples is $\mathcal{N}=3\times 10^7$ for every system size $L$.
    The constant term is evaluated using the SRE at $L=\qty{4,6,\ldots,14}$, fitted to $M_\alpha=m_\alpha L -c_\alpha + rL^{-1}$.
    Statistical errors are small enough compared to the size of the markers.
    The data show excellent agreement with the analytical prediction $c_\alpha=\ln\sqrt{\alpha}/(\alpha-1)$, which is indicated by the dashed curve.}
  \label{fig:c_alpha_ising}
\end{figure}

We use the first method, the replica-Pauli MPS method, to calculate the full-state SRE as a function of transverse field strength $\lambda$ for $\alpha=2,3$ and different system sizes $L$.
We then perform least squares fitting using the functions $M_2=m_2L-c_2+rL^{-1}$ and $M_3=m_3L -c_3$ to determine the constant terms $c_2$ and $c_3$ as a function of $\lambda$ near the critical point.
The results are presented in Fig.~\ref{fig:c_2_ising} and Fig.~\ref{fig:c_3_ising}, together with finite-size scaling analyses that confirm the universal behavior.
At the critical point $\lambda=1$, we find that the numerical values converge precisely to our analytical predictions $c_2=\ln\sqrt{2}$ and $c_3=\frac{1}{2}\ln\sqrt{3}$.

To further investigate the universal behavior of the SRE, we implement the second method, the perfect Pauli sampling method, to calculate the full-state SRE across a continuous range of (noninteger) R\'{e}nyi indices $\alpha$.
This method leverages the fact that once a representative sample of the Pauli strings is generated, the SRE for any value of $\alpha$ can be efficiently computed from the same dataset. In this manner, we systematically extract the constant contribution $c_\alpha$ as a function of the R\'{e}nyi index $\alpha$ in the interval $\alpha \in [0.5,4.0]$.

The resulting data, presented in Fig.~\ref{fig:c_alpha_ising}, demonstrates remarkable agreement with the analytical prediction given by $c_\alpha = \ln \sqrt{\alpha} / (\alpha-1)$.
This result provides compelling evidence that our replica-based approach accurately captures the universal features of $c_\alpha$ for a broad range of $\alpha$.
We note that this behavior contrasts with the participation entropies of the TFIM in the computational basis, which undergo a phase transition at $\alpha=1$~\cite{stephan2010renyi}.
These results confirm the universality of the SRE and its consistency with our field-theoretical framework~\footnote{Recent work~\cite{rajabpour2025stabilizershannon} predicts a boundary transition in the SRE of the Ising model at $\alpha=4$. However, the numerical precision required to observe such transitions exceeds that achievable with our current sample size, and we leave the numerical verification to future work.}.

\subsubsection{Mutual SRE}
To numerically calculate the mutual SRE, we employ a Markov chain Monte Carlo sampling approach as described in Ref.~\cite{tarabunga2023manybody}.
The mutual SRE $W_2(A:B)$ can be estimated as
\begin{equation}\label{eq:estimator_W}
  W_2(A:B) = -\ln\left\langle\frac{\Tr^4[\sigma^{\vec{m}_A}\rho_A]\,\Tr^4[\sigma^{\vec{m}_B}\rho_B]}{\Tr^4[\sigma^{\vec{m}}\rho_{AB}]}\right\rangle_{\tilde{\Pi}(\vec{m})},
\end{equation}
where the expectation value is taken with respect to the probability distribution $\tilde{\Pi}(\vec{m})\propto\Tr^4[\sigma^{\vec{m}}\rho_{AB}]$, and $\vec{m}=\vec{m}_A\oplus\vec{m}_B$.
Note that the perfect sampling cannot sample the Pauli string drawn from this probability distribution.
Similarly, the mutual information $I_2(A:B)$ can be estimated as
\begin{equation}\label{eq:estimator_I}
  I_2(A:B) = -\ln\left\langle\frac{\Tr^2[\sigma^{\vec{m}_A}\rho_A]\,\Tr^2[\sigma^{\vec{m}_B}\rho_B]}{\Tr^2[\sigma^{\vec{m}}\rho_{AB}]}\right\rangle_{\Pi(\vec{m})},
\end{equation}
where $\Pi(\vec{m})\propto\Tr^2[\sigma^{\vec{m}}\rho_{AB}]$.
The samples of Pauli strings are obtained through standard algorithms such as the Metropolis algorithm, which require only relative probability ratios between consecutive steps.

\begin{figure}[tb]
  \centering
  \includegraphics[width=\linewidth, clip]{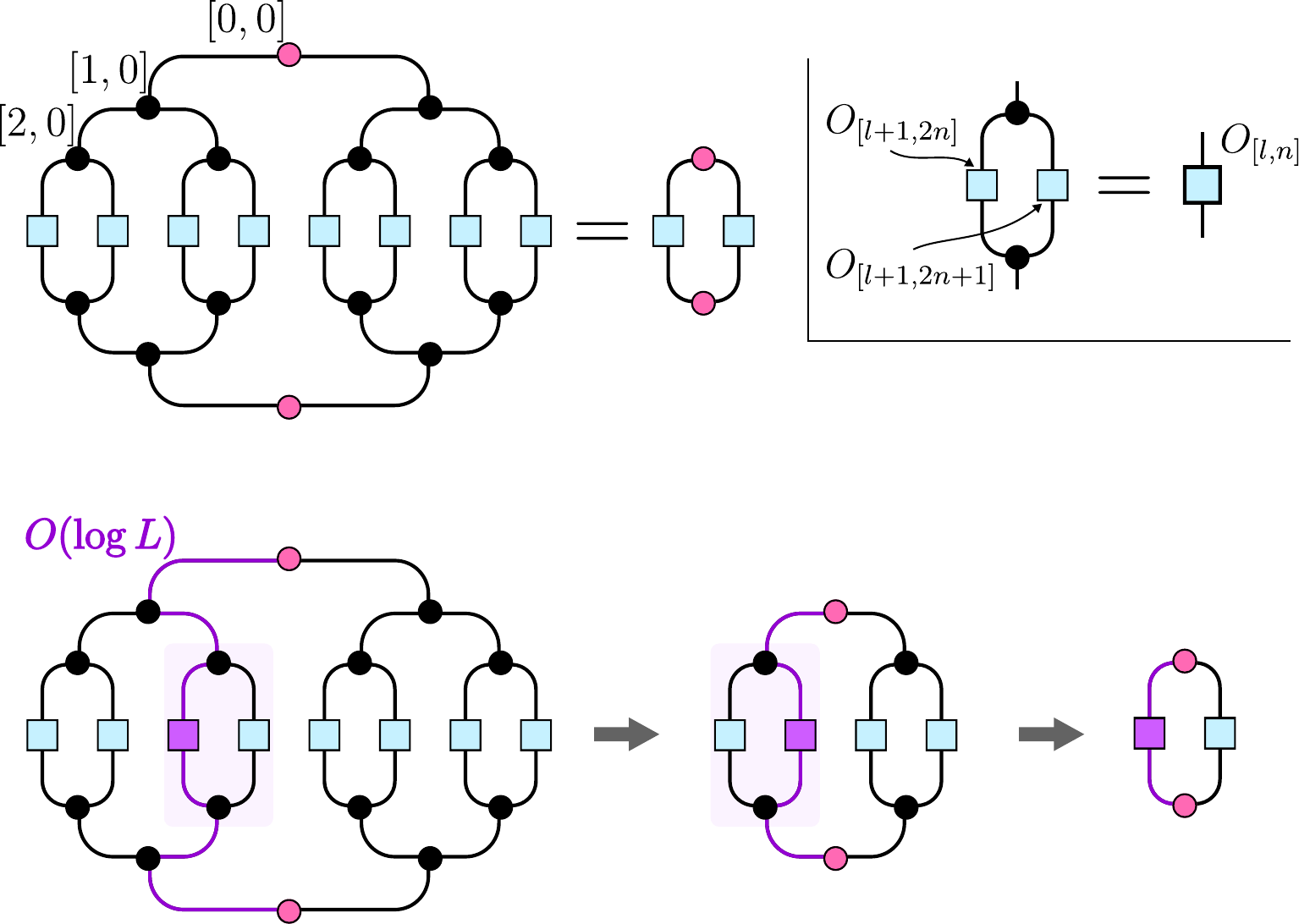}
  \caption{Schematic illustration of efficiently calculating Pauli string expectation values using a tree tensor-network (TTN).
    (Top)~The TTN representation of the ground state $\ket{\psi}$ with site tensors labeled by $(l,n)$, where $l \in \{0,1,\ldots,\log_2(L)\}$ indicates the layer and $n \in \{0,1,\ldots,2^l-1\}$ indicates the position within each layer.
    The evaluation of $\mel{\psi}{\sigma^{\vec{m}}}{\psi}$ is simplified to contracting just four tensors: the two root tensors and the link operators $O_{[1,0]}$ and $O_{[1,1]}$.
    (Bottom)~The efficient update procedure when calculating a new Pauli string that differs from the previous one at a single site.
    This approach requires updating only $O(\log L)$ tensors along the path from the modified site to the root, rather than recomputing the entire contraction, resulting in substantial computational savings for large systems.}
  \label{fig:pauli_markov}
\end{figure}

To calculate the expectation values of Pauli strings efficiently, we implemented a tree tensor-network (TTN) approach that offers significant advantages for large system sizes.
The TTN not only provides an efficient representation of the ground state but also enables efficient updates during the Markov chain process.
The key computational advantage emerges when proposing a new Pauli string that differs from the current one at only few sites.
In this case, we can reuse most of the previously computed tensor contractions, requiring updates only along the logarithmic-depth path from the modified site to the root of the tree.
This hierarchical structure reduces the computational complexity from $O(L)$ (as required in MPS approaches) to $O(\log L)$, resulting in substantial performance improvements.
The procedure consists of the following steps:
\begin{enumerate}
  \item Construct the TTN representation of the ground state $\ket*{\psi}$ using adaptive variational optimization~\cite{gerster2014unconstrained}. In this representation, tensors are organized hierarchically with each tensor labeled by $(l,n)$, where $l \in \{0,1,\ldots,\log_2(L)\}$ indicates the layer and $n \in \{0,1,\ldots,2^l-1\}$ specifies the position within that layer.
  \item For the initial Pauli string $\sigma^{\vec{m}_1}$, evaluate $\Tr[\sigma^{\vec{m}_1}\psi]$ by contracting the tensor-network from the physical indices to the root. During this contraction, we store the intermediate ``link operators'' $O_{[l,n]}$ created by combining the site tensor at $[l,n]$ with its child operators $O_{[l+1,2n]}$ and $O_{[l+1,2n+1]}$.
  \item At each Markov chain step, propose a modified Pauli string that differs from the current one at a minimal number of sites (typically one or two) while preserving any required symmetry constraints, e.g., the $\mathbb{Z}_2$ symmetry in the TFIM.
  \item Calculate the expectation value of the proposed Pauli string by updating only the link operators along the path from the modified sites to the root tensor, while reusing all other link operators.
\end{enumerate}

\begin{figure}[tb]
  \centering
  \includegraphics[width=0.95\linewidth, clip]{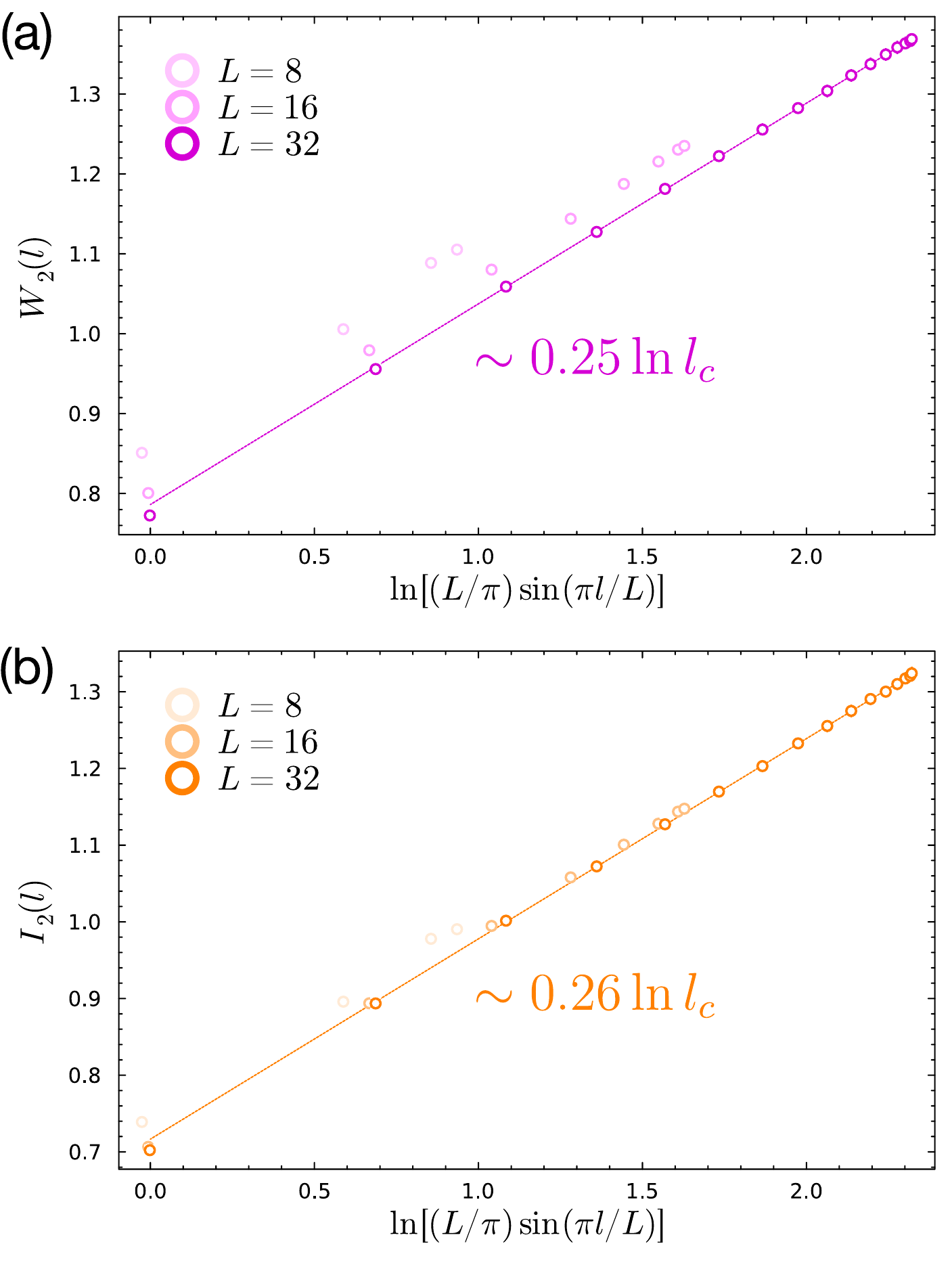}
  \caption{Numerical calculations of (a)~the mutual SRE $W_2(l)$ and (b)~the R\'{e}nyi-$2$ mutual information $I_2(l)$ for the TFIM at different system sizes $L=8,16,32$.
    The fitting lines for $L=32$ are shown in the plots.
    Statistical errors are small enough compared to the size of the markers.
    The data agree well with the analytical results $\sim(1/4)\ln l_c$.}
  \label{fig:mutual_ising}
\end{figure}

Figure~\ref{fig:mutual_ising} presents the numerical results for the mutual $2$-SRE $W_2$ and the R\'{e}nyi-2 mutual information $I_2$ in the TFIM at criticality, evaluated for three different system sizes $L = 8,16,32$.
As shown in Fig.~\ref{fig:mutual_ising}(a), the computed values of the mutual SRE $W_2$ exhibit a clear logarithmic scaling as a function of the system size, in direct agreement with the field-theoretical prediction in Eq.~\eqref{eq:result4} with $\alpha=2$.
Specifically, the coefficient of the logarithmic term in the mutual SRE is consistent with the scaling dimension $\Delta_{2\alpha}=1/16$ of the BCCO in the replicated BCFT, which confirms that nonstabilizerness of critical states can obey universal scaling laws analogous to entanglement in CFT.
Moreover, the mutual R\'{e}nyi-2 information $I_2$, which quantifies long-distance correlations between subsystems, also exhibits a logarithmic scaling as shown in Fig.~\ref{fig:mutual_ising}(b). The extracted coefficients again match the analytical expression in Eq.~\eqref{eq:result5} derived from BCFT.
These results provide a robust benchmark for our theoretical analysis and elucidate the connection between nonstabilizerness and quantum correlations in critical many-body states.

\section{\label{sec:summary}Summary and Discussions}
\subsection{Summary}
We have uncovered the universal behaviors of nonstabilizerness in many-body systems and elucidated its origin by constructing a field-theoretical framework for calculating the stabilizer R\'{e}nyi entropy (SRE) in critical states using boundary conformal field theory (BCFT).
Our key findings show that the SRE exhibits universal behavior through two features (cf. Eqs.~\eqref{eq:result1} and~\eqref{eq:result2}): (i)~the size-independent term in the full-state SRE determined from the $g$-factor of a boundary condition in the replicated theory, and (ii)~the logarithmic scaling term in the mutual SRE with the coefficient given by the scaling dimension of a boundary-condition-changing operator (BCCO).
We have demonstrated these general findings in the case of the Ising criticality through analytical calculations of the size-independent term in the full-state SRE (cf. Eq.~\eqref{eq:result3}) and the coefficient of the logarithmic scaling in the mutual SRE (cf. Eq.~\eqref{eq:result4}).
These results have been numerically confirmed with high accuracy by tensor-network calculations for the critical ground state of the transverse field Ising model (cf. Figs.~\ref{fig:c_2_ising},~\ref{fig:c_3_ising},~\ref{fig:c_alpha_ising}, and~\ref{fig:mutual_ising}).

\subsection{\label{sec:unified}Toward a unified understanding of many-body magic resources}
Our analysis of the SRE may be extended to other resource measures beyond nonstabilizerness.
Resource theories for quantum computation can be developed in other physical systems than qudit systems, such as the resource theory of non-Gaussianity for fermion/boson systems.
In fact, for any given Hilbert space, there exist some resourceful gates in order to achieve universal quantum computation~\cite{eastin2009restrictions}, giving the concept of nonstabilizerness a far more flexible and ubiquitous nature.

The measures for these resources should ideally be invariant under the corresponding classically simulable unitary subgroups $\mathcal{U}_{\mathrm{cl}}$, such as Clifford unitaries or Gaussian unitaries.
A common definition of such measures takes the form $\mathcal{M}(\psi) = \Tr[\Omega \psi^{\otimes k}]$, where $\Omega$ is an element of the commutant of the subgroup $\mathcal{U}_{\mathrm{cl}}$.
The commutant is defined as the linear space of operators
\begin{equation}
  \text{Comm}_k(\mathcal{U}_{\mathrm{cl}}) := \qty{O\;|\; [O,U^{\otimes k}]=0,\; \forall U\in\mathcal{U}_{\mathrm{cl}}}.
\end{equation}
Notably, entanglement (Rényi) entropy can also be formulated in this way, as the commutant of the full unitary group is spanned only by the permutation operators (cf. Schur-Weyl duality).
This family of resource measures includes the SRE for nonstabilizerness and anti-flatness for fermionic non-Gaussianity~\cite{sierant2025fermionic}.

Alternatively, there exist measures based on entanglement generated through convolution, such as magic entropy.
For instance, a measure for non-Gaussianity can be defined by taking the key unitary in the convolution to be a beam splitter operation.
Specifically, for fermions~\cite{lyu2024fermionic,coffman2025measuring}, we have
\begin{equation}
  \boxtimes_2\rho = \Tr_2[V(\rho\otimes\rho)V^\dag],\quad V = \exp(\frac{\pi}{8}\sum_{j}\gamma_j^{(1)}\gamma_j^{(2)}),
\end{equation}
where $\gamma_j$ denotes the Majorana operator at the $j$-th site, and for bosons~\cite{bu2025efficient,hahn2025measuring},
\begin{align}
  \boxtimes_2\rho & = \Tr_2[V(\rho\otimes\rho)V^\dag], \notag                                            \\
  V               & =\exp(\frac{\pi}{4}\sum_{j}(a_j^{(1)\,\dag} a_j^{(2)} - a_j^{(1)} a_j^{(2)\,\dag})),
\end{align}
where $a_j$ is the annihilation operator for the $j$-th mode.
Crucially, the entropic measures defined through these convolutions satisfy invariance under the subgroups $\mathcal{U}_{\mathrm{cl}}$.
If we choose the Rényi-2 entropy (purity) for the measure, it can be expressed as follows:
\begin{align}
  S_2(\boxtimes_K\psi)
   & = -\ln\Tr[\mathbb{F}(\boxtimes_K\psi)^{\otimes 2}]\notag                \\
   & = -\ln\Tr[(\boxtimes_K^\dag)^{\otimes 2}(\mathbb{F})\psi^{\otimes 2K}].
\end{align}
Then, from the unitary invariance, the dual of the convolution applied to the swap operator $(\boxtimes^\dag)^{\otimes 2}(\mathbb{F})$ necessarily belongs to the commutant.
The same applies to higher orders of the Rényi entropy, where the swap operator is replaced with a cyclic permutation operator.
This reveals a general relationship between commutant-based measures and convolution-based measures.

In fact, the analysis of all these resource measures in CFT reduces to determining a certain line defect inserted in the partition function of the replicated theory.
The two general strategies we propose in this paper can be summarized as follows.
For commutant-based measures:
\begin{enumerate}
  \item Move to the doubled Hilbert space so that the commutant $\Omega$ is expressed as a projection operator.
  \item Represent the projection operator as boundary perturbations in the Euclidean path-integral formalism.
  \item Identify the conformal boundary condition imposed by the projection, and calculate the corresponding $g$-factor.
\end{enumerate}
For convolution-based measures:
\begin{enumerate}
  \item On the basis of a key unitary $V$ generating the convolution, compute $\ket*{\Psi(\vec{x}_1,\vec{y}_1)}=(V^\dag\otimes V^{\mathsf{T}})\ket*{\vec{x}_1,\vec{y}_1}\ket*{\Phi\cdots\Phi}$ (cf. Eq.~\eqref{eq:projection_V_basis} in the case of qubits).
  \item From the lattice configuration of the projection operator $\sum_{\vec{x}_1,\vec{y}_1}\ketbra*{\Psi(\vec{x}_1,\vec{y}_1)}{\Psi(\vec{x}_1,\vec{y}_1)}$, infer the line defect configuration in the continuum limit, and calculate the corresponding $g$-factor.
\end{enumerate}
The former approach can be done with the help of knowledge of the lattice-CFT correspondence.
The latter approach could provide an intuitive way to determine line defect configurations.
While the classification of conformal defects/boundaries in replicated CFTs with central charge greater than unity remains a challenging problem, our approach might allow for a systematic identification of universal features of general resource measures.
This will require further numerical calculations on a variety of lattice models and their comparison with CFT calculations, which we leave to future studies.

\subsection{Future directions}
Our findings shed light on the intricate relationship between entanglement and nonstabilizerness, a largely unexplored area in quantum many-body physics.
The key advancement presented in this work is that, within our field-theoretical framework, both the entanglement entropy (EE) and the SRE can now be understood on the same footing.
On the one hand, the EE is calculated from a partition function of the replicated theory with sewing boundary conditions on one subsystem and the trivial boundary condition on the other subsystem, which can be expressed as a correlation function of BCCOs~\cite{calabrese2009entanglement,hoshino2024entanglement}.
On the other hand, the SRE corresponds to a partition function of the replicated theory involving interlayer boundary conditions imposed by the Bell-state measurements (see Appendix~\ref{appendix:comparison_ee} for further discussions).
Given this connection, several interesting directions for future research emerge.

\begin{figure}[tb]
  \centering
  \includegraphics[width=0.6\linewidth, clip]{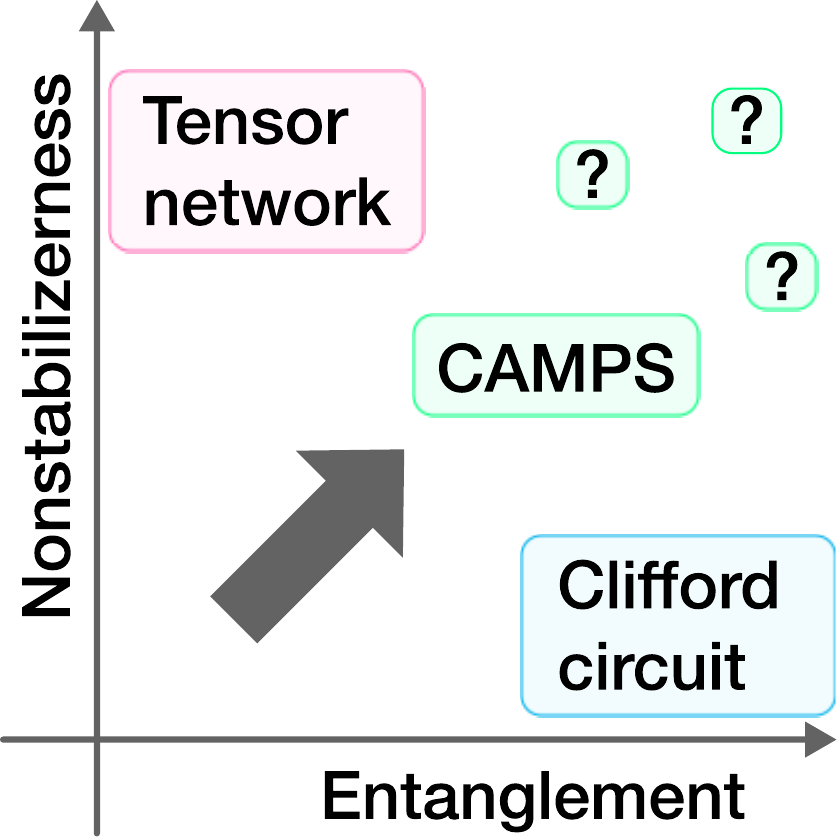}
  \caption{Complementary relationship between entanglement and nonstabilizerness.
    Clifford circuits efficiently encode entanglement without nonstabilizerness, while tensor-networks can represent nonstabilizer states with limited entanglement.
    Classical simulation of quantum many-body states containing both high entanglement and high nonstabilizerness poses a key challenge.
    In this context, clifford augmented matrix product states (CAMPS) may serve as a one possible approach that extends toward such computationally demanding region.}
  \label{fig:camps}
\end{figure}

First, it would be valuable to understand how Clifford unitaries acting on the lattice spin translate to operations in the field theory that describes the long-distance bahavior.
One of the ultimate goals in this direction is to construct a resource theory of nonstabilizerness without references to the microscopic models at UV scales~\cite{hahn2025bridging}.
Recently, Clifford circuit augmented matrix product states (CAMPS) have emerged as a way to enhance the power of MPS by encoding entanglement in the Clifford circuit rather than in the MPS bond indices~\cite{mello2024hybrid,qian2024augmenting,fux2024disentangling} (see Fig.~\ref{fig:camps}).
In this context, Ref.~\cite{frau2024stabilizer} has shown that the entanglement Clifford circuits cannot remove is linked to the nonlocal nonstabilizerness of the state (see also Refs.~\cite{huang2024nonstabilizerness,qian2025quantum} for related discussions).
Moreover, when disentangling the Ising model with open boundary conditions using Clifford circuits, the transformation from ground-state MPS to CAMPS was shown to be equivalent to a Kramers-Wannier duality transformation~\cite{fan2025disentangling}.
In the long-distance limit, this transformation manifests itself as a deformation of the spatial boundary from free to fixed boundary conditions, with the removed entanglement entropy being exactly calculable using BCFT.
This is one example of how Clifford unitaries act as deformations of the boundary in field theory.
Also, exact relations among the SRE, the EE, and the participation entropy in the computational basis have been discussed in Refs.~\cite{turkeshi2023measuring,tirrito2024quantifying,magni2025anticoncentration}, where the Clifford unitary plays a fundamental role.
These findings may provide valuable insights into a field-theoretical description of Clifford unitaries.

Second, it would be worthwhile to consider the universal properties of the SRE in higher-dimensional systems.
In particular, two-dimensional systems may include contributions to the SRE reflecting topological order, analogous to topological entanglement entropy~\cite{kitaev2006topological}.
From a field-theoretical perspective, the SRE in two dimensions can be calculated from a replicated theory with two-dimensional defects (or surfaces) from the Bell-state measurements.
Notable contributions in this direction include Refs.~\cite{misguich2017finitesize,ellison2021symmetryprotected}.

Third, it would be interesting to interpret our results through the AdS/(B)CFT correspondence~\cite{maldacena1999largen,takayanagi2011holographic}.
The AdS/CFT correspondence states that a $d$-dimensional CFT is equivalent to a $(d{+}1)$-dimensional theory of gravity in the anti-de Sitter (AdS) space.
This correspondence enables the computation of various quantities between CFT and gravitational theory, including entanglement entropy~\cite{ryu2006holographic}.
Consequently, it seems natural to explore whether nonstabilizerness in CFT can be computed through the AdS/CFT correspondence.
In this context, we note that Refs.~\cite{white2021conformal,goto2022probing} have used numerical simulations to reveal aspects of nonstabilizerness in AdS/CFT.

Finally, it would be valuable to consider nonstabilizerness in dynamical phenomena, particularly in the context of measurement-induced phenomena~\cite{skinner2019measurementinduced,fisher2023random}.
The entanglement in a monitored random circuit has been known to exhibit phase transitions from a highly entangled volume-law phase to a less entangled area-law phase by varying the strength of mid-circuit measurements, where the critical point can be described by CFT.
Universal behaviors at the critical point have also been observed for the participation entropy in the computational basis~\cite{sierant2022universal}, suggesting that monitored Haar random circuits might exhibit criticality akin to what we found in the SRE (expressions for the SRE of random Haar states without measurements can be found in Ref.~\cite{turkeshi2025pauli}).
Similar transitions of the nonstabilizerness have been observed in random Clifford circuits doped by $T$ gates~\cite{bejan2024dynamical,fux2024entanglement,ahmadi2024mutual} or by other unitaries~\cite{niroula2024phase} and measurement-only circuits~\cite{tarabunga2024magica}.
If these critical points realize some (possibly nonunitary) CFT, we can anticipate that the corresponding SRE will display universal behavior similar to what we found.
Moreover, it would be interesting to consider the MIPT critical point of Clifford random unitaries.
This critical point arises from the competition between Clifford random unitaries and computational basis measurements~\cite{bao2020theory,li2021conformal,buznachahituv2025unveiling}, both of which are free operations within magic-state resource theory.
Thus, we expect a vanishing universal term $c_{\alpha}=0$ at this critical point corresponding to the $g$-factor $g_1=1$, though the line defect must have nonzero energy density for the SRE to vanish after subtracting the normalization term $(\ln 2)L$.
These problems represent an interesting direction for future investigation.

We hope that our study stimulates further studies in these directions.


\begin{acknowledgments}
  We are grateful to Shunsuke Furukawa and Kenya Tasuki for valuable discussions.
  We used the ITensor package~\cite{ITensor,ITensor-r0.3} for tensor-network calculations.
  M.H. was supported by FoPM, WINGS Program, the University of Tokyo.
  Y.A. acknowledges support from the Japan Society for the Promotion of Science through Grant No.~JP19K23424 and from JST FOREST Program (Grant No.~JPMJFR222U, Japan).
  The work of M.O. was partially supported by the JSPS KAKENHI Grants No.~JP23K25791 and No.~JP24H00946.
\end{acknowledgments}

\appendix

\section{\label{appendix:theta-functions}Theta functions}
In this section, we present an overview of the Dedekind eta function and theta functions, which are useful for our calculations of the transition amplitudes presented in the main text.
The Dedekind eta function and the three theta functions are defined as follows:
\begin{align}
  \eta(q)
   & = q^{1/24}\prod_{n=1}^{\infty}(1 - q^n),                                                \\
  \theta_2(q)
   & = \sum_{n\in\mathbb{Z}} q^{(n+1/2)^2/2} = 2q^{1/8}\prod_{n=1}^{\infty}(1-q^n)(1+q^n)^2, \\
  \theta_3(q)
   & = \sum_{n\in\mathbb{Z}}q^{n^2/2} = \prod_{n=1}^{\infty}(1-q^n)(1+q^{n-1/2})^2,          \\
  \theta_4(q)
   & = \sum_{n\in\mathbb{Z}}(-1)^n q^{n^2/2} = \prod_{n=1}^{\infty}(1-q^n)(1-q^{n-1/2})^2.
\end{align}
The infinite product representations of the theta functions can be derived from the Jacobi triple product identity:
\begin{equation}
  \sum_{n\in\mathbb{Z}}q^{n^2/2}t^n
  =  \prod_{n=1}^{\infty}(1-q^n)(1+q^{n-1/2}t)(1+q^{n-1/2}/t).
\end{equation}
A relationship connecting these functions is given by
\begin{equation}
  2\eta(q)^3 = \theta_2(q)\theta_3(q)\theta_4(q).
\end{equation}
Under the modular transformation $\tau\to-1/\tau$ and correspondingly $q\to\tilde{q}$, these functions transform as
\begin{align}
  \eta(\tilde{q})
   & = \sqrt{-i\tau}\eta(q),     \\
  \theta_2(\tilde{q})
   & = \sqrt{-i\tau}\theta_2(q), \\
  \theta_3(\tilde{q})
   & = \sqrt{-i\tau}\theta_3(q), \\
  \theta_4(\tilde{q})
   & = \sqrt{-i\tau}\theta_4(q).
\end{align}
Additional useful identities are
\begin{align}
  \theta_2(q^2) & = \sqrt{(\theta_3(q) - \theta_4(q))/2}, \\
  \theta_3(q^2) & = \sqrt{(\theta_3(q) + \theta_4(q))/2}, \\
  \theta_4(q^2) & = \sqrt{\theta_3(q)\theta_4(q)}.
\end{align}

We demonstrate the practical application of these functions by verifying the mutual consistency between the DBC state $\ket*{D(\phi_D)}$ and the NBC state $\ket*{N(\theta_D)}$ of a single-component $S^1$ free-boson CFT.
The calculation proceeds as follows:
\begin{align}
  z_{DN}
   & = \mel*{D(\phi_D)}{e^{-\frac{\beta}{2}H}}{N(\theta_D)}\notag                                                                                                        \\
   & = g_Dg_N q^{-1/24}\prod_{n=1}^{\infty}\frac{1}{1+q^n}\notag                                                                                                         \\
   & = \sqrt{\frac{\eta(q)}{\theta_2(q)}}\notag                                                                                                                          \\
   & = \sqrt{\frac{\eta(\tilde{q})}{\theta_4(\tilde{q})}}\quad (\because q\to\tilde{q})\notag                                                                            \\
   & = \frac{\theta_2(\tilde{q}^{1/2})}{2\eta(\tilde{q})}\notag                                                                                                          \\
   & = \frac{1}{\eta(\tilde{q})}\sum_{n=1}^{\infty}\tilde{q}^{\frac{1}{4}\qty(n-\frac{1}{2})^2} = \sum_{n=1}^{\infty}\chi_{\frac{1}{4}\qty(n-\frac{1}{2})^2}(\tilde{q}).
\end{align}
We substitute the $g$-factors $g_D$ and $g_N$ in Eqs.~\eqref{eq:$g$-factor-dbc},~\eqref{eq:$g$-factor-nbc}, which will be derived in the following section.
This calculation reveals that the amplitude can indeed be expressed as a sum of Virasoro characters, with the lowest conformal weight $h=1/16$ appearing due to the presence of a BCCO.

\section{\label{appendix:derivation-$g$-factor}Derivation of the $g$-factor of Dirichlet and Neumann boundary states}
We now derive the $g$-factors $g_D$ and $g_N$ in Eqs.~\eqref{eq:$g$-factor-dbc},~\eqref{eq:$g$-factor-nbc} by requiring that the boundary states satisfy Cardy's consistency conditions.
First, we calculate the amplitude between two general boundary states $\ket*{\mathcal{R}(\vec{\phi}_D,\vec{\theta}_D)}$ and $\ket*{\mathcal{R}'(\vec{\phi}_D',\vec{\theta}_D')}$:
\begin{align}\label{eq:general-amplitude}
    & \mel*{\mathcal{R}'(\vec{\phi}_D',\vec{\theta}_D')}{e^{-\frac{\beta}{2}H}}{\mathcal{R}(\vec{\phi}_D,\vec{\theta}_D)}\notag                                                                                                                                                                                                                                        \\
  = & g_{\mathcal{R}'}g_{\mathcal{R}}\sum_{\substack{\vec{R}\in\Lambda_{\mathcal{R}'}\cap\Lambda_{\mathcal{R}}                                      \\ \vec{K}\in\Lambda^\ast_{\mathcal{R}'}\cap\Lambda^\ast_{\mathcal{R}}}} e^{i\vec{R}\cdot\Delta\vec{\theta}_D + i\vec{K}\cdot\Delta\vec{\phi}_D}q^{\frac{\kappa\vec{R}^2}{4}+\frac{\vec{K}^2}{4\kappa}-N/24}\notag \\
    & \times \mel*{0}{S^\dag(\mathcal{R}')q^{\sum\frac{n}{2}\qty((\vec{a}_n^L)^\dag\vec{a}_n^L +(\vec{a}_n^R)^\dag\vec{a}_n^R)} S(\mathcal{R})}{0},
\end{align}
where $\Delta\vec{\phi}_D=\vec{\phi}_D'-\vec{\phi}_D$ and $\Delta\vec{\theta}_D=\vec{\theta}_D'-\vec{\theta}_D$.
The oscillator mode contribution factorizes into amplitudes between the subspaces of the DBC and the NBC:
\begin{align}\label{eq:oscillator-amplitude}
    & \mel*{0}{S^\dag(\mathcal{R}')q^{\sum\frac{n}{2}\qty((\vec{a}_n^L)^\dag\vec{a}_n^L +(\vec{a}_n^R)^\dag\vec{a}_n^R)} S(\mathcal{R})}{0}\notag \\
  = & q^{N/24}\qty(\frac{1}{\eta(q)})^d\qty(\sqrt{\frac{2\eta(q)}{\theta_2(q)}})^{N-d},
\end{align}
where $d=\mathrm{dim}(\mathcal{V}_D'\cap\mathcal{V}_D)+\mathrm{dim}(\mathcal{V}_N'\cap\mathcal{V}_N)$.
This formula was used to calculate the amplitude between $\ket*{D(\vec{\phi}_D)}$ and $\ket*{\Gamma_1}_{c}$ as in Eq.~\eqref{eq:d_gamma1_amplitude} in the main text.
Using the above equations~\eqref{eq:general-amplitude} and~\eqref{eq:oscillator-amplitude} with $d=N$, the amplitudes of the boundary states $\ket*{D(\vec{\phi}_D)}$ and $\ket*{N(\vec{\theta}_D)}$ with themselves reads
\begin{align}
  \mel*{D(\vec{\phi}_D)}{e^{-\frac{\beta}{2}H}}{D(\vec{\phi}_D)}
   & = \frac{g_D^2}{(\eta(q))^N}\sum_{\vec{K}\in\Lambda^\ast}q^{\frac{1}{4\kappa}\vec{K}^2}, \\
  \mel*{N(\vec{\theta}_D)}{e^{-\frac{\beta}{2}H}}{N(\vec{\theta}_D)}
   & = \frac{g_N^2}{(\eta(q))^N}\sum_{\vec{R}\in\Lambda}q^{\frac{\kappa}{4}\vec{R}^2}.
\end{align}
The modular transformation $q\to\tilde{q}$ of these amplitudes follows from the Poisson resummation formula:
\begin{equation}
  \sum_{\vec{y}\in\Lambda} f(\vec{y}) = \frac{1}{v_0(\Lambda)} \sum_{\vec{x}\in\Lambda^\ast}\hat{f}(\vec{x}),
\end{equation}
where $\Lambda$ is a lattice, $\Lambda^\ast$ is its dual, and $\hat{f}$ is the Fourier transform:
\begin{equation}
  \hat{f}(\vec{x}) = \int_V \mu(\vec{y}) e^{-2\pi i\vec{y}\cdot\vec{x}} f(\vec{y}).
\end{equation}
Here $V$ is the vector space containing the lattice $\Lambda$ and $\mu(\vec{y})$ is the measure in $V$, which in our case, $V=\mathbb{R}^N$ and $\mu(\vec{y})=d^Ny$.
For the Gaussian function $f(\vec{y})=q^{a\vec{y}^2}$, the Fourier transform is $\hat{f}(\vec{x})=(-2i\tau a)^{-N/2}\tilde{q}^{\vec{x}^2/4a}$.
Thus, the amplitudes become:
\begin{align}
  \mel*{D(\vec{\phi}_D)}{e^{-\frac{\beta}{2}H}}{D(\vec{\phi}_D)}
   & = \frac{g_D^2(2\kappa)^{N/2}}{(\eta(\tilde{q}))^N v_0(\Lambda^\ast)}\sum_{\vec{R}\in\Lambda}\tilde{q}^{\kappa\vec{R}^2},                 \\
  \mel*{N(\vec{\theta}_D)}{e^{-\frac{\beta}{2}H}}{N(\vec{\theta}_D)}
   & = \frac{g_N^2 (2\kappa^{-1})^{N/2}}{(\eta(\tilde{q}))^N v_0(\Lambda)}\sum_{\vec{K}\in\Lambda^\ast}\tilde{q}^{\frac{1}{\kappa}\vec{K}^2}.
\end{align}
Since the Virasoro character for the $N$-component free boson is $\chi_h(\tilde{q})=\tilde{q}^h/(\eta(\tilde{q})^N)$, the self-consistency condition requires:
\begin{equation}
  \frac{g_D^2(2\kappa)^{N/2}}{v_0(\Lambda^\ast)} = 1,\;\;\frac{g_N^2(2\kappa^{-1})^{N/2}}{v_0(\Lambda)}=1,
\end{equation}
ensuring that the $h=0$ Virasoro character appears exactly once.
Using $v_0(\Lambda)v_0(\Lambda^\ast)=1$, we obtain the results~\eqref{eq:$g$-factor-dbc} and~\eqref{eq:$g$-factor-nbc} in the main text.

\section{\label{appendix:consistency-check}Consistency check of the \texorpdfstring{$S^1\!/\mathbb{Z}_2$}{orbifold} free-boson boundary states}
We verify the consistency conditions for the boundary states in the $S^1\!/\mathbb{Z}_2$ free-boson CFT through the following steps:
\begin{enumerate}
  \item Determine the coefficient $1/\sqrt{\abs{G}}$ of the regular boundary states from self-consistency.
  \item Fix the coefficient $1/\sqrt{\abs{G}}$ for the untwisted sector of the fixed point boundary states through mutual consistency with the regular states.
  \item Determine the coefficient $2^{-N/4}$ for the twisted sector of the fixed point boundary states from self-consistency.
\end{enumerate}

\subsection{Self-consistency of the regular boundary states}
For regular boundary states defined as
\begin{equation}
  \ket*{\mathcal{R}(\vec{\phi}_D,\vec{\theta}_D)}_{\mathrm{orb}} = \frac{1}{\sqrt{\abs{G}}}\sum_{a\in G}D(a)\ket*{\mathcal{R}(\vec{\phi}_D,\vec{\theta}_D)},
\end{equation}
self-consistency requires that the amplitude
\begin{align}
    & {}_{\mathrm{orb}}\mel*{\mathcal{R}(\vec{\phi}_D,\vec{\theta}_D)}{e^{-\frac{\beta}{2}H}}{\mathcal{R}(\vec{\phi}_D,\vec{\theta}_D)}_{\mathrm{orb}}\notag \\
  = & \frac{1}{\abs{G}}\sum_{a,a'}\mel*{\mathcal{R}(\vec{\phi}_D,\vec{\theta}_D)}{D(a')e^{-\frac{\beta}{2}H}D(a)}{\mathcal{R}(\vec{\phi}_D,\vec{\theta}_D)}
\end{align}
contains exactly one $h=0$ Virasoro character and nonnegative integer numbers of $h\neq0$ characters.
This follows from the consistency of the original $S^1$ free-boson boundary states $\ket*{\mathcal{R}(\vec{\phi}_D,\vec{\theta}_D)}$.
The $h=0$ character appears only when $a=a'$, while other characters appear in multiples of $\abs{G}$ due to the relation
\begin{align}
    & \mel*{\mathcal{R}'(\vec{\phi}_D',\vec{\theta}_D')}{D(a')e^{-\frac{\beta}{2}H}D(a)}{\mathcal{R}(\vec{\phi}_D,\vec{\theta}_D)}\notag  \\
  = & \mel*{\mathcal{R}'(\vec{\phi}_D',\vec{\theta}_D')}{D(a'a'')e^{-\frac{\beta}{2}H}D(aa'')}{\mathcal{R}(\vec{\phi}_D,\vec{\theta}_D)},
\end{align}
where $a''\in G$.
Together with the factor $\abs{G}^{-1}$, the self-consistency is confirmed.
The same argument can be applied to confirm mutual consistency among the regular boundary states.

\subsection{Mutual consistency between regular and fixed point states}
For mutual consistency between the regular states $\ket*{\mathcal{R}(\vec{\phi}_D,\vec{\theta}_D)}_{\mathrm{orb}}$ and the fixed point states $\ket*{\mathcal{R}'(\vec{\phi}_E',\vec{\theta}_E')}_{\mathrm{orb}}$, the amplitude
\begin{align}
    & {}_{\mathrm{orb}}\mel*{\mathcal{R}(\vec{\phi}_D,\vec{\theta}_D)}{e^{-\frac{\beta}{2}H}}{\mathcal{R}'(\vec{\phi}_E',\vec{\theta}_E')}_{\mathrm{orb}}\notag                                                                                                                                         \\
  = & \frac{1}{\abs{G}}\sum_{\substack{a\in G                                                                                                                   \\b\in G_0}}\mel*{\mathcal{R}(\vec{\phi}_D,\vec{\theta}_D)}{D(a)e^{-\frac{\beta}{2}H}D(b)}{\mathcal{R}'(\vec{\phi}_E',\vec{\theta}_E')}
\end{align}
must contain nonnegative integer numbers of Virasoro characters.
Note that amplitudes between untwisted and twisted sectors vanish.
For terms in $\sum_{a,b}$, pairs $(a,b)$ and $(-a,b)$ give identical contributions, yielding $\abs{G}$ pairs $\qty{(ab',bb'),(-ab',bb')}_{b'\in G_0}$ with equal contributions.
Thus, the mutual consistency is established.

\subsection{Self-consistency of the fixed point states}
To be concrete, we first explicitly construct the twisted sector boundary states.
The boson field $\vec{\phi}$ in the twisted sector can be mode expanded as follows:
\begin{align}
   & \vec{\phi}(x,t)\notag                                                                                                                                                      \\
   & =\vec{\phi}_0 + \sum_{\substack{r\in\mathbb{Z}{+}\frac{1}{2} \\ r>0}}\frac{1}{\sqrt{2\kappa r}}\qty(\vec{b}_r^L e^{-ik_r(x+t)} + \vec{b}_r^R e^{ik_r(x-t)} + \text{H.c.}),
\end{align}
with oscillator modes satisfying $[(\vec{b}_r^s)_i,(\vec{b}_u^t)_j]=\delta_{st}\delta_{ru}\delta_{ij}$.
Here $\vec{\phi}_0$ takes values in $\pi\Lambda_0$, where $\Lambda_0$ is the unit-cell of $\Lambda$:
\begin{equation}
  \Lambda_0=\qty{\vec{x}\;|\;\vec{x}=\sum_{i=1}^{N}\varepsilon_i\vec{e}_i,\;\varepsilon_i=0,1}.
\end{equation}
The Hamiltonian in the twisted sector is:
\begin{equation}
  H_t = \frac{2\pi}{L}\qty(\sum_{\substack{r\in\mathbb{Z}{+}\frac{1}{2}\\ r>0}} r[(\vec{b}_r^L)^\dag\cdot\vec{b}_r^L + (\vec{b}_r^R)^\dag\cdot\vec{b}_r^R] + \frac{N}{24}).
\end{equation}
The oscillator vacua are labeled either by the eigenvalues of $\vec{\phi}_0$ as $\ket*{\pi\vec{R}}_t\,(\vec{R}\in\Lambda_0)$ or by its dual $\vec{\theta}_0$ as $\ket*{\pi\vec{K}}_t\,(\vec{K}\in\Lambda^\ast_0)$, where $\Lambda_0^\ast$ is defined similarly to $\Lambda_0$.
These vacua are related by:
\begin{align}
  \ket*{\pi\vec{R}}_t & = \frac{1}{\sqrt{2^N}}\sum_{\vec{K}\in\Lambda^\ast_0}e^{-i\pi\vec{K}\cdot\vec{R}}\ket*{\pi\vec{K}}_t,\notag \\
  \ket*{\pi\vec{K}}_t & = \frac{1}{\sqrt{2^N}}\sum_{\vec{R}\in\Lambda_0}e^{i\pi\vec{K}\cdot\vec{R}}\ket*{\pi\vec{R}}_t.
\end{align}
The boundary states in the untwisted sector are eigenstates of the zero-mode operators $\mathcal{P}_D\vec{\phi}_0,\mathcal{P}_N\vec{\theta}_0$, so we wish to obtain the corresponding boundary states also in the twisted sector.
Since the compactification lattice $\Lambda$ is a square lattice, the state $\ket*{\pi\vec{R}}_t$ is in the form of a tensor product of the boundary states for each component: $\ket*{\pi\vec{R}}_t=\ket*{\pi R_1\varepsilon_1}\ket*{\pi R_2\varepsilon_2}_t\cdots\ket*{\pi R_N\varepsilon_N}_t$.
We can exchange the basis of this Hilbert space so that the state $\ket*{\pi\vec{R}}_t$ is expressed as a tensor product $\ket*{\pi\mathcal{P}_D\vec{R}}_t\ket*{\pi\mathcal{P}_N\vec{R}}_t$.
Then, we rotate the state in $\mathcal{V}_N$ to create the eigenstate of the operator $\mathcal{P}_N\vec{\theta}_D$ as
\begin{equation}
  \ket*{\pi\mathcal{P}_N\vec{K}}_t = \frac{1}{\sqrt{2^{\mathrm{dim}\mathcal{V}_N}}}\sum_{\vec{R}\in\mathcal{P}_N\Lambda_0}e^{i\pi\vec{K}\cdot\vec{R}}\ket*{\pi\mathcal{P}_N\vec{R}}_t.
\end{equation}
The boundary state $\ket*{\pi\mathcal{P}_D\vec{R}}_t\ket*{\pi\mathcal{P}_N\vec{K}}_t$ is now an eigenstate of the two operators $\mathcal{P}_D\vec{\phi}_0,\mathcal{P}_N\vec{\theta}_0$ as we wished.
Then, the boundary state $\ket*{\mathcal{R}(\vec{\phi}_E,\vec{\theta}_E)}_t$ in the twisted sector corresponding to the fixed point boundary state $\ket*{\mathcal{R}(\vec{\phi}_E,\vec{\theta}_E)}$ ($\vec{\phi}_E\in\pi\Lambda_0\cap\mathcal{V}_E,\vec{\theta}_E\in\pi\Lambda_0^\ast\cap\mathcal{V}_N$) in the untwisted sector is expressed as
\begin{equation}
  \ket*{\mathcal{R}(\vec{\phi}_E,\vec{\theta}_E)}_t = S_t(\mathcal{R})\ket*{\vec{\phi}_E}_t\ket*{\vec{\theta}_E}_t,
\end{equation}
where
\begin{equation}
  S_t(\mathcal{R}) = \exp(-\sum_{r}(\vec{b}_r^L)^\dag\cdot\mathcal{R}(\vec{b}_r^R)^\dag)
\end{equation}
is the squeezing operator in the twisted sector.
The group $G$ acts on the twisted sector in the same way as it acts on the untwisted sector boundary states: $D(a)\ket*{\mathcal{R}(\vec{\phi}_E,\vec{\theta}_E)}_t = \ket*{a\mathcal{R}a(a\vec{\phi}_E,a\vec{\theta}_E)}_t$.

We now confirm that the fixed point boundary states of the $S^1\!/\mathbb{Z}_2$ free boson (Eq.~\eqref{eq:orbifold_boundary_fixed}) satisfy the self-consistency condition.
The amplitude of the twisted sector boundary state with itself reads
\begin{align}
    & {}_t\mel*{\mathcal{R},\vec{\phi}_E,\vec{\theta}_E}{e^{-\frac{\beta}{2}H_t}}{\mathcal{R},\vec{\phi}_E,\vec{\theta}_E}_t\notag \\
  = & q^{N/48}\qty(\prod_{n=1}^{\infty}\frac{1}{1-q^{n-1/2}})^N\notag                                                              \\
  = & \qty(\frac{\theta_2(q^{1/2})}{2\eta(q)})^N\notag                                                                             \\
  = & \qty(\frac{\theta_4(\tilde{q}^2)}{\sqrt{2}\eta(\tilde{q})})^N = 2^{-N/2}\chi_0(\tilde{q})+\cdots .
\end{align}
Since this amplitude contains the $h=0$ Virasoro character $2^{-N/2}$ times, the amplitude of the fixed point boundary state~\eqref{eq:orbifold_boundary_fixed} with itself contains the $h=0$ Virasoro character $\abs{G}^{-1}\times \abs{G_0}=1/2$ times within the untwisted sector and $2^{-N/2}\times \abs{G_0}\times 2^{-N/2}=1/2$ times within the twisted sector.
Therefore, the contributions from both sectors sum to unity, and the self-consistency is confirmed.

\begin{figure}[tb]
  \centering
  \includegraphics[width=\linewidth, clip]{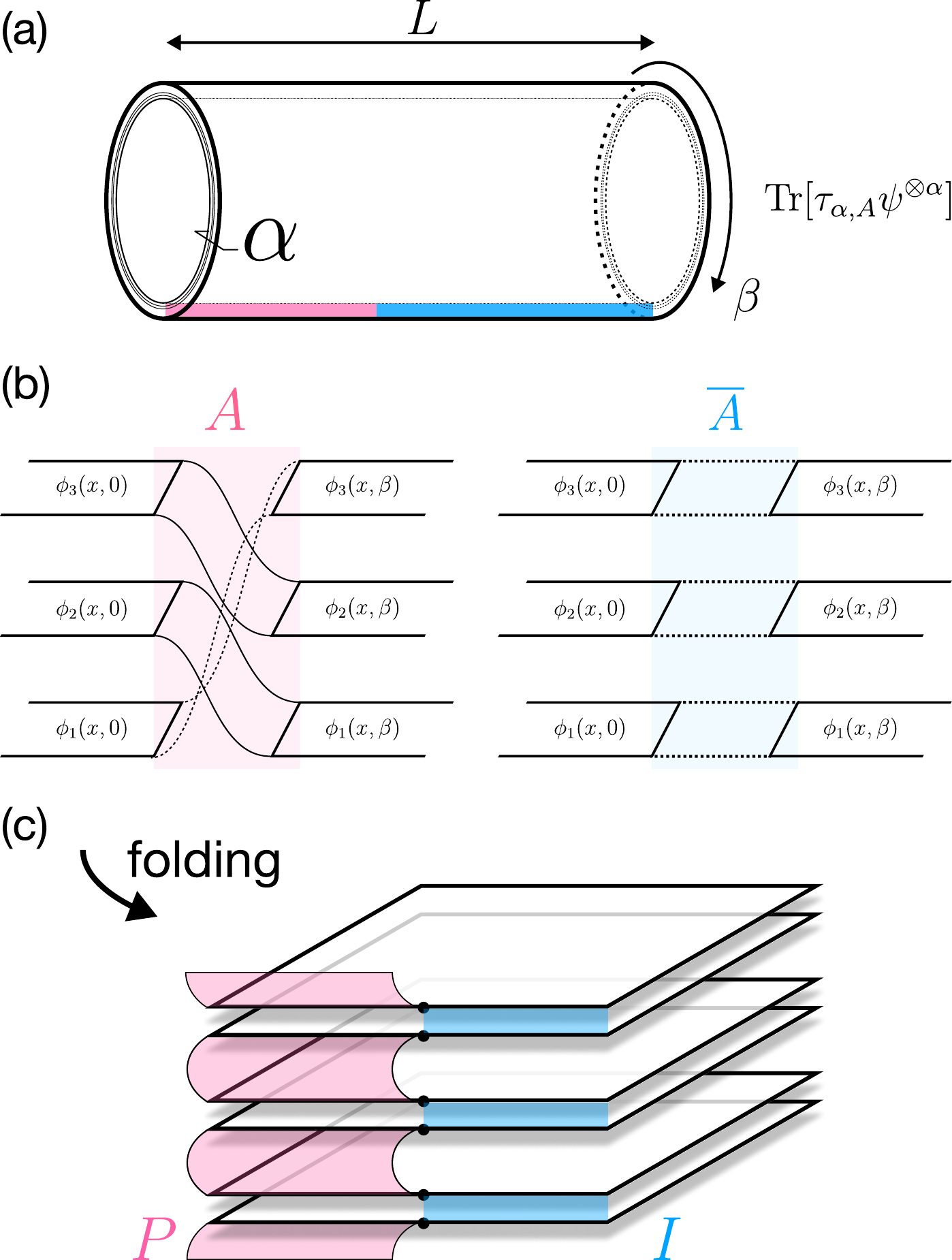}
  \caption{(a)~The partition function of the $\alpha$-component CFT with the line defect imposed by the permutation operator. The figure only shows periodicity in the imaginary time direction. (b)~The permutation operator creates a topological defect by connecting neighboring components. (c)~After the folding trick, the line defect becomes a boundary. We name the boundary from the nontrivial topological defect in subsystem $A$ as $P$, and the trivial one as $I$.}
  \label{fig:ee_defect}
\end{figure}

\section{\label{appendix:comparison_ee}Comparison with entanglement entropy}
As mentioned in Sec.~\ref{sec:unified}, the entanglement entropy (EE) and the stabilizer Rényi entropy (SRE) share similar structural features in their definitions.
This section provides a detailed comparison between these two quantities.

The EE can be calculated using the replica trick as follows:
\begin{equation}\label{eq:ee_def}
  S_{A}^{(\alpha)}(\psi) = \frac{1}{1-\alpha}\ln\Tr[\tau_{\alpha,A}\psi^{\otimes \alpha}].
\end{equation}
Here, $\tau_{\alpha,A}$ denotes a cyclic permutation operator that acts exclusively on subsystem $A$.
Permutation operators are unitary transformations and, according to Schur-Weyl duality, constitute the unique class of operators that commute with arbitrary unitary operations.
Consequently, permutation operators form a complete basis for the commutant of the full unitary group.

The universal behavior of the EE is determined by the line defect introduced by the permutation operator.
Since the permutation operator $\tau_{\alpha,A}$ commutes with the Hamiltonian of the replicated theory $\psi^{\otimes\alpha}$, the resulting line defect is a topological defect, meaning that the energy-momentum tensors are continuous across the defect interface~\cite{quella2007reflection}.
Based on physical considerations, we can identify the topological defect in the infrared limit as the configuration depicted in Fig.~\ref{fig:ee_defect}.
The defect can be expressed in terms of local field operators $\phi_i$ $(i=1,2,\ldots,\alpha)$ through the boundary condition $\phi_i(x,\beta)=\phi_{i+1}(x,0)$.
It simply connects the neighboring components by an identity map.
The topological character of this defect is confirmed by the relation $T_i(x,\beta)=T_{i+1}(x,0)$, where $T_i$ denotes the energy-momentum tensor of the $i$-th replica, leading to
\begin{equation}
  T_{\text{tot}}(x,\beta) := \sum_iT_i(x,\beta) = \sum_iT_i(x,0) =: T_{\text{tot}}(x,0).
\end{equation}
Under the folding trick, this topological defect maps to what is known as a permutation brane~\cite{recknagel2003permutation}.

\begin{figure}[tb]
  \centering
  \includegraphics[width=0.9\linewidth, clip]{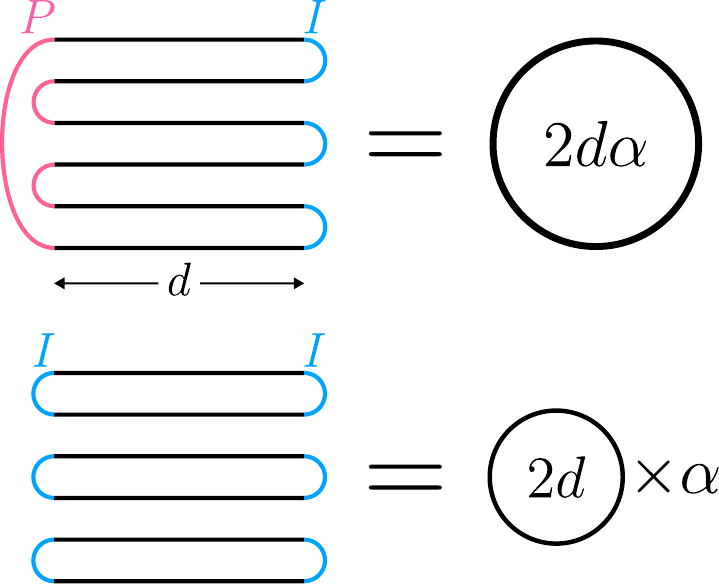}
  \caption{(Top)~The ground-state energy $E_{PI}^0(d)$ of the $2\alpha$-component CFT on a strip of width $d$ with boundaries $P$ and $I$ on each end is equivalent to the Casimir energy of a single-component CFT on a cylinder of circumference $2d\alpha$. (Bottom)~The ground-state energy $E_{II}^0(d)$ on a strip of width $d$ with boundaries $I$ on both ends is equivalent to $\alpha$ times the Casimir energy of a single-component CFT on a cylinder of circumference $2d$.}
  \label{fig:bcco}
\end{figure}

After identifying the conformal boundary induced by the permutation operator $\tau_{\alpha,A}$, we can determine the universal scaling of the EE by calculating the scaling dimension of the boundary-condition-changing operator (BCCO) between the permutation brane ($P$) and the trivial boundary ($I$).
The scaling dimension $\Delta_{PI}$ of the BCCO is related to the ground-state energy of the CFT on a strip geometry with boundary conditions $P$ and $I$ imposed on opposite sides.
Specifically, we employ the following relation:
\begin{equation}\label{eq:bcco}
  \frac{\pi\Delta_{PI}}{d} = E_{PI}^0(d) - E_{II}^0(d),
\end{equation}
where $d$ denotes the width of the strip geometry, and $E_{PI}^0(d)$ ($E_{II}^0(d)$) represents the ground-state energy of the CFT on the strip geometry with boundary conditions $P,I$ ($I,I$) at the two ends~\cite{affleck1997boundary}.
The energy $E_{PI}^0(d)$ reduces to the Casimir energy of a cylinder geometry with circumference $2d\alpha$, as illustrated in Fig.~\ref{fig:bcco}.
This reduction occurs because the boundaries $P$ and $I$ connect the replicas, effectively transforming the strip into a cylinder.
Similarly, $E_{II}^0(d)$ equals $\alpha$ times the Casimir energy of a cylinder with circumference $2d$.
Therefore, the scaling dimension can be calculated from Eq.~\eqref{eq:bcco} as
\begin{equation}\label{eq:bcco_scalingdim}
  \Delta_{PI} = \frac{d}{\pi}\qty(-\frac{\pi}{6}\frac{c}{2d\alpha} -\qty(-\frac{\pi}{6}\frac{c\alpha}{2d})) = \frac{c}{12}\qty(\alpha-\frac{1}{\alpha}).
\end{equation}
Here, the Casimir energy of a cylinder with circumference $L_{\text{cyl}}$ is given by $E^0 = -(\pi/6)(c/L_{\text{cyl}})$.
Since the partition function can be expressed as a two-point correlation function of the BCCO, the logarithmic scaling of the EE follows as:
\begin{align}
  \Tr[\tau_{\alpha,A}\psi^{\otimes \alpha}] & \propto \langle \mathcal{B}_{PI}(l)\mathcal{B}_{PI}(0)\rangle \sim l^{-2\Delta_{PI}},          \\
  S_A^{(\alpha)}(\psi)                      & \sim \frac{1}{1-\alpha}\times (-2\Delta_{PI})\ln l = \frac{c}{6}\qty(1+\frac{1}{\alpha})\ln l.
\end{align}
Here, $l$ represents the length of subsystem $A$, and we have used Eqs.~\eqref{eq:ee_def} and \eqref{eq:bcco_scalingdim} in the final expression.

We now compare the above results for the EE with the case of the SRE, which can also be formulated using the replica trick:
\begin{equation}
  M_{\alpha}(\psi) = \frac{1}{1-\alpha}\ln\Tr[\Omega_{2\alpha}\psi^{\otimes 2\alpha}].
\end{equation}
Here, $\Omega_{2\alpha} = 2^{-L}\sum_{\vec{m}}(\sigma^{\vec{m}})^{\otimes 2\alpha}$ belongs to the $2\alpha$-order commutant of the Clifford group.
It has been proven that the complete commutant of the Clifford group is generated by this operator $\Omega_{2\alpha}$ (with $\alpha=2,3$) together with the permutation operators that also span the commutant of the full unitary group~\cite{bittel2025complete}.
Therefore, the SRE can be viewed as a natural extension of the EE obtained by restricting the unitary group with which the operators must commute to the Clifford group, thereby revealing more detailed structure through the nontrivial commutant $\Omega_{2\alpha}$.
In contrast to the permutation operators, the Clifford commutant $\Omega_{2\alpha}$ generally does not commute with the Hamiltonian, and consequently the resulting line defect is not topological.
As demonstrated in Sec.~\ref{sec:sre-in-criticalstates}, the line defect is also not factorizing.
This represents the key contribution of our work: it provides practical methods for determining the nontrivial line defect imposed by the Clifford commutant (or equivalently, the Bell-state measurements), which exhibits neither topological nor factorizing character.

\bibliography{HOA2025}

\end{document}